\pdfoutput=1 
\documentclass[ 
 aps,%
 twocolumn,%
 reprint,%
 prb,%
 groupedaddress,%
 superscriptaddress,%
 showpacs,%
 lengthcheck,%
 amsfonts,%
 floatfix,%
 citeautoscript
]{revtex4-1} 

\RequirePackage{amsmath,amssymb,bm}
\RequirePackage{graphicx}
\RequirePackage[normalem]{ulem}
\RequirePackage[]{subfigure}
\RequirePackage{hyperref}
\hypersetup{%
  breaklinks = {true},
  citecolor = {blue},
  colorlinks = {true},
  linkcolor = {red},
  pdfauthor = {\textcopyright\ F\'{a}bio Hip\'{o}lito },
  pdfcreator = {\LaTeX\ and \flqq hyperref\frqq},
  pdffitwindow = {true},
  pdfmenubar = {true},
  pdfpagelayout = {SinglePage},
  pdfstartview = {Fit},
  pdftoolbar = {true},
  plainpages = {false},
}

\allowdisplaybreaks

\newcommand{\ie}{i.e.~}
\newcommand{\eg}{e.g.~}

\newcommand{\cf}{cf.~}

\newcommand{\hatH}{\hat{H}}

\newcommand{\Fref}[1]{Fig.~\ref{#1}}
\newcommand{\Fsref}[1]{Figs.~\ref{#1}}
\newcommand{\Pref}[1]{(\ref{#1})}
\newcommand{\Eqref}[1]{Eq.~(\ref{#1})}
\newcommand{\Eqsref}[1]{Eqs.~(\ref{#1})}
\newcommand{\Tbref}[1]{Tab.~\ref{#1}}

\newcommand{\bAcal}{\boldsymbol{\mathcal{A}}}

\newcommand{\bNabla}{\boldsymbol{\nabla}}

\newcommand{\teq}{{\,=\,}}

\newcommand{\tequiv}{{\,\equiv\,}}

\DeclareMathOperator{\Tr}{Tr}

\newcommand{\sump}{\sideset{}{'}\sum}

\newcommand{\Rcite}[1]{Ref.~\onlinecite{#1}}
\newcommand{\Rscite}[1]{Refs.~\onlinecite{#1}}

\usepackage[usenames,dvipsnames]{xcolor}

\graphicspath{ {./Figs/} } 

\begin{document}

\title{Nonlinear optical response of doped mono- and bilayer graphene: length 
gauge tight-binding model}

\author{F. Hipolito}

\email{fh@nano.aau.dk}

\affiliation{Department of Physics and Nanotechnology,
Aalborg University, DK-9220 Aalborg {\O}st, Denmark}

\author{Alireza Taghizadeh}

\affiliation{Department of Physics and Nanotechnology,
Aalborg University, DK-9220 Aalborg {\O}st, Denmark}

\author{T. G. Pedersen}
\email{tgp@nano.aau.dk}
\affiliation{Department of Physics and Nanotechnology,
Aalborg University, DK-9220 Aalborg {\O}st, Denmark}
\affiliation{Center for Nanostructured Graphene (CNG), 
DK-9220 Aalborg {\O}st, Denmark}

\begin{abstract}
We compute the nonlinear optical response of doped mono- and bilayer graphene 
using the full dispersion based on tight-binding models.
The response is derived with the density matrix formalism using the length 
gauge and is valid for any periodic system, with arbitrary doping.
By collecting terms that define effective nonlinear response tensors, we 
identify all nonlinear Drude-like terms (up to third-order) and show that all 
additional spurious divergences present in the induced current vanish.
The nonlinear response of graphene comprises a large Drude-like divergence and 
three resonances that are tightly connected with transitions occurring in the 
vicinity of the Fermi level.
The analytic solution derived using the Dirac approximation captures accurately 
the first- and third-order responses in graphene, even at very high doping 
levels.
The quadratic response of gapped graphene is also strongly
enhanced by doping, even for systems with small gaps such as commensurate 
structures of graphene on SiC.  
The nonlinear response of bilayer graphene is significantly richer, combining 
the resonances that stem from doping with its intrinsic strong low-energy 
resonances.
\end{abstract}



\maketitle

\section{Introduction}

The interaction of intense light with matter encompasses a wide range of 
phenomena with many applications in nonlinear optics \cite{Shen2002, Boyd2008}, 
over a large portion of the energy spectrum.
Recent developments in the production and characterization of 2D materials 
have led to intense experimental study of nonlinear optical phenomena, 
including multiple wave mixing processes \cite{Gu2012, Hendry2010, 
Mikhailov2007, Ishikawa2010, Avetissian2012, Wu2015, Chizhova2016}, 
harmonic generation \cite{Mikhailov2008, Hong2013, Kumar2013, Li2013a, 
Zeng2013, Avetissian2013, Janisch2014, Wang2015a, Yin2014a, Clark2014, 
Al-Naib2015, Chizhova2017, Hipolito2017}, and optical rectification (OR) 
\cite{McIver2012}.
From a theoretical point of view, several methods can be used to evaluate 
the nonlinear optical response functions, but frequently generate contradictory 
results, as discussed in \Rcite{Taghizadeh2017a, Taghizadeh2018}.
Several of these differences can be traced to approximations in the 
calculations, such as the truncation of the Hamiltonian basis size, or the 
choice of gauge. \cite{Aversa1995, Taghizadeh2017a, Taghizadeh2018}
Methods derived using the length gauge (LG) have been shown to be 
least sensitive to the truncation of the basis set compared to the conventional
velocity gauge (VG) \cite{Taghizadeh2017a}.
Hence, the LG offers an accurate estimate of the nonlinear response 
even in calculations truncated to just two bands.
In addition, the choice of response function can also lead to different 
results, as identified in \Rscite{Aversa1995, Taghizadeh2017a}.
For instance, \Rcite{Aversa1995} shows that a direct evaluation of the 
third-order current density response function of cold insulators is plagued by 
several unphysical divergences, while the polarization density counterpart is 
regular.

In the present work, we compute the linear and nonlinear optical response of
doped mono- and bilayer graphene using the LG formalism.
For the third-order nonlinearity in monolayer graphene (MLG), 
we compare results obtained from the full band structure to those found within 
the Dirac approximation.
We begin by deriving expressions for the current density response showing that 
the unphysical divergences in the response of cold insulators \cite{Aversa1995} 
are spurious and can be removed for all nonlinear processes up to third-order, 
independently of the symmetry of the crystal.
Moreover, we expand our previous formalism in \Rcite{Hipolito2018} at several
levels.
First, we consider the single-particle nonlinear response valid for general
crystalline systems with an arbitrary number of bands, rather than particular 
solutions valid for two-band models.
Second, the expressions shown in the present work are valid for any wave mixing 
process up to third-order, rather than just the particular case of third
harmonic generation (THG).
Third, we demonstrate how, for arbitrary doped or intrinsic systems, it is
possible to remove all spurious divergences previously observed in the 
evaluation of the conductivity tensor for cold insulators \cite{Aversa1995} and 
identify nonlinear features analogous to the Drude peak that should be present 
in the response of doped semiconductors or metallic systems.
Within the Dirac approximation, the first- and third-order LG results for MLG 
can be evaluated analytically
and reproduce the previously identified logarithmic divergences 
\cite{Cheng2014, Mikhailov2014, Soavi2017}.
By comparison with full dispersion tight-binding (TB), the Dirac approximation 
is shown to accurately capture the third-order response, even for highly doped 
systems, with Fermi level up to $\mu = 1.5$ eV, that exceed current 
experimental reports.

We provide general expressions for nonlinear optical response using the LG and 
gauge invariant generalized derivatives, that 
are not limited by:
i)   particular solutions tailored for two-band systems, such as the Dirac 
approximation for MLG, truncated Hamiltonians for biased bilayer graphene (BBG);
ii)  cold semiconductors approximations;
iii) lattice symmetry restrictions.
Considering full dispersion TB models, our expressions can be
used to probe the optical response at higher energies, including transitions 
with bands farther away from the Fermi level, of 
paramount importance to the nonlinear response of BBG, even at the 
energy scale of $\hbar\omega\sim 150$ meV.
All even-order nonlinear response functions vanish (in the dipole 
approximation) in centrosymmetric structures. A dipole-allowed even-order 
response in graphene-based systems can be obtained by \eg rolling the material 
into a chiral nanotube \cite{Pedersen2009}.
Instead, in the present study, we consider commensurate structures of graphene
on SiC or hBN substrates \cite{Zhou2007, Riedl2009, Woods2014}. Such 
substrates break the sublattice symmetry of the two atoms in the graphene 
unit-cell leading to broken centrosymmetry and opening of a band gap at the
Dirac points. However, regardless of the broken symmetry the Dirac 
approximation still predicts vanishing even-order responses due to the full 
rotation symmetry of the Dirac Hamiltonian\cite{Brun2015}.
In contrast, by using TB models that capture the reduced symmetry, our 
expressions can be used to correctly evaluate the quadratic response of such 
gapped systems.

We present results for the optical conductivity, THG, optical Kerr effect, and 
second harmonic generation (SHG).
Regarding the latter, we consider the effect of doping on the quadratic 
response of non-centrosymmetric systems, such as graphene on SiC or hBN 
substrates \cite{Zhou2007, Riedl2009, Woods2014} and hydrogenated 
graphene \cite{Elias2009, Balog2010, Son2016}.

\section{Theoretical framework}

The interaction of light with the electrons is treated in the 
dipole approximation and we, therefore, ignore the position-dependence 
of the electromagnetic field.
We do not consider electron-electron interaction, \ie 
excitonic effects, and therefore the many-body effects arise from the 
Fermi-Dirac statistics only.
Hence, the total single-particle Hamiltonian reads
\begin{equation}
\hat{\mathcal{H}} = \hat{\mathcal{H}}_0 + e \, \mathbf{r} \cdot \mathbf{E}(t)\,,
\end{equation}
where $\hat{\mathcal{H}}_0$ denotes the unperturbed Hamiltonian of the 
crystal and $e>0$ is the elementary charge.
The electromagnetic field $\mathbf{E}(t)$ is a linear combination of
monochromatic fields restricted to propagate along the $z$ axis (normal to the 
crystal plane)
\begin{equation}
\label{eq:E}
 \mathbf{E}(t) = \sum_{\alpha, \omega_i} \Big[
  E_{\omega_i}^\alpha e^{-i\bar\omega_i t}
 +E_{-\omega_i}^\alpha e^{ i\bar\omega_i^* t } \Big] 
 \, \mathbf{e}_\alpha / 2  \, ,
\end{equation}
where the polarization plane is taken as the $xOy$ plane.
Throughout this paper, sub- or superscripts using the greek alphabet 
$\{ \alpha, \beta, \lambda, \phi \}$ represent the spatial coordinates 
$\{x,y,z\}$.
Furthermore, the adiabatic coupling of the interaction is ensured by the 
analytic continuation of the photon frequency $\bar\omega \equiv \omega +i\eta$ 
\cite{Fetter1971}.
The diagonalization of the unperturbed periodic Hamiltonian provides the 
crystal 
band dispersions $\epsilon_m(\mathbf{k})$ and respective eigenstates 
$|m\mathbf{k}\rangle$, which serve as the basis for the calculation of the 
response function. Here, $m$ and $\mathbf{k}$ denote band index and electron
wave vector, respectively. 
The calculation is based on the time-dependent density 
matrix, $\hat\rho (t) \tequiv \sum_\mathbf{k} \sum_{mn} \rho_{mn}(\mathbf{k}) 
|m\mathbf{k}\rangle\langle n\mathbf{k}|$, that obeys the quantum Liouville 
equation
$i\hbar \, \partial \hat \rho / \partial t \teq \big[ \hatH, \hat\rho \big ]$, 
which lends itself to a perturbative expansion.
Excitonic effects have been shown to play an important role in the optical 
response of 2D materials, particularly in systems with large gaps such as
hexagonal boron-nitride (hBN) \cite{Wirtz2006, Gruning2014, Pedersen2015}, the 
vast class of transition-metal dichalcogenides \cite{Wagoner1998, 
Splendiani2010, Zeng2012, Jiang2014, Gruning2014, Wang2015c, Trolle2015}, 
few-layered black phosphorus, \cite{Wang2015c, Li2016a, Zhang2017} and many 
others.
Excitonic effects in the optical response usually manifest themselves
as a red-shift of the response onset and significant transfer of spectral 
weight to bound excitons.
In pristine suspended graphene, the electron-electron interaction can be 
important due to low screening of interactions, leading to potentially 
significant excitonic effects on the spectrum of graphene and on its low-energy 
linear \cite{Elias2011, Park2009, Hwang2011, Grushin2009, Peres2010b, 
Stroucken2011, Breusing2011, Sun2012, Malic2011, Sule2014, Vasko2012} and 
nonlinear \cite{Avetissian2018, Avetissian2018a} optical response.
Conversely, the presence of substrates or encapsulation of graphene increases 
the screening of interactions rendering the excitonic effects negligible for 
MLG and in commensurate systems of graphene on SiC or hBN.
Moreover, electron-electron interactions can cause additional effects, such as 
the renormalization of the low-energy band structure, \cite{Elias2011, Yu2013, 
Stauber2017} leading to further corrections to the low energy optical response. 
\cite{Grushin2009, Peres2010b, Malic2011, Stroucken2011, Breusing2011, 
Sun2012a, Peres2010, Stauber2017}

In BBG, the potentially larger gaps can give rise to moderate 
manifestations of excitonic coupling \cite{Park2010, Ju2017} in charge neutral 
systems.
The presence of free carriers in doped systems prompts strong screening of 
electron-electron interactions, and consequently the manifestations of 
excitonic effects should be at least softened, if not removed altogether.
Therefore, single-particle calculations of the response of doped MLG and 
BBG offer a sound description of the optical response 
\cite{Horng2011, Stauber2008, Stauber2008a, Stauber2008b, Peres2008a, 
Zhang2008}.

\subsection{Optical response of multi-band systems}
We evaluate the optical response to an external electromagnetic field based on 
the current density response
$\mathbf{J}(t) = \Tr\big[ \hat{\rho}(t) \hat{\mathbf{j}} \big] $.
The current density operator
$\hat{\mathbf{j}} \equiv -ge \, \hat{\mathbf{v}}/\Omega$ is then defined in 
terms of the single-particle velocity
$ \hat{ \mathbf{v} } = i[ \hat{H}_0, \hat{ \mathbf{r} } ]/\hbar$, spin 
degeneracy $g=2$, and the $D$-dimensional volume of the system $\Omega$.
The integration of the equation of motion of the density matrix is based on the 
LG formalism proposed in \Rcite{Aversa1995}.
In this approach, the equation of motion of the density matrix reads
\begin{align}
i\hbar \frac{ \partial \rho_{mn} }{ \partial t } &= 
\epsilon_{mn} \rho_{mn} +i e( \rho_{mn} )_{;\mathbf{k}} \cdot \mathbf{E}(t)
\nonumber\\ &+
e\sum_l \big[ 
 \bar\delta_{ml} \bAcal_{ml} \rho_{ln} 
- \bar\delta_{ln} \bAcal_{ln} \rho_{ml} \big] \cdot \mathbf{E}(t) \, ,
\end{align}
where we make use of the ``generalized derivative''
$ ( S_{mn} )_{;k_\alpha} = ( S_{mn} )_{;\alpha} \equiv 
\partial S_{mn} /\partial k_\alpha
-i S_{mn} ( \mathcal{A}_{mm}^\alpha -\mathcal{A}_{nn}^\alpha ) $,
$\bar\delta_{mn} \equiv 1-\delta_{mn} $, and the energy dispersion 
differences $\epsilon_{mn} \equiv \epsilon_m -\epsilon_n$.
To simply notation, we omit the explicit $\mathbf{k}$ dependence on all 
variables, \eg the density matrix $\rho_{mn} \equiv \rho_{mn}(\mathbf{k})$. 
Moreover, the matrix elements of the Berry connection in periodic systems read
\begin{align}
\bAcal_{mn} = \frac{ i }{ \Omega_C } \int_{\Omega_C} \mathrm{d} \mathbf{r}
u_{m\mathbf{k}}^*( \mathbf{r} ) \bNabla_\mathbf{k}
u_{n\mathbf{k}}  ( \mathbf{r} )
\end{align}
with cell-periodic functions $u_{m\mathbf{k}}( \mathbf{r} )$ \cite{Aversa1995, 
Hipolito2016} and cell volume $\Omega_C$.
For details regarding calculation of the perturbative solution for the density 
matrix we refer to the extensive literature \cite{Aversa1995, 
Mikhailov2014, Cheng2014, Pedersen2015, Cheng2015b, Hipolito2016, 
Mikhailov2016, Hipolito2018, Mikhailov2017, Ventura2017, Taghizadeh2017a, 
Dimitrovski2017a, McGouran2017}.
We follow the procedure and notation outlined in \Rcite{Hipolito2016} and 
present the relevant results for the first-, second-, and third-order terms of 
the density matrix in \Eqsref{eq:rho:1}, \eqref{eq:rho:2}, and 
\eqref{eq:rho:3}, respectively.
We then find the optical response (linear and nonlinear) by evaluating the 
$n^\mathrm{th}$-order current density
\begin{align}
\label{eq:T}
j_\phi^{(n)} (t) &=
\sum_{\omega_n\ldots\omega_1} \sum_{\lambda\ldots\alpha}
\sigma_{\phi\lambda\ldots\alpha}^{(n)}(\omega_n+\ldots+\omega_1)
\nonumber \\ &\times
E_{\omega_n}^\lambda \ldots E_{\omega_1}^\alpha
e^{-i(\bar\omega_n +\ldots+ \bar\omega_1)t} \, .
\end{align}
The final expressions for the response functions are rather cumbersome
containing various combinations of intraband ($i$) and interband ($e$) 
transitions.
We consequently relegate the full expressions to the appendix, in which 
conductivities up to third-order can be found in \Eqsref{eq:s1}, \eqref{eq:s2}, 
and \eqref{eq:s3}.

\subsection{\texorpdfstring{$\pi$}{}-electron tight-binding}
\label{sec:TB}
\begin{figure}
\includegraphics[width=1.00\linewidth]{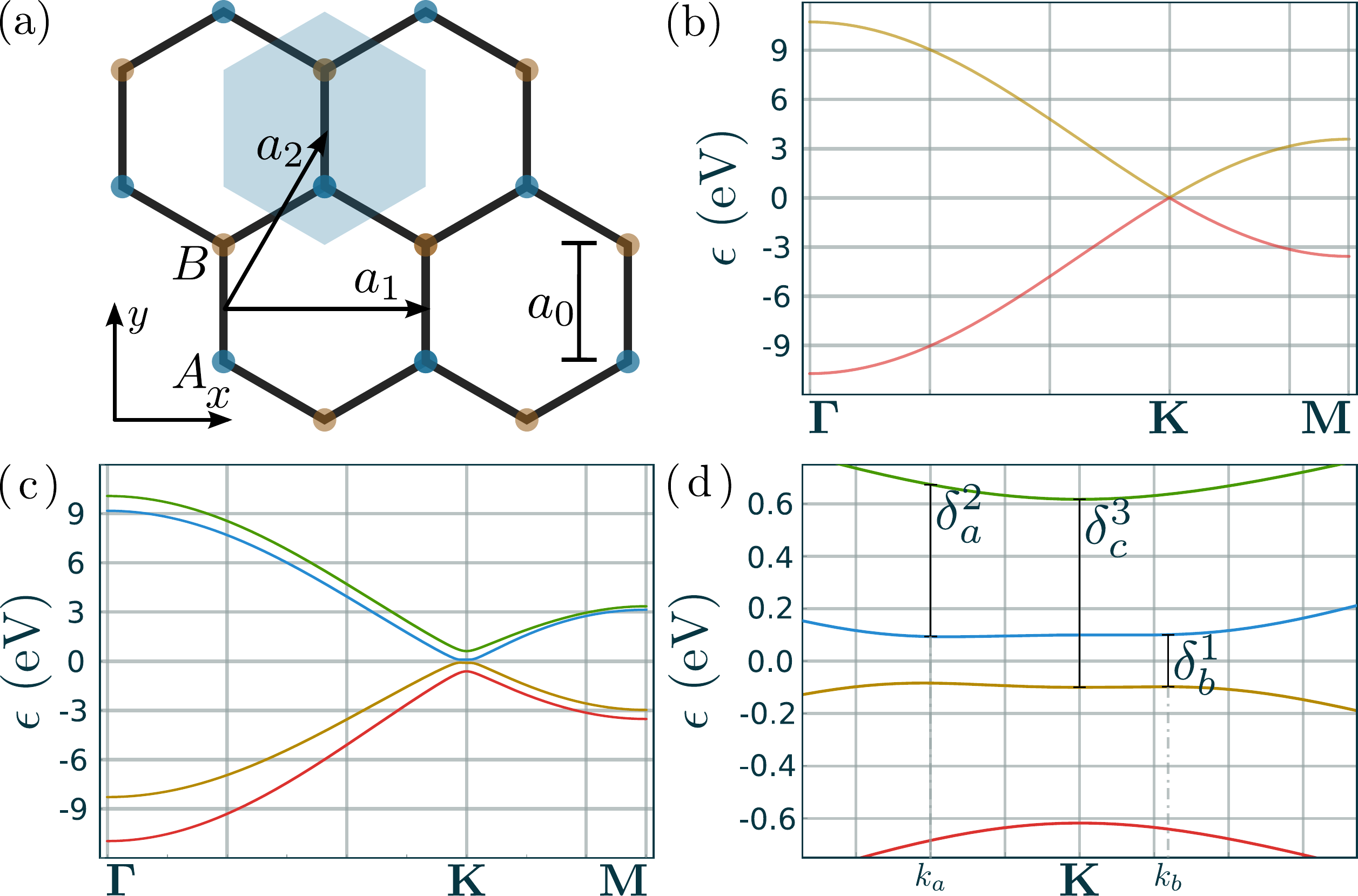}
\caption[Lattice representation and energy dispersions]{\label{fig:1}%
(a) Representation of the monolayer lattice, including the primitive vectors 
and Wigner-Seitz cell (light blue hexagon).
Energy dispersions for MLG (b), and BBG (c) and (d).
Plots (c) and (d) consider BBG with external bias potential $\Delta = 200$ 
meV. 
In (d), we indicate some of the energy differences $\delta_l^i$ occurring at 
the van Hove singularities (vHs) between the $i$th pair of bands, as listed in 
\Tbref{tab:AB:res}.}
\end{figure}

The low-energy electronic properties of graphene systems with an underlying
honeycomb lattice, see \Fref{fig:1}a, can be characterized by orthogonal TB 
models that include a $p_z$ orbital per atom in the unit-cell.
%
%
%
%
In this context, the Hamiltonian operator $ \hat H = \sum_\mathbf{k} 
\Psi_\mathbf{k}^\dagger h_\mathbf{k} \Psi_\mathbf{k}$ lends itself to a simple 
representation in crystal momentum, where $\Psi_\mathbf{k}^\dagger $ represents 
the Fourier-transformed electron creation operators at different sites in the 
unit cell.
The Fourier transforms of the Hamiltonians for MLG and AB-stacked BBG read
\begin{subequations}
\label{eq:H}
\begin{align}
\label{eq:H:ML}
h_\mathbf{k}^\mathrm{MLG} &\equiv
\begin{pmatrix}
 -\delta/2      & \gamma_0 f \\
  \gamma_0 f^*  & \delta/2
\end{pmatrix} \, ,
\\
\label{eq:H:AB}
h_\mathbf{k}^\mathrm{BBG} &\equiv
\begin{pmatrix}
 -\Delta/2      &  \gamma_0 f    & \gamma_4 f    & \gamma_3 f^* \\
  \gamma_0 f^*  & -\Delta/2      & \gamma_1      & \gamma_4 f   \\
  \gamma_4 f^*  &  \gamma_1      & \Delta/2      & \gamma_0 f   \\
  \gamma_3 f    &  \gamma_4 f^*  & \gamma_0 f^*  & \Delta/2
\end{pmatrix} \, ,
\end{align}
\end{subequations}
where $f \equiv \exp( i k_y a_0 )
+2\exp( -i k_y a_0/2 ) \cos( \sqrt{3} k_x a_0/2 )$, and nearest neighbor 
distance $a_0 = 1.42$ \AA{}.
The hopping integral for graphene is taken as $\gamma_0 = -3.0$ eV 
\cite{Mucha-Kruczynski2008, Zhang2008, Gruneis2008, Kuzmenko2009a}
while parametrization of TB models for the AB stacked bilayer graphene has been 
an issue of intense research \cite{Partoens2006, McCann2006, Malard2007, 
Mucha-Kruczynski2008, Zhang2008, Gruneis2008, Kuzmenko2009, Kuzmenko2009a, 
Li2009, CastroNeto2009, McCann2013, Cheng2015b}.
In this work, we follow the parametrization obtained from ARPES data in 
\Rcite{Cheng2015b}: $\gamma_0 = -3.21$  eV, $\gamma_1 = 0.61$  eV,
$\gamma_3 = 0.39$ eV, and $\gamma_4 = 0.15$ eV, where $\gamma_1$, $\gamma_3$ 
and $\gamma_4$ are the interlayer hopping integrals.
The presence of an on-site potential $\delta$ in MLG and an
interlayer bias $\Delta$ in BBG opens a gap in the respective energy 
dispersions.
Estimates range from very small gaps for graphene-hBN structures and 
substrates, $E_g =31$ meV, to larger $\sim 260$ meV in SiC substrates and 
$\sim500$ in hydrogenated graphene \cite{Zhou2007,Balog2010}.
Hence, for pristine MLG we take $\delta = 0$, while $\delta = E_g$ in the 
gapped case.
In BBG, the gap can be tuned by electrostatic gating, \cite{Castro2007, 
CastroNeto2009} providing an additional mechanism to control the optical 
properties of the material.
In \Fsref{fig:1}b and~\ref{fig:1}c, we show the band structures along the 
relevant high symmetry paths.
For the bilayer, we display the energy dispersion for a biased system ($\Delta 
= 200$ meV), highlighting the possible vertical transitions occurring at the 
low-energy van Hove singularities (vHs).
The trigonal warping of the dispersion is amplified by the presence of finite 
interlayer hopping integrals $\gamma_3$ and $\gamma_4$ that shift the band gap 
along the high symmetry path $\overline{\boldsymbol{\Gamma}\mathbf{K}}$.
As a reference, the bias $\Delta$ dependent energy differences highlighted in 
\Fref{fig:1}d are listed 
in \Tbref{tab:AB:res}.
%
%
\begin{table}[t]
\caption[Resonances in BBG]{\label{tab:AB:res}%
Energy differences $\delta_l^i$ between the $i$th pair of bands $(m,n)$ 
at the $k_l$ vHs in the BBG, with external bias potential set at $\Delta = 
200$ meV.
The three relevant low-energy vHs are found along the high symmetry paths 
$\overline{\boldsymbol{\Gamma}\mathbf{K}}$ and 
$\overline{\mathbf{K}\mathbf{M}}$, respectively $k_a$ and $k_b$, and at the 
high symmetry point $\mathbf{K} = k_c$, as shown in 
\Fref{fig:1}d.}
\begin{ruledtabular}
\begin{tabular}{cccc}
 $i$  & $1$ & $2$ & $3$ \\
$(m,n)$ & $(3,2)$ & $(4,3)$ & $(4,2)$ \\ \hline
 $\delta _a^i $(eV) & 0.178 & 0.585 & 0.763 \\ 
 $\delta _b^i $(eV) & 0.198 & 0.534 & 0.731 \\ 
 $\delta _c^i $(eV) & 0.200 & 0.518 & 0.718
\end{tabular}
\end{ruledtabular}
\end{table}

%
The adoption of the crystal momentum representation simplifies the 
evaluation of matrix elements of the velocity operator that reduce to
$\mathbf{v}_{mn} = \hbar^{-1} \langle m\mathbf{k} | 
\bNabla_\mathbf{k} h_\mathbf{k} | n \mathbf{k}\rangle$.
Finally, the Berry connection in periodic systems \cite{Xiao2010, Hipolito2016} 
reads
 $\bAcal_{mn} = i \langle m\mathbf{k} | 
 \bNabla_\mathbf{k} | n \mathbf{k}\rangle  $.
%
The details regarding the numerical implementation of the derivatives present 
in the Berry connection are discussed in \Rcite{Hipolito2018} and references 
therein.

\subsection{Effective rank-2 tensors for the nonlinear response}
\label{sec:effTen}

Here, we show that the spurious divergences found in the current density 
response \cite{Aversa1995} naturally vanish when considering an \emph{effective 
rank-2 tensor} for the nonlinear conductivity, rather than the conductivity 
tensor as defined in \Eqref{eq:T} or its susceptibility counterpart.
The definition of effective tensors is not new \emph{per se}, in fact, it has 
been used extensively in nonlinear optics \cite{Boyd2008, Haussuhl2007, 
Yang1995}, but is frequently not taken into consideration in theoretical 
calculations of the nonlinear response, where authors tend to only consider 
individual tensor elements.
By combining and adding contributions according to the dependence on the 
fields and the output frequency, it is possible to define an effective rank-2 
tensor $\bar\sigma _{\phi\nu} ^{(n)} (\omega_s) $ for the $n^\mathrm{th}$-order 
conductivity
\begin{align}
\label{eq:sEff}
j_\phi^{(n)}( t ) &=
\sum_{\omega_n\ldots\omega_1} \sum_{\lambda\ldots\alpha}
\sigma _{\phi\lambda\ldots\alpha} ^{(n)} (\omega_n, \ldots, \omega_1)
E_{\omega_n}^\lambda \ldots E_{\omega_1}^\alpha
\nonumber \\ &\times
\exp \big[ -i(\bar\omega_n+\ldots+\bar\omega_1)t \big]
\nonumber\\ &\equiv
\sum_{\omega_s} \sum_\nu
\bar\sigma _{\phi\nu} ^{(n)} (\omega_s) 
\mathcal{E}_{\omega_s}^{(\nu)}
e^{-i\bar\omega_s t}
\, ,
\end{align}
where the index $\nu$ contains all combinations of $\lambda\ldots\alpha$ that 
preserve equal powers of the Cartesian components of the electric field, 
$\mathcal{E}_{\omega_s}^{(\nu)} \equiv E_{\omega_n}^\lambda \ldots 
E_{\omega_1}^\alpha $, with $\omega_s \equiv \omega_n +\ldots +\omega_1$ 
defining the sum of all input frequencies.
Several labeling conventions are possible for the effective tensor elements 
\cite{Shen2002, Boyd2008, Haussuhl2007, Yang1995}.
We map the effective tensor index $\nu$ to combinations of indices of rank-3 
and rank-4 tensors as listed in \Tbref{tab:nu}, and remap the index 
$\phi = \{ x,y,z \} \to \{ 1,2,3 \}$.
%
%
\begin{table}[t]
\caption[Mapping of $\nu$ to $\lambda\alpha $ and $ 
\lambda\beta\alpha$]{\label{tab:nu}%
Mapping of index $\nu$ to combinations of Cartesian indices for the 
second-order ($\lambda\alpha$) and third-order ($\lambda\beta\alpha$) 
responses, respectively.}
\begin{ruledtabular}
\begin{tabular}{ccc}
$\nu$ & $ \lambda\alpha $ & 
$ \lambda\beta\alpha $ \\ \hline
 1 & $xx$          & $xxx$ \\ 
 2 & $yy$          & $yyy$ \\ 
 3 & $zz$          & $zzz$ \\
 4 & $xy +yx$ & $yzz +zyz +zzy$ \\ 
 5 & $yz +zy$ & $yyz +yzy +zyy$ \\ 
 6 & $zx +xz$ & $zzx +zxz +xzz$ \\
 7 &          & $zxx +xzx +xxz$ \\
 8 &          & $xyy +yxy +yyx$ \\ 
 9 &          & $xxy +xyx +yxx$ \\ 
 0 &          & $xyz +xzy +yzz +yxz +zxy +zyx$
\end{tabular}
\end{ruledtabular}
\end{table}
%
Moreover, the summation over $\omega_s$ includes the combinations of all 
external frequencies that generate the same output frequency.
As an example, consider the element $14$ of the effective tensor for the OR 
process (that requires the combination of spatial indices and different 
frequency components)
\begin{align}
\bar\sigma_{14}^{(2)}(0) &\equiv
 \sigma_{xxy}^{(2)}( \omega,-\omega)
+\sigma_{xxy}^{(2)}(-\omega, \omega)
\nonumber\\&+
 \sigma_{xyx}^{(2)}( \omega,-\omega)
+\sigma_{xyx}^{(2)}(-\omega, \omega).
\end{align} 

In Sec.~\ref{sec:reg}, we show that all spurious divergences present in the 
direct evaluation of the nonlinear conductivity tensors 
$\sigma_{\phi\lambda\dots\alpha}^{(n)}(\omega_n, \dots, \omega_1)$  
vanish when considering the relevant effective rank-2 tensor 
$\bar\sigma_{\phi\nu}^{(n)}(\omega_s)$, for second- and third-order processes, 
namely \Eqsref{eq:s2:eff} and \eqref{eq:s3:eff}.
In addition, we identify the remaining physical divergences occurring at zero 
frequency.
The origin of these divergences can be traced to the diagonal elements of the 
density matrix, \ie $\rho_{mm}^{(n)}$.
The linear order the density matrix $\rho_{mm}^{(1)}$ introduces a divergent 
term in the linear optical response
\begin{align}
\label{eq:s1:Drude}
\sigma_{\phi\alpha}^{i}(\omega) =
-\frac{ 2 i g \sigma_1 }{ \Omega } 
\frac{ 1 }{ \omega +i \eta }
\sum_\mathbf{k} \sum_{m} 
v_{mm}^\phi
\frac{ \partial f_m }{ \partial k_\alpha }
\, .
\end{align}
By keeping the adiabatic parameter finite, rather than taking the formal 
adiabatic limit $\eta \to 0^+$, the response remains finite and captures the 
so-called \emph{Drude} term, \cite{Ashcroft1976, Mahan2000, Marder2010}
with the finite $\eta$ representing the scattering rate.
Similar divergences are found in terms beyond the linear order, defining 
nonlinear Drude-like processes.
The quadratic response contains two such terms, a linear divergence in 
\Eqref{eq:s2:ei} and a quadratic in \Eqref{eq:s2:ii}, whereas the cubic 
response spawns a total of four terms that define a linear and a cubic 
divergence. 
The cubic term emerges from \Eqref{eq:s3:iii} and the linear divergence stems 
from \Eqsref{eq:s3:divs:B:iee}, \eqref{eq:s3:divs:b2}, and 
\eqref{eq:s3:divs:B:iie}.
By the same token, $\eta$ should be kept finite and mapped to the scattering 
rate of each nonlinear Drude-like process.
To avoid confusion with the spurious divergences, we refer to these as 
nonlinear \emph{Drude-like} terms.
Spurious divergences are found in almost all contributions that involve 
intraband processes, with the exception of the purely intraband processes 
and also those comprised by a single interband transition ($e$) followed by 
multiple intraband processes ($i$), \ie processes labeled as $ie$, $iie$, and 
so forth.
In contrast to the intensive study of the linear optical response of weakly
disordered MLG, \cite{Horng2011, Stauber2008, Stauber2008a, Stauber2008b, 
Peres2008a} little progress has been made on the characterization of the 
nonlinear Drude terms.
To qualitatively identify all contributions from the Drude-like terms, we 
consider a simple model based on a fixed scattering rate approximation.
The choice of such method restricts the characterization to pristine or weakly 
disordered systems, as in disordered systems the scattering rates are dependent 
on doping level \cite{DasSarma2011}.
In the absence of a proper estimate for the nonlinear scattering rates, 
\cite{Soavi2017} we take the rate for the linear process in weakly disordered 
MLG as a reference figure for all scattering rates, thereby setting 
$\hbar\eta = 10$ meV for two reasons.
First, it represents the finite scattering rate of charge carries in the 
system, used successfully in the analysis of the linear response of graphene in 
the presence of weak disorder. \cite{Horng2011, Stauber2008, Stauber2008a, 
Stauber2008b, Peres2008a}
Second, it facilitates the convergence of the calculation over the 
entire spectrum, particularly for the contributions associated with interband 
transitions.

The lattice symmetry determines the number of independent and finite elements 
in the optical conductivity and susceptibility tensors, which in turn 
determine the properties of the respective effective tensors.
Restricting the external fields to normal incidence limits the response to the 
in-plane motion of electrons in the crystal.
For threefold symmetric crystals, such as graphene-based systems, the 
third-order effective tensor contains only one independent tensor element 
$\bar\sigma_{11}^{(3)} =\bar\sigma_{18}^{(3)} 
=\bar\sigma_{29}^{(3)} =\bar\sigma_{22}^{(3)}$.
Breaking the inversion center of the honeycomb lattices, while preserving the 
threefold symmetry, allows for second-order effects, \eg SHG and OR, that are 
governed by $\bar\sigma_{14}^{(2)}/2 = \bar\sigma_{21}^{(2)} = 
-\bar\sigma_{22}^{(2)}$.
%
%

\section{Results}
\label{sec:Res}

Throughout this section we address several linear and nonlinear optical 
response functions of MLG and BBG.
Although the linear optical conductivity has been studied intensively
\cite{Horng2011, Stauber2008, Stauber2008a, Stauber2008b, Peres2008a}, we 
briefly discuss it here to serve as a basis for our analysis of the nonlinear 
response.
We restrict our analysis to nonlinear interactions with a monochromatic 
field, namely THG 
$\bar\sigma_{\phi\nu}^{(3)}( \omega_s = 3\omega )$, optical Kerr conductivity 
$\bar\sigma_{\phi\nu}^{(3)}( \omega_s =  \omega )$, and SHG 
$\bar\sigma_{\phi\nu}^{(2)}( \omega_s = 2\omega )$.
For THG in graphene we show that our spectra are in 
agreement with previous results computed within the Dirac approximation
\cite{Mikhailov2014, Cheng2014, Cheng2015b, Mikhailov2016, Soavi2017}.

\subsection{Optical conductivity}
\begin{figure}
\includegraphics[width=1.00\linewidth]{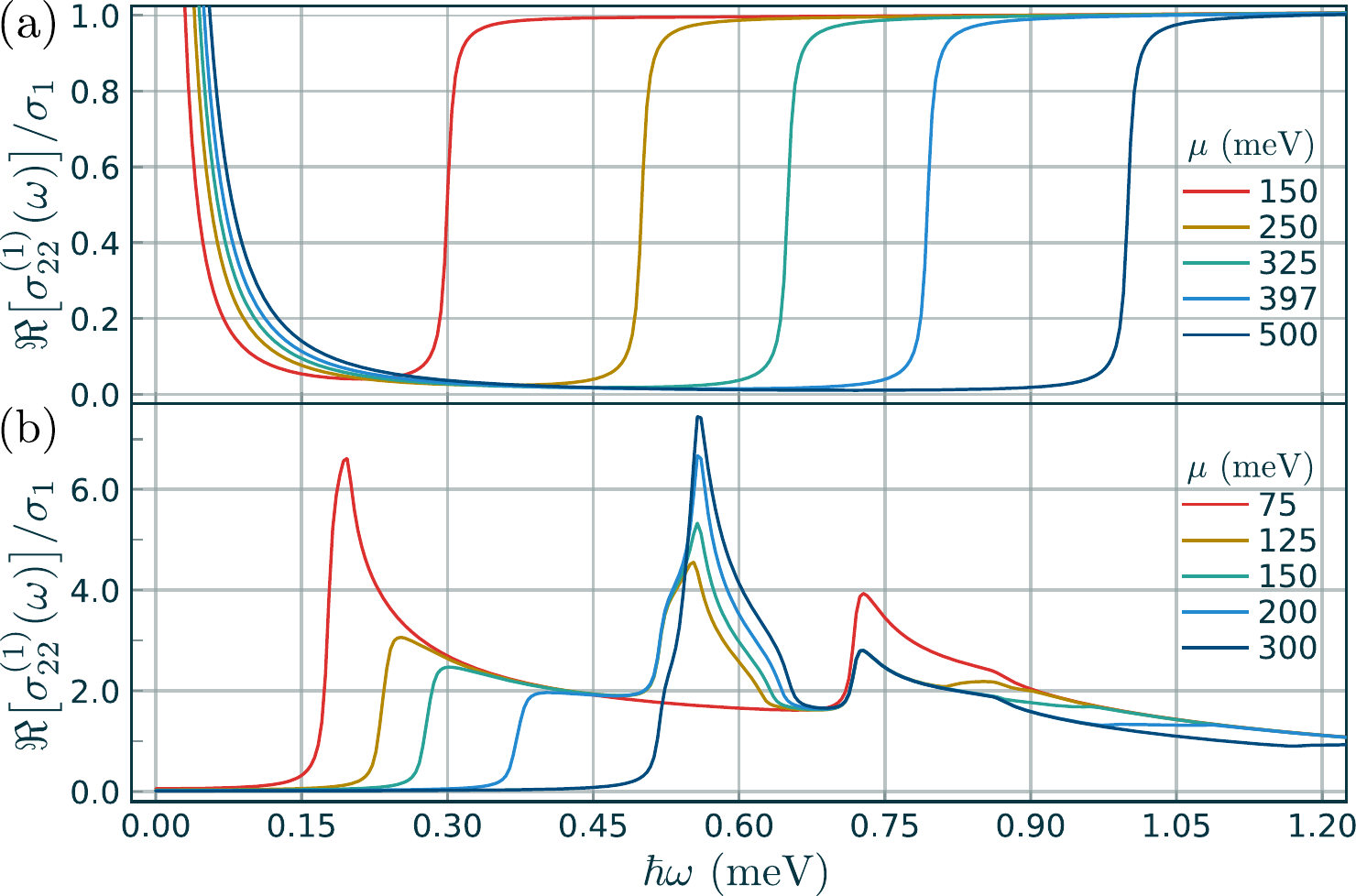}
\caption[Optical conductivity]{\label{fig:2}%
Linear optical conductivity (in units of $\sigma_1=e^2/4\hbar$) for MLG (a) and 
BBG (b) as a function of energy for several $\mu$ taking 
$T=1$ K and $\Delta = 200$ meV.}
\end{figure}

The effect of doping on the linear optical response of MLG has been 
discussed extensively \cite{Horng2011, Stauber2008, Stauber2008a, Stauber2008b, 
Peres2008a}.
It manifests itself as a combination of Pauli blocking and a Drude 
low-frequency peak.
Pauli blocking suppresses the interband optical response below the chemical 
potential $|\mu|$, \ie $\hbar\omega < 2|\mu|$, as shown in \Fref{fig:2}a, while
the intraband motion [governed by the Drude peak, \Eqref{eq:s1:Drude}] is 
characterized by the finite scattering rate of charge carriers.
Note that we only consider $n$-doping, i.e. $\mu > 0$.
In the low-energy regime, the energy dispersions of MLG and BBG are nearly 
electron-hole symmetric, even considering next-nearest neighbors hopping 
in the Hamiltonian.
Hence, $p-$doping of an equal magnitude would lead to essentially identical 
optical response.
%
%
%
A doping level $\mu = 397$ meV sets the Pauli blocking threshold at 
$\hbar\omega = 2\mu = 793$ meV, corresponding to the reference wavelength 
$\lambda\sim 1560$ nm \cite{Saynatjoki2013, Autere2017}.
The response of BBG in \Fref{fig:2}b is significantly richer than that of the
monolayer.
Manifestations of Pauli blocking are still present, but the tuneable chemical 
potential enables and disables several transitions associated with the gap (for 
BBG) and the interlayer hopping $\gamma_1$.
In contrast to the featureless response of MLG, the transitions 
associated with the low-energy vHs introduce several resonances that dominate 
the optical response of BBG.
In \Fref{fig:2}b, we show the linear response of BBG ($\Delta = 200$ meV) as a
function of photon energy for several $\mu$.
Unlike MLG, BBG supports a rich optical response that is highly 
sensitive to doping, \eg the large and tuneable resonance that emerges in the 
vicinity of $\hbar\omega\sim 0.6$ eV, whenever doping is large enough to 
populate the first conduction band with electrons or to introduce holes in the 
top valence band that provide an additional set of allowed resonant 
transitions.

\subsection{Nonlinear response of MLG}

Regarding the nonlinear response of graphene, it has been shown 
\cite{Mikhailov2014, Cheng2014, Cheng2015b, Alexander2017, Soavi2017} that the 
THG can be strongly enhanced and tuned by controlling the doping via 
electrostatic gating.
Our calculations and those of \Rscite{Cheng2014, Soavi2017} start from the 
evaluation of the response function using the LG \cite{Aversa1995}, 
but differ at several levels.
Alternative approaches have been used in \Rscite{Mikhailov2014, Mikhailov2016} 
and references therein, where the electric field is assumed to be 
space-dependent.
More recently, non-perturbative methods have been used to compute the nonlinear 
response of graphene using the Dirac Hamiltonian \cite{Mikhailov2017}.

The present evaluation of the nonlinear response incorporates the effects of 
finite temperature from the onset, rather than estimating finite temperature
conductivity from zero temperature calculations \cite{Cheng2015b}.
It naturally extends beyond the particular solutions for the Dirac model 
\cite{Mikhailov2014, Cheng2014, Cheng2015b, Mikhailov2016, Soavi2017}, as none 
of the initial expressions used, \Eqsref{eq:s3}, are derived for a specific 
Hamiltonian or fixed number of bands.
Therefore, they are valid for more complex systems that require more 
elaborate 
Hamiltonians.
Although the Dirac approximation has proven useful for the characterization of 
the low-energy linear- and third-order optical response of graphene 
\cite{Mikhailov2014, Cheng2015b, Soavi2017, Cheng2014, Cheng2015b, 
Mikhailov2016, Dimitrovski2017a}, it is of little use for the study of any 
quadratic response, as all even-order processes vanish in the presence of full 
rotation symmetry, at least within the dipole approximation.

Temperature plays an important role in nonlinear optics. It may not only 
soften the resonant features present in the optical response by modifying the 
effective electron distribution, \cite{Ruzicka2010, Sun2012a, Sun2012, 
Tielrooij2013} but also change the scattering rates.
The leading order effects on the second- \cite{Hipolito2016, Hipolito2017} and 
third-order\cite{Soavi2017, Cheng2014} responses are the broadening of the 
resonant features.
In addition, temperature can switch on transitions that would otherwise be 
forbidden due to Pauli blocking, but this is more frequently than not a minor 
effect, when compared with the strong resonances associated with large $\mu$
\cite{Hipolito2016, Hipolito2017}.
For the sake of brevity we refrain from discussing the effects of temperature 
in detail. Unless stated otherwise, results shown in this paper were computed 
for $T=10$ K, such that thermal broadening remains minimal and manifestations 
of Pauli blocking are preserved, thus allowing for a clear identification of 
all processes involved in the optical response.
In realistic experimental scenarios, \cite{Soavi2017} the effective carrier 
temperature lies in the range $T \sim 1000-1500$ K and scattering rates are 
likely to be different.

\subsubsection{Third harmonic generation}
%
\begin{figure}
\includegraphics[width=1.00\linewidth]{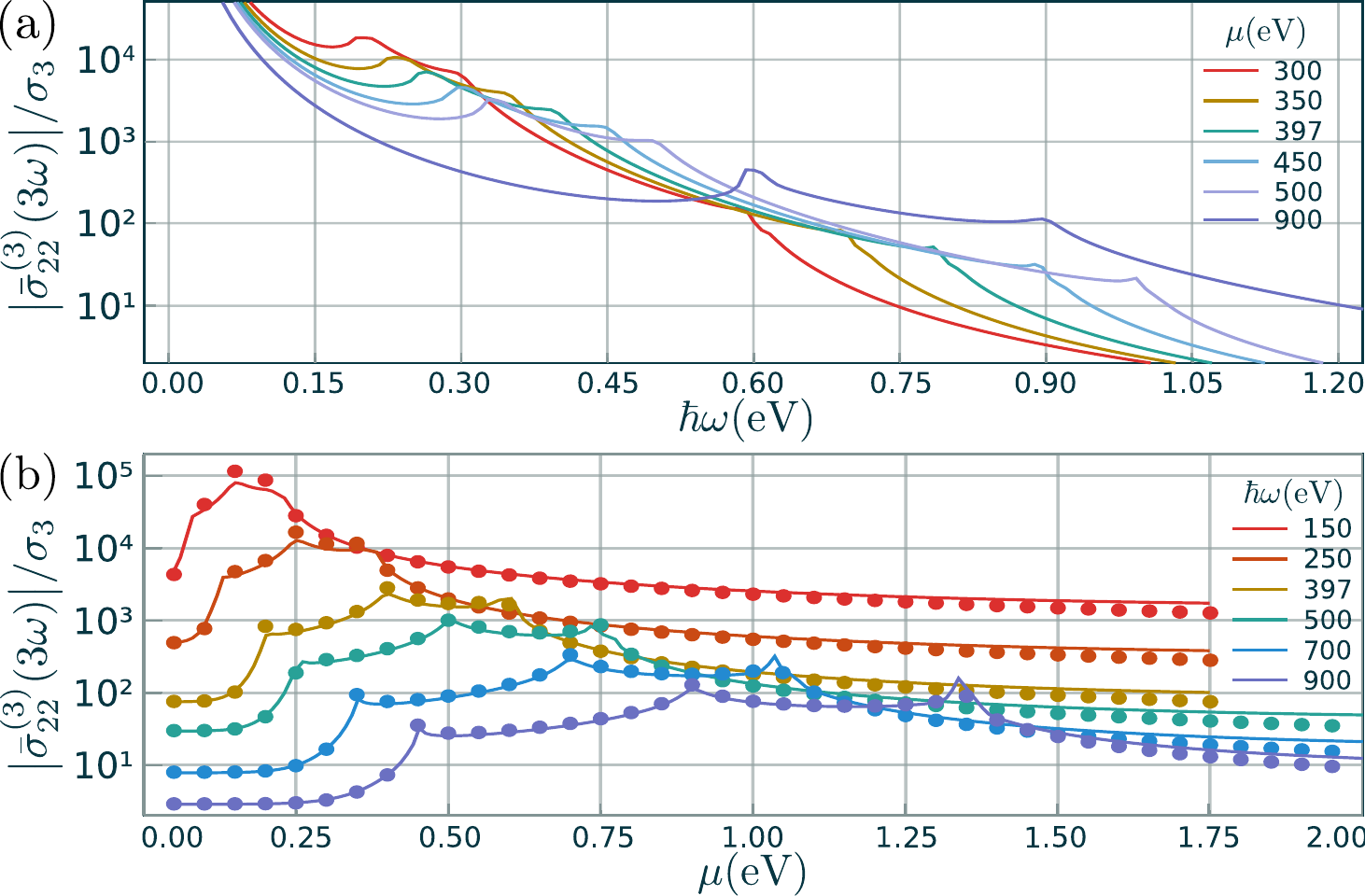}
\caption[THG conductivity of doped graphene]{\label{fig:3}%
Magnitude of THG conductivity of doped graphene in units of $\sigma_3 \equiv
e^4a_0^2/(8\gamma_0^2 \hbar) = 6.28 \times 10^{-26}\, \mathrm{Sm^2/V^2}$,
computed from full dispersion TB using \Eqsref{eq:s3:eff}.
In (a), we plot the frequency-dependent THG conductivity at several finite 
chemical potentials.
(b) shows the dependence on the chemical potential for several photon 
frequencies, computed with full dispersion TB (solid lines) and Dirac 
approximation at zero temperature (dots) using \Eqref{eq:s3:Dirac}.
}
\end{figure}%
%

In \Fref{fig:3}a, we plot the magnitude of the THG conductivity 
$\bar\sigma_{22}^{(3)}(3\omega)$ at finite doping, computed from full 
dispersion TB using \Eqsref{eq:s3:eff}.
Similarly to the linear response, the Drude-like nonlinear response, comprised 
by \Eqsref{eq:s3:divs:B:iee}, \Pref{eq:s3:divs:b2} and \Pref{eq:s3:divs:B:iie},
has a  smooth featureless power law decay, that dominates the THG response 
below the first resonance associated with the chemical potential.
In the present model, with a single nonlinear scattering rate, the magnitude of 
the Drude-like response at the DC limit and the broadening of the resonances 
associated with the Fermi level, \ie $\hbar\omega/\mu = \{ 2/3,1,2 \}$, are 
both controlled by the same parameter.
Therefore, in the presence of sufficiently large scattering rates $\hbar\eta 
> 50$ meV the resonances are strongly suppressed and the response reduces to 
the Drude-like contributions.
For moderate nonlinear scattering rate, the regular part clearly surpasses the 
Drude-like terms in the vicinity of the resonances associated with the Fermi 
level.
As in previous studies \cite{Mikhailov2014, Cheng2014, Mikhailov2016, 
Soavi2017, Dimitrovski2017a}, results show that the nonlinear response can be 
tuned by controlling $\mu$, as the resonances are shifted by the 
chemical potential. Temperature and the nonlinear scattering rate play a 
large role and can significantly soften the resonances.
Moreover, it is worth noting that the resonance $\hbar\omega/\mu = 2/3$ lies 
within the energy range of minimal absorption due to Pauli blocking.
Also, this resonance should not be disturbed by the optical Kerr effect on 
absorption, as the latter can only generate resonances at 
$\hbar\omega  /\mu = \{ 1, 2 \}$ as discussed in detail below.

The general solutions for the third-order response, \Eqsref{eq:s3:eff}, can be 
integrated analytically for THG by taking the zero temperature limit and making 
use of the Dirac dispersion
\begin{align}
\label{eq:s3:Dirac}
\frac{ \bar\sigma_{22}^{(3)} }{ \sigma_3 }  & =
\frac{-3i}{1024\pi} \frac{\gamma_0^4}{(\hbar\bar\omega)^4}
\bigg[
 45 \log\bigg( \frac{ 2\mu -3\hbar\bar\omega }{ 2\mu +3\hbar\bar\omega } \bigg)
\nonumber\\&
-64 \log\bigg( \frac{ 2\mu -2\hbar\bar\omega }{ 2\mu +2\hbar\bar\omega } \bigg)
+17 \log\bigg( \frac{ 2\mu - \hbar\bar\omega }{ 2\mu + \hbar\bar\omega } \bigg)
\bigg] ,
\end{align}
in agreement with previous results. \cite{Cheng2014}
Likewise the divergent terms present in THG can be determined straightforwardly 
from the general solutions for the Drude-like terms. The linear divergence 
stems from \Eqsref{eq:s3:divs:b2} and \Pref{eq:s3:divs:B:iie}
\begin{widetext}
\begin{subequations}
\label{eq:s3:drude}
\begin{align}
\frac{ \bar\sigma_{22}^{\{iee\}+\{iie\}} }{ \sigma_3 } &= 
\frac{ -\hbar^3 C_3 }{ 4 \hbar\bar\omega_s }
\sum_{m\neq n} \sum_\mathbf{k}
\bigg\{
\frac{2}{\hbar}
\frac{\bar\omega_s \bar\omega_3}{\bar\omega_2 \bar\omega_1}
\frac{ \partial^2 }{ \partial k_y^2 }\bigg[ 
\frac{ 2|v_{nm}^2| }{ \epsilon_{mn}^3 }
\bigg]
%
+
2\frac{ \partial }{ \partial k_y } \bigg[
 \frac{ \partial }{ \partial k_y } \bigg(
  \frac{ v_{nm}^y }{ \epsilon_{mn} }
 \bigg)
 \frac{ v_{mn}^y }{ \epsilon_{mn}^2 }
+\frac{ \partial }{ \partial k_y } \bigg(
  \frac{ v_{mn}^y }{ \epsilon_{mn} }
 \bigg)
 \frac{ v_{nm}^y }{ \epsilon_{mn}^2 }
\bigg]
\nonumber\\&+
 \frac{ (\bar\omega_2 +\bar\omega_1)^2 
 }{ \bar\omega_2\bar\omega_1 }
 \frac{ \partial }{ \partial k_y } \bigg[ 
 \frac{ v_{nn}^y -v_{mm}^y }{ \epsilon_{mn}^2 }
 \frac{ 2| v_{nm}^y |^2 }{ \epsilon_{mn}^2 }
 \bigg] 
%
+
 \frac{ \partial (v_{nn}^y -v_{mm}^y) }{ \partial k_y }
 \frac{ 2|v_{nm}^2| }{ \epsilon_{mn}^4 }
\bigg\} f_{nm}
%
\simeq
\frac{   45i \gamma_0^4        }{   256 \pi \mu^3  \hbar \bar\omega } \, ,
\end{align}
and the cubic term follows from \Eqref{eq:s3:divs:D:iii}
\begin{align}
\frac{ \bar\sigma_{22}^{iii} }{ \sigma_3 } &=
\frac{ \hbar C_3 }{
\hbar^3 \bar\omega_s( \bar\omega_2 +\bar\omega_1) \bar\omega_1 }
\sum_n \sum_\mathbf{k}
\frac{ \partial^3 v_{nn}^y }{ \partial k_y^3 } f_n
%
\simeq
\frac{    9i \gamma_0^4        }{   128 \pi \mu (\hbar \bar\omega)^3 } \, .
\end{align}
\end{subequations}
\end{widetext}
In VG, spurious divergences appear for gapped systems if a truncated band 
structure is applied \cite{Taghizadeh2017a}.
For gapless two-band systems neither LG nor VG exhibit spurious divergences in 
the DC limit.
Therefore, the Drude-like terms can always be derived from the full 
non-regularized expressions \Eqsref{eq:s3} by means of a Taylor series.
\Eqsref{eq:s3:drude} are in agreement with the Taylor expansion of 
\Eqref{eq:s3:Dirac}, which reads
\begin{align}
\frac{ \bar\sigma_{22}^{(3)} }{ \sigma_3 }  & =
 \frac{    9i \gamma_0^4             }{   128 \pi \mu   (\hbar\bar\omega)^3 }
+\frac{   45i \gamma_0^4             }{   256 \pi \mu^3  \hbar\bar\omega }
\nonumber\\ &
+\frac{ 3339i \gamma_0^4 \hbar\bar\omega }{ 10240 \pi \mu^5 }
+\mathcal{O}(\bar\omega^3).
\end{align}
The divergence with doping $\mu$ is a mere artifact of the Taylor series 
computed in the limit $\hbar\omega \ll \mu$.
This becomes evident by considering opposite limit $\mu \ll \hbar\omega$
\begin{align}
\frac{ \bar\sigma_{22}^{(3)} }{ \sigma_3 }  & =
 \frac{ 3 \gamma_0^4       }{ 512   (\hbar \bar\omega )^4 }
-\frac{ i \gamma_0^4 \mu^3 }{ 6 \pi (\hbar \bar\omega )^7 }
+\mathcal{O}(\mu^4) \,.
\end{align}
Our result not only offers a clear separation between the regular and divergent 
terms, it also
offers a general model for the nonlinear Drude-like terms, that can be used to 
characterize the nonlinear scattering rate beyond the constraints of the Dirac 
approximation.
In \Fref{fig:3}b, we show the dependence of the THG response in MLG as 
a function of doping level, evaluated using two different models, namely:
(i) full TB dispersions using \Eqsref{eq:s3:eff};
(ii) the analytic result derived using the Dirac approximation at zero 
temperature \Eqref{eq:s3:Dirac}.
The agreement between the exact solution and the numerically integrated 
expressions \Eqsref{eq:s3:eff} is remarkable, even for large doping levels 
$\mu\sim1.5$ eV, indicating that the closed form expressions in 
\Eqref{eq:s3:Dirac} can be used to accurately characterize the THG response of 
graphene, even in cases of large photon energy and substantial doping.
The agreement between the exact solution and the numerically integrated 
expression extends to the Drude-like features.

Data shown in \Fref{fig:3} highlight the strongly varying nature of the 
nonlinear response in graphene with respect to the photon energy.
We present ``figure of merit'' estimates for the nonlinear susceptibility based 
on the finite temperature $T=10$ K results.
Additional enhancement of the magnitude of the response can be achieved by 
considering smaller nonlinear scattering rates.
The 3D nonlinear susceptibility is evaluated from the 2D nonlinear conductivity 
using $|\chi^{(3)}(\omega_s)| = |\sigma^{(3)}(\omega_s)|/( c_0 \varepsilon_0 
\omega_s )$, with $\varepsilon_0$ being the vacuum permittivity, and $c_0 = 
3.35$
\AA{} the interplanar distance in graphite \cite{Partoens2006}.
Given the absence of an accurate estimate for the nonlinear scattering rate, we 
refrain from showing data obtained in the region where the Drude-like terms 
dominate.
Considering the main THG resonances at $\mu = \{300,900\}$ meV, \ie 
$\hbar\omega \sim \{ 200, 600 \}$ meV, the response can be tuned up to 
$|\chi^{(3)}( 3\omega )| = \{ 504, 4.2 \} \times10^{-18}\, \mathrm{m^2/V^2}$.
Our estimates provide figures comparable with some previous 
experimental reports for THG in graphene that range from $\sim10^{-19}$ to 
$\sim10^{-15}\, \mathrm{m^2/V^2}$ \cite{Hendry2010, Kumar2013, Saynatjoki2013, 
Woodward2016}.
It is worth noting that the THG susceptibility of doped graphene 
can significantly exceed that of other materials, \eg gold 
\cite{Boyd2014} $|\chi^{(3)}|\sim 10^{-19}\, \mathrm{m^2/V^2}$ and AlGaAs 
\cite{Le1991, Pu2016a} $|\chi^{(3)}|\sim 10^{-18}\, \mathrm{m^2/V^2}$.

\subsubsection{Optical Kerr effect}

In addition to the $n^\mathrm{th}$ harmonic generation, the interaction of an 
intense monochromatic electromagnetic field with a crystal also generates 
nonlinearities associated with combinations of positive and negative 
frequency components of the field.
At third-order, it generates a nonlinear current density with the frequency of 
the driving field. This optical Kerr conductivity 
$\bar\sigma_{22}^{(3)}(\omega)$ gives rise to intensity-dependent effects 
on the refractive index \cite{Shen2002, Boyd2008}.
%
%
\begin{figure}
\includegraphics[width=1.00\linewidth]{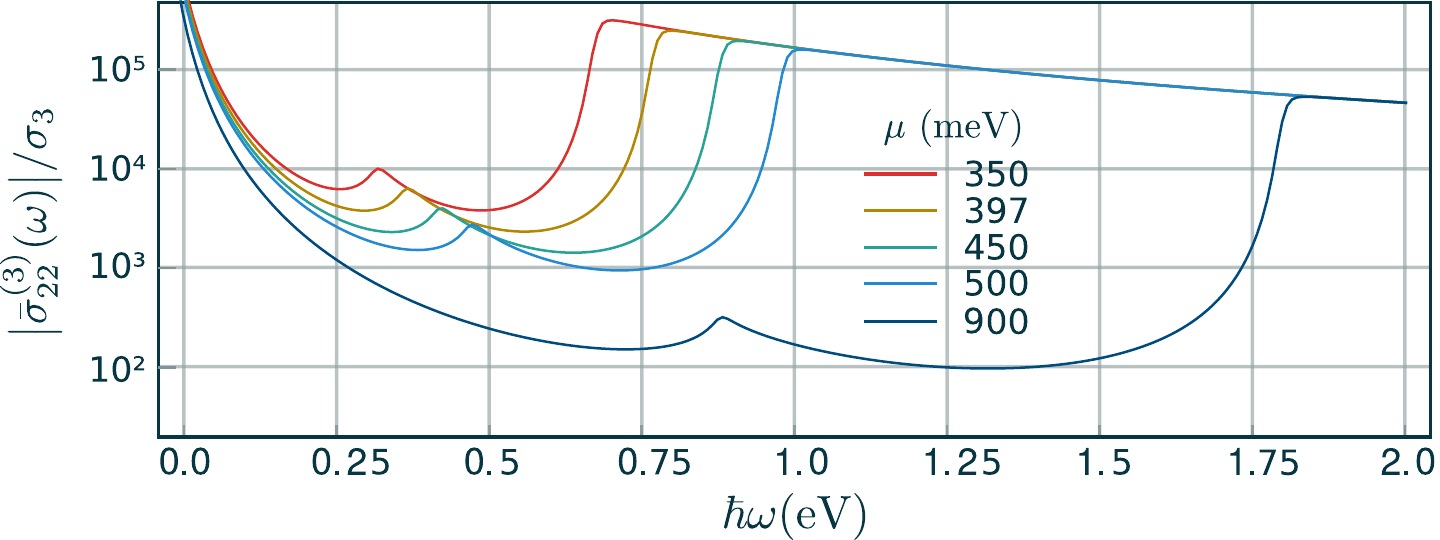}
\caption[Optical Kerr conductivity in graphene]{\label{fig:4}%
Frequency dependence of the optical Kerr conductivity 
conductivity of MLG in units of $\sigma_3$ for several 
chemical potentials $\mu$.}
\end{figure}%
%
%
Following the analysis of the THG, we show results for the optical Kerr 
conductivity $|\sigma^{(3)}(\omega)|$ in \Fref{fig:4}.
As expected from the expression for the third-order conductivity tensor, 
\Eqsref{eq:s3}, this response contains two resonances, $\hbar\omega = 2\mu \{ 
1/2, 1 \}$.
The former stems from the two photon resonance $\propto 1/( \bar\omega_2 
+\bar\omega_1 -\epsilon_{mn}/\hbar)$, whenever $\omega_2=\omega_1$.
The latter originates from the single and three photon resonances. 
All terms involving the three photon resonance $\propto 1/(\bar\omega_3 
+\bar\omega_2 +\bar\omega_1 -\epsilon_{mn}/\hbar)$ reduce to a single photon 
resonance due to the presence of frequencies with opposite sign. 

The contrast between portions of the spectra dominated by Drude-like and 
regular parts of the optical Kerr conductivity is stronger than in the THG 
response.
Even at moderate doping levels, $\mu \sim 200$ meV, contributions from the 
regular part to the first resonance $\hbar\omega = \mu$, clearly surpass the 
Drude-like peak.
Moreover, from the onset of the second resonance $\hbar\omega = 2\mu$, the 
regular part of the response exhibits a strong step-like feature and becomes 
the leading term driving the response.
The Drude-like terms show little dependence on the details of the band 
structure and, similarly, the Fermi level only affects the overall magnitude 
of the Drude response but not the shape.
The robust increase of the response at the second resonance $\hbar\omega=2\mu$, 
indicates that the dependence of the optical Kerr effect on doping could be 
experimentally probed in highly doped graphene samples, $\mu \sim 397$ meV.
For such samples, the $\hbar\omega=2\mu$ resonance should lie within range of 
high intensity lasers frequently used for nonlinear optics experiments 
$\hbar\omega \sim 793$ meV, \ie $\lambda\sim 1560$ nm\cite{Saynatjoki2013, 
Autere2017}.
It is worth noting that the low-energy results are in agreement with previous 
estimates using the Dirac model \cite{Mikhailov2014, Cheng2014, 
Mikhailov2016}. 
Using data from \Fref{fig:4}b, the magnitude the optical
Kerr susceptibility at $\mu = 350$ meV for $\hbar\omega = \{ 350, 700 
\}$ meV reads 
$ |\chi^{(3)}( \omega )| \sim \{ 43, 4.3 \} \times10^{-15} 
\;\mathrm{m^2/V^2}$,
indicating significantly larger third-order susceptibilities than those 
observed in THG.

\subsection{Response of gapped graphene}

%
Below, we consider the nonlinear response in non-centrosymmetric systems, such 
as commensurate structures of graphene on hBN substrates, where the electronic 
dispersion of graphene is gapped.

For doped systems, the contributions to the third-order response that emerge 
from features associated with the gap, $E_g\sim 30$ meV, are dwarfed by the 
Drude-like terms.
Moreover, in the case of large doping $\mu \gg E_g$, the interband 
transitions are strongly suppressed by Pauli blocking, while the features 
associated with resonances at the Fermi level nearly match the resonances found 
in the gapless dispersion.
Therefore, we only consider the implications of symmetry breaking on the 
quadratic response.

In previous studies \cite{Brun2015, Hipolito2016}, the authors addressed the 
quadratic response of non-centrosymmetric honeycomb lattices in the regime 
where transitions between the top valence and bottom conduction bands dominate.
Here, we focus on the regime where the chemical potential is significantly  
larger than the band gap $\mu\gg E_g$.
Considering typical values for the Fermi level in graphene $\mu\sim 250$ meV, 
it is feasible to reach this regime in systems such as graphene on hBN with 
$E_g\sim 31$ meV \cite{Woods2014}.
%
%
\begin{figure}
\includegraphics[width=1.00\linewidth]{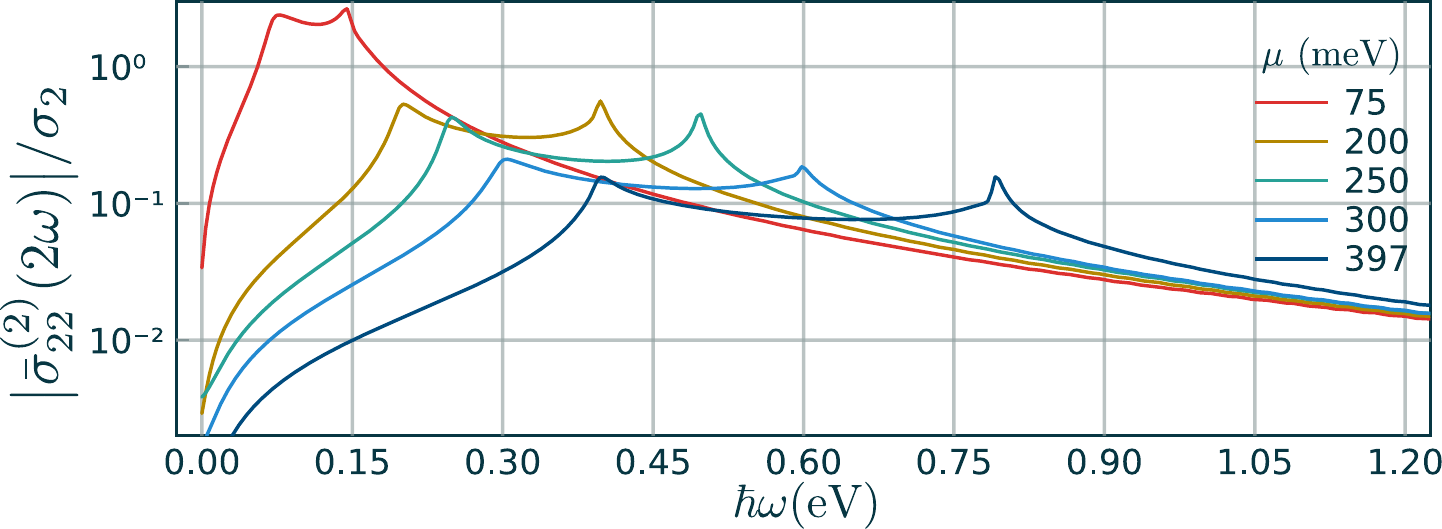}
\caption[SHG in non-centrosymmetric graphene]{\label{fig:5}%
SHG conductivities $\bar\sigma_{22}^{(2)}(2\omega)$ of gapped graphene in 
units of $\sigma_2 \equiv e^3 a_0/(4|\gamma_0|\hbar) = 2.88\times 10^{-15} \, 
\mathrm{Sm/V}$ with $E_g = \delta = 30$ meV.
}
\end{figure}%
%
%
The Fermi level suppresses the resonances associated with the band 
gap, $\hbar\omega \sim E_g\{1/2,1\}$ and the response is then controlled by the 
mixed inter-intraband processes occurring at the energy scale of the Fermi 
level $\hbar\omega = \{ \mu,2\mu \} $.
Figure \ref{fig:5} shows the magnitude of the SHG conductivity for several 
$\mu$ as a function of photon energy, indicating that the response is 
essentially confined to resonances with energy associated with the Fermi level.
Moreover, for sufficiently large doping levels, $\mu\sim 150$ meV, the lowest 
energy resonance in the SH spectrum is sufficiently energetic to avoid the 
large absorption associated with the Drude peak, i.e. it remains 
within the Pauli blocked region of the spectrum.
The magnitude of the SH features at high doping is significantly smaller than 
for weak doping. Nonetheless, it remains comparable to the estimates of SHG in 
BBG \cite{Brun2015}.
Therefore, SHG should be detectable in commensurate structures of graphene on 
SiC or hBN, and also tuneable by doping.
Similar results are found for the OR process, where the magnitude of the 
response at large doping is comparable to that of BBG in the same energy range 
\cite{Hipolito2016}.

It is worth noting that, in systems with identical occupation of the
$\mathbf{K}$ and $\mathbf{K'}$ valleys, the quadratic response emerges solely 
from the mixed $ie$ and purely interband $ee$ processes, as the remaining 
processes, $ei$ and $ii$, cancel out upon integration over the full Brillouin 
zone.
Yet, in systems out-of-equilibrium such as valley polarized honeycomb lattices, 
the $ei$ and $ii$ processes no longer vanish \cite{Hipolito2017} and may 
enhance the response even further.

\subsection{Nonlinear response of BBG}

As discussed in the context of the linear response, the presence of several vHs 
in the low-energy dispersion of BBG introduces additional resonances.
%
\begin{figure}
\includegraphics[width=1.00\linewidth]{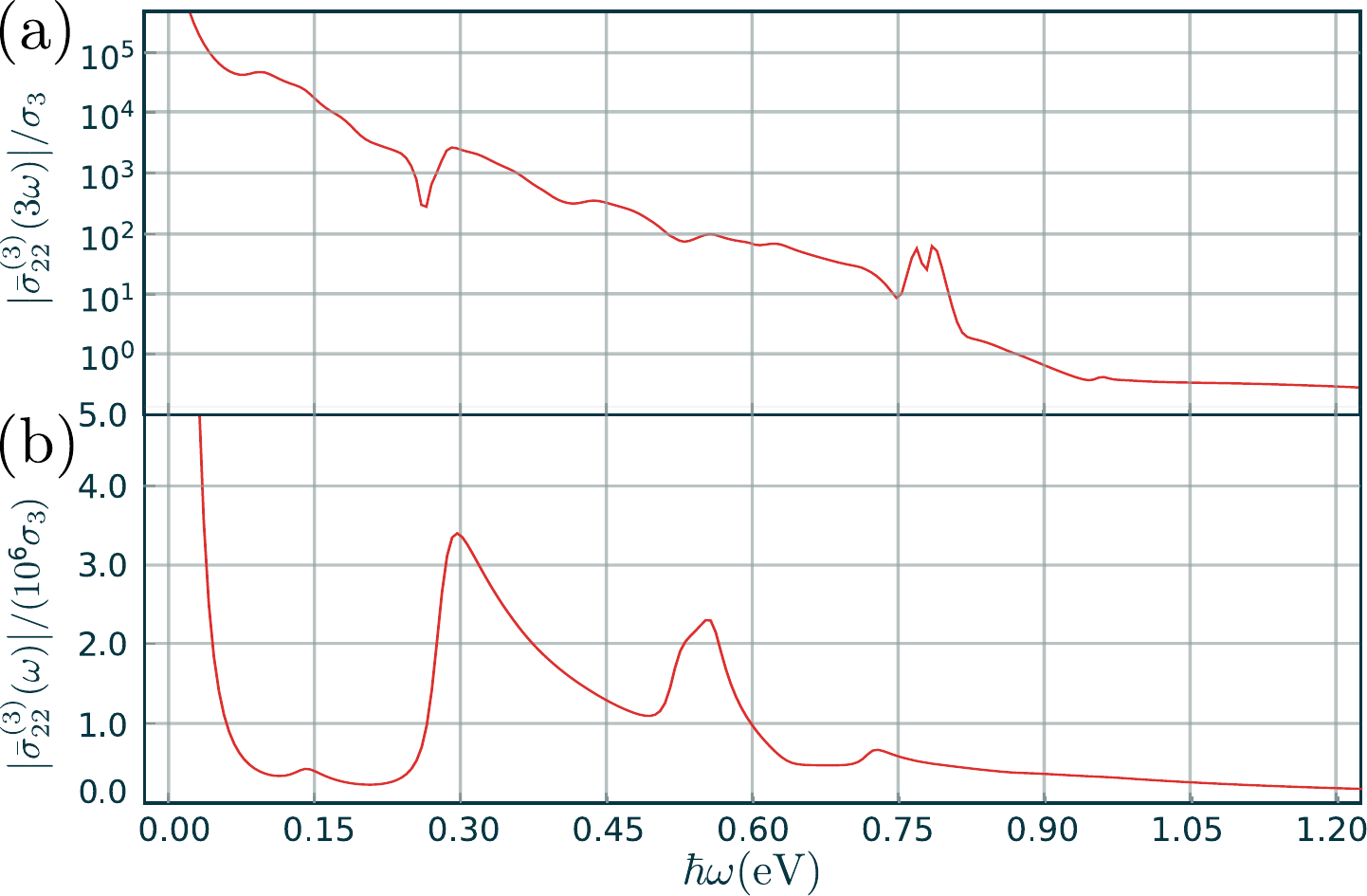}
\caption{\label{fig:6}%
Third-order response of BBG (in units of $\sigma_3$) with
$\Delta = 200$ meV, $\mu = 150$ meV, showing (a) the THG response and (b) the 
optical Kerr conductivity.}
\end{figure}%
%
The presence of broadening softens these features, converting otherwise 
resonances to small bumps, \cf \Fref{fig:6}a.
In addition, it also contains resonances that stem from the intraband motion at 
the Fermi level.
Considering doped BBG, $\mu=150$ meV and $\Delta=200$ meV, the 
resonances at lowest energy $\hbar\omega \sim \{ 91, 142\} $ meV arise 
mainly from transitions associated with the Fermi level.
In contrast with the THG response, the resonant features present in the optical 
Kerr effect are robust and significantly larger than the contributions from the 
Drude-like terms, \Fref{fig:6}b.
The features arise from transitions occurring at the low-energy vHs and 
intraband motion at the Fermi level.
The lowest energy resonances that stem from intraband motion are found at 
$\hbar\omega \sim \{ 137, 274 \}$ meV, while the dominant resonance at 
$\hbar \omega \sim 300$ meV arises from the doping cut-off of the vHs 
resonances that are located at $\hbar\omega \sim \{ \delta_a^2, \delta_b^2, 
\delta_c^2 \}/2 \sim \{ 293, 267, 259 \}$ meV.
The results shown in \Fref{fig:6}b indicate that the largest resonance stems 
from the combination of both processes, yet it is important to note that this 
combination is accidental, as the energies of the resonances depend on the 
details of the band structure and, also, on the Fermi level.

The amplitude of the THG and optical Kerr conductivities in BBG is 
significantly larger than in MLG.
The THG nonlinear susceptibility at the lowest resonances associated with 
doping and the vHs, $\hbar\omega \sim \{ 91, 274 \}$ meV, read $|\chi^{(3)}( 
3\omega )| \sim \{ 202, 9.1 \} \times10^{-16}\, \mathrm{m^2/V^2}$, 
respectively.
Regarding the optical Kerr effect in BBG the amplitude of the most intense 
resonance $\hbar\omega \sim 274$ meV reads $|\chi^{(3)}( \omega )| \sim 
1.7\times10^{-13}\, \mathrm{m^2/V^2}$.
Akin to the results for MLG, doping plays a crucial role as the resonances 
associated with the intraband motion can be displaced and, furthermore, it can 
suppress the resonances associated with the low energy vHs of BBG.

\section{Concluding remarks}

We have derived length gauge optical response functions up to third-order 
for a general periodic system.
The expressions for the effective nonlinear conductivity tensors are not only 
valid for any periodic system at finite temperature and doping, but are also 
free of any spurious divergences and all Drude-like terms are identified.
The spurious divergences are proven to vanish by considering effective rank-2 
tensors, which comprise all contributions to the relevant physical observable, 
as introduced in Sec.~\ref{sec:effTen} and explicitly shown in the appendix.
We identify all terms that contribute to the nonlinear Drude-like response in 
the DC limit, providing the basis for a comprehensive study of the scattering 
rates in nonlinear processes in 2D materials.

The expressions for the effective conductivity tensors are then used to 
evaluate the optical response of doped MLG and BBG.
We compute the optical conductivity and several nonlinear response functions, 
namely THG, optical Kerr effect, and SHG.
All results are strongly dependent on doping, showing that the nonlinear 
third-order susceptibility can be tuned over several orders of magnitude, in 
line with experimental reports for THG in graphene \cite{Hendry2010, Kumar2013, 
Saynatjoki2013, Woodward2016}.
By comparing the exact solutions derived with the Dirac approximation, with 
numerical integration using full dispersion TB models, our results show that 
the Dirac approximation provides remarkably accurate results, for THG in MLG 
even at very large doping.
The results show that the third-order response can be highly dependent on 
contributions that arise from nonlinear Drude-like terms, stressing the 
necessity for further studies probing the nonlinear optical response in the 
low-frequency regime, and also more elaborate theoretical models that 
can describe accurately scattering in nonlinear processes.

Proceeding beyond pristine MLG, we show that the second-order response of 
systems with small gaps can be strongly enhanced by the presence of finite 
doping, especially in the regime where the chemical potential is significantly 
larger than the energy gap.
The nonlinear response of BBG exhibits many resonant features that arise from 
two sources:
transitions occurring at the vHs in the vicinity of the Dirac points, 
present even in charge neutral systems, and the contributions associated with 
intraband motion occurring at the Fermi level.
This leads to richer nonlinear conductivity spectra that deviate strongly 
from the simple and well-localized resonances in MLG occurring 
at $\hbar\omega/\mu = \{ 2/3, 1, 2 \}$ \cite{Cheng2014, Mikhailov2016, 
Soavi2017}.

\acknowledgments 

This work was supported by the QUSCOPE center sponsored by the Villum 
Foundation, and 
TGP is supported by the CNG center under the Danish National 
Research Foundation, project DNRF103. 

\bigskip

\appendix

\section{Length gauge response}

In this appendix, we present all relevant results for the evaluation of 
the current density response.
We start by listing the elements of the power series expansion of the density 
matrix $\rho_{mn}(t)$ in the electromagnetic field, up to third-order, followed 
by the respective conductivity tensors elements.
We then proceed with a detailed description of the procedure used to isolate 
all the divergences present in the expressions for the conductivity tensor, 
terminating with the separation of the physical Drude-like terms from the 
spurious divergences that are proven to vanish.

The power series solution of the quantum Liouville equation is determined using 
the procedure outlined in \Rscite{Aversa1995, Hipolito2016}.
Using the notation defined in \Rcite{Hipolito2016}, the perturbative expansion 
of the density matrix up to third-order reads
\begin{widetext}
\begin{subequations}
\label{eq:rho}
\begin{align}
\label{eq:rho:1}
\rho_{mn}^{(1)}( t )  &\equiv \rho_{mn}^{e}(t) +\rho_{mn}^{i}(t) 
%
=
%
\frac{ e }{ 2\hbar }
\sum_{\omega_1} \bigg[
\bar\delta_{mn} \mathcal{A}_{mn}^\alpha f_{nm}
-i \delta_{mn} \frac{ \partial f_n }{ \partial k_\alpha } \bigg]
\frac{ E_{\omega_1} ^\alpha e^{-i \bar\omega_1 t}
}{ \bar\omega_1 -\omega_{mn} } \, ,
\\
\label{eq:rho:2}
\rho _{mn} ^{ (2) } ( t ) & \equiv 
 \rho_{mn}^{ee}(t) +\rho_{mn}^{ei}(t) 
+\rho_{mn}^{ie}(t) +\rho_{mn}^{ii}(t)
%
\nonumber \\ &=
%
\frac{ e^2 }{ 4\hbar^2 } \sum_{\omega_2 \omega_1} \sum_{l}
\frac{ E_{\omega_2}^\lambda E_{\omega_2}^\alpha 
e^{ -i( \bar\omega_2 +\bar\omega_1 )t }
}{ \bar\omega_2 +\bar\omega_1 -\omega_{mn} } \Bigg[
%
%
\bar\delta_{lm} \bar\delta_{ln} \bigg( 
 \frac{ \mathcal{A}_{ml}^\lambda \mathcal{A}_{ln}^\alpha f_{nl}
}{ \bar\omega_1 -\omega_{ln} }
-\frac{ f_{lm} \mathcal{A}_{ml}^\alpha \mathcal{A}_{ln}^\lambda
}{ \bar\omega_1 -\omega_{ml} }  \bigg)
%
%
-i \, \delta_{lm} \bar\delta_{mn} \frac{\mathcal{A}_{mn} }{ \bar\omega_1 }
\frac{ \partial f_{nm} }{ \partial k_\alpha }
%
\nonumber \\ &
%
%
-i\, \delta_{lm} \bar\delta_{mn} \bigg( 
\frac{ \mathcal{A}_{mn}^\alpha f_{nm} }{ \bar\omega_1 -\omega_{mn} }
\bigg)_{;\lambda}
%
%
-\frac{ \delta_{lm} \delta_{mn} }{ \bar\omega_1 }
\frac{ \partial^2 f_n }{ \partial k_\lambda \partial k_\alpha }
\Bigg] ,
\\
\label{eq:rho:3}
\rho_{mn}^{(3)}(t) &\equiv
 \rho_{mn}^{eee}(t) +\rho_{mn}^{eie}(t) +\rho_{mn}^{eei}(t) 
+\rho_{mn}^{eii}(t) +\rho_{mn}^{iee}(t) +\rho_{mn}^{iie}(t)
+\rho_{mn}^{iei}(t) +\rho_{mn}^{iii}(t) 
\nonumber \\ &=
\frac{ -e
^3 }{ 8\hbar^3 } \sum_{\omega_3 \omega_2 \omega_1}
\frac{ E_{\omega_3}^\lambda E_{\omega_2}^\beta E_{\omega_1}^\alpha
e^{ -i \bar\omega_s t } }{ \bar\omega_s -\omega_{mn} 
}
\bigg\lbrace 
%
%
\sum_{pl} 
\bigg[ 
\frac{ \mathcal{A}_{mp}^\lambda \bar\delta_{mp} \bar\delta_{lp} \bar\delta_{ln} 
}{ \bar\omega_2+\bar\omega_1 -\omega_{pn} } \bigg(
 \frac{ \mathcal{A}_{pl}^\beta \mathcal{A}_{ln}^\alpha f_{nl}
}{ \bar\omega_1 -\omega_{ln} }
-\frac{ f_{lp} \mathcal{A}_{pl}^\alpha \mathcal{A}_{ln}^\beta
}{ \bar\omega_1 -\omega_{pl} } \bigg)
\nonumber \\ &-
\frac{ \bar\delta_{lm} \bar\delta_{lp} \bar\delta_{pn} \mathcal{A}_{pn}^\lambda
}{ \bar\omega_2 +\bar\omega_1 -\omega_{mp} } \bigg(
 \frac{ \mathcal{A}_{ml}^\beta \mathcal{A}_{lp}^\alpha f_{pl}
}{ \bar\omega_1 -\omega_{lp} }
-\frac{ f_{lm} \mathcal{A}_{ml}^\alpha \mathcal{A}_{lp}^\beta
}{ \bar\omega_1 -\omega_{ml} } \bigg)
\bigg] \nonumber\\ &+
%
%
i\sum_l \bar\delta_{ml} \bar\delta_{ln} \bigg[
\frac{ \mathcal{A}_{ml}^\lambda }{ \bar\omega_2 +\bar\omega_1 -\omega_{ln} }
\bigg( \frac{ \mathcal{A}_{ln}^\alpha f_{nl} }{  \bar\omega_1 -\omega_{ln} } 
\bigg)_{;\beta}
-
\frac{ \mathcal{A}_{ln}^\lambda }{ \bar\omega_2 +\bar\omega_1 -\omega_{ml} }
\bigg( \frac{ \mathcal{A}_{ml}^\alpha f_{lm} }{  \bar\omega_1 -\omega_{ml} } 
\bigg)_{;\beta} \bigg]
\nonumber \\ &
%
%
+i\sum_l \bar\delta_{ml} \bar\delta_{ln} \bigg[
 \frac{ \mathcal{A}_{ml}^\lambda \mathcal{A}_{ln}^\beta / \bar\omega_1
}{ \bar\omega_2 +\bar\omega_1 -\omega_{ln} }
\frac{ \partial f_{nl} }{ \partial k_\alpha }
-\frac{ \mathcal{A}_{ml}^\beta \mathcal{A}_{ln}^\lambda / \bar\omega_1
}{ \bar\omega_2 +\bar\omega_1 -\omega_{ml} }
\frac{ \partial f_{lm} }{ \partial k_\alpha }
\bigg]
%
%
%
- \frac{ \bar\delta_{mn} \mathcal{A}_{mn}^\lambda /\bar\omega_1
}{ \bar\omega_2 +\bar\omega_1 }
\frac{ \partial^2 f_{nm} }{ \partial k_\beta \partial k_\alpha }
\nonumber \\ &
%
%
%
-i\sum_l \bigg(
 \frac{ \bar\delta_{ml} \bar\delta_{ln} }{ \bar\omega_2 +\bar\omega_1 
-\omega_{mn} }
\bigg( \frac{ \mathcal{A}_{ml}^\beta \mathcal{A}_{ln}^\alpha f_{nl}
}{ \bar\omega_1 -\omega_{ln} }
-\frac{ f_{lm} \mathcal{A}_{ml}^\alpha \mathcal{A}_{ln}^\beta
}{ \bar\omega_1 -\omega_{ml} } \bigg) \bigg)_{;\lambda}
\nonumber \\ &
%
%
-\bigg( \frac{ \bar\delta_{mn} }{ \bar\omega_2 +\bar\omega_1 -\omega_{mn} }
\bigg( \frac{ \mathcal{A}_{mn}^\alpha f_{nm} }{ \bar\omega_1 -\omega_{mn} }
\bigg)_{;\beta} \bigg)_{;\lambda}
%
%
%
-\bigg( \frac{ \bar\delta_{mn} \mathcal{A}_{mn}^\beta/ \bar\omega_1
}{ \bar\omega_2 +\bar\omega_1 -\omega_{mn} }
\frac{ \partial f_{nm} }{ \partial k_\alpha } \bigg)_{;\lambda}
%
%
-i \frac{ \delta_{mn}/ \bar\omega_1 }{ \bar\omega_2 +\bar\omega_1 }
\frac{ \partial^3 f_n }{\partial k_\lambda \partial k_\beta \partial k_\alpha }
\bigg\rbrace \, ,
\end{align}
\end{subequations}
where $\omega_s \equiv \omega_3 +\omega_2 +\omega_1$ and 
$\omega_{mn}=\epsilon_{mn}/\hbar$.
The perturbative expansion of the density matrix is then used to define the 
current density response as a power series in the electromagnetic field.
From the current density, we define the linear and nonlinear conductivity 
tensors, namely:
$\sigma_{\phi\alpha}^{(1)}$ the optical conductivity; 
$\sigma_{\phi\lambda\alpha}^{(2)}$ the quadratic response; 
$\sigma_{\phi\lambda\beta\alpha}^{(3)}$ the third-order response.
The Fourier components of the optical conductivity tensor read
\begin{align}
\label{eq:s1}
\sigma_{\phi\alpha}^{(1)}(\omega_1) \equiv 
 \sigma_{\phi\alpha}^{e}(\omega_1) 
+\sigma_{\phi\alpha}^{i}(\omega_1) =
\frac{ 4 i g \sigma_1 }{ \Omega } \hbar^2
\sum_\mathbf{k} \sum_{mn} 
\bigg( 
\frac{ v_{nm}^\phi v_{mn}^\alpha  }{ \hbar\bar\omega_1 -\epsilon_{mn}  }
\frac{ \bar\delta_{mn} f_{nm} }{ \epsilon_{mn} }
-\frac{ \delta_{mn} }{ \hbar }
\frac{ v_{nn}^\phi }{ \hbar\bar\omega_1 }
\frac{ \partial f_n }{ \partial k_\alpha }
\bigg) \, ,
\end{align}
where the conductivity scale is defined, for 2D systems, by the quantum of 
conductance $\sigma_1 = e^2/4\hbar = \pi e^2/ 2h$ and the summation of all 
wave vectors $\mathbf{k}$ represents the integration over the entire Brillouin 
zone.
In addition, the nondiagonal $(m\neq n)$ position matrix elements can be 
transformed to velocity matrix elements with $\mathcal{A}_{mn}^\alpha \equiv
-i\hbar \, v_{mn}^\alpha / \epsilon_{mn}$. \cite{Aversa1995}
To identify the inter- or intraband nature of the processes involved in any 
given tensor element, the interband processes are labeled by superscripts $e$, 
while the intraband motion is labeled by superscripts $i$.
By the same token, we separate the contributions to quadratic response 
according to the nature of the two interactions. At second-order this spawns 
$4$ processes: a purely interband $ee$ process, two mixed inter-intraband 
processes $ei$ and $ie$, and finally a purely intraband $ii$ process.
The respective Fourier components read
\begin{align}
\label{eq:s2}
 \sigma _{\phi\lambda\alpha} ^{(2)} ( \omega_2, \omega_1 ) &\equiv 
 \sigma _{\phi\lambda\alpha} ^{ee}  ( \omega_2, \omega_1 )
+\sigma _{\phi\lambda\alpha} ^{ei}  ( \omega_2, \omega_1 )
+\sigma _{\phi\lambda\alpha} ^{ie}  ( \omega_2, \omega_1 )
+\sigma _{\phi\lambda\alpha} ^{ii}  ( \omega_2, \omega_1 )
\nonumber\\ & =
\frac{ g \sigma_2 }{ \Omega } \frac{ \hbar \gamma_0 }{ a }
\sum_\mathbf{k} \sum_{mn} 
\frac{ v_{nm}^\phi }{ \hbar( \bar\omega_2 + \bar\omega_1) -\epsilon_{mn} } 
\bigg[ 
%
%
\hbar^2 \sum_l
\frac{ \bar{\delta}_{lm} \bar{\delta}_{ln} }{ \epsilon_{ml} \epsilon_{ln} }
\bigg( \frac{ v_{ml}^\lambda v_{ln}^\alpha f_{nl}
       }{ \hbar\bar\omega_1 -\epsilon_{ln} }
      -\frac{ f_{lm} v_{ml}^\alpha v_{ln}^\lambda
       }{ \hbar\bar\omega_1 -\epsilon_{ml} } \bigg) 
\nonumber \\ &
%
%
-\frac{ \hbar \, \bar{\delta}_{mn}  }{ \hbar\bar\omega_1 }
 \frac{ v_{mn}^\lambda }{ \epsilon_{mn} }
 \frac{ \partial f_{nm} }{ \partial k_\alpha }
%
%
-\hbar\, \bar{\delta}_{mn}
  \bigg( \frac{ v_{mn}^\alpha f_{nm}/ \epsilon_{mn} 
  }{ \hbar\bar\omega_1 -\epsilon_{mn} } \bigg)_{;\lambda}
%
%
+\frac{ \delta_{mn} }{ \hbar\bar\omega_1 }
\frac{ \partial^2 f_n }{ \partial k_\lambda \partial k_\alpha } \bigg]\, .
\end{align}
In contrast to linear response, the second-order conductivity scale, 
$\sigma_2$, depends explicitly on the physical properties of the system, namely 
the hopping energy $\gamma_0$ and the carbon-carbon bond length $a_0$.
For the second-order conductivity of 2D systems, the scale is set by $\sigma_2 
= e^3 a_0/ 4 |\gamma_0| \hbar$.
At third-order, we obtain eight terms involving inter- and intraband processes, 
namely
\begin{subequations}
\label{eq:s3}
\begin{align}
\label{eq:s3:eee}
\sigma_{\phi\lambda\beta\alpha}^{eee}( \omega_3, \omega_2, \omega_1 )
&=
-\frac{ g \sigma_3 }{ i \Omega } \frac{ \hbar^4 \gamma_0^2 }{ a_0^2 }
\sum_{plmn} \sum_\mathbf{k}
\frac{ \bar\delta_{ml} \bar\delta_{lp} \bar\delta_{pn}
 }{ \epsilon_{ml} \epsilon_{lp} \epsilon_{pn} }
\bigg[
\frac{ v_{nm}^\phi }{ \hbar\bar\omega_s -\epsilon_{mn} }
\bigg(
 \frac{ v_{ml}^\beta v_{lp}^\alpha f_{pl} }{ \hbar\bar\omega_1 -\epsilon_{lp} }
-\frac{ f_{lm} v_{ml}^\alpha v_{lp}^\beta }{ \hbar\bar\omega_1 -\epsilon_{ml} } 
\bigg)
\frac{ v_{pn}^\lambda }{ \hbar(\bar\omega_2 +\bar\omega_1) -\epsilon_{mp} }
\nonumber\\ &+
\frac{ v_{np}^\lambda }{
\hbar(\bar\omega_2 +\bar\omega_1 ) +\epsilon_{mp} } \bigg(
 \frac{ v_{pl}^\beta v_{lm}^\alpha f_{ml} }{ \hbar\bar\omega_1 -\epsilon_{lm} }
-\frac{ f_{lp} v_{pl}^\alpha v_{lm}^\beta }{ \hbar\bar\omega_1 -\epsilon_{pl} } 
\bigg)
\frac{ v_{mn}^\phi }{ \hbar\bar\omega_s +\epsilon_{mn} }
\bigg] \, ,
\\
%
%
\label{eq:s3:iee}
\sigma_{\phi\lambda\beta\alpha}^{\{iee\}}( \omega_3, \omega_2, \omega_1 ) &=
\frac{ g \sigma_3 }{ i \Omega } \frac{ \hbar^3 \gamma_0^2 }{ a_0^2 }
\sum_{lmn} \sum_\mathbf{k}
\bigg\{
\bigg( \frac{ v_{nm}^\phi }{ \hbar\omega_s -\epsilon_{mn} }
\bigg)_{;\lambda}
\frac{ \bar\delta_{ml} \bar\delta_{ln} / \epsilon_{ml} \epsilon_{ln}
}{ \hbar(\bar\omega_2 +\bar\omega_1) -\epsilon_{mn} }
\bigg(
 \frac{ v_{ml}^\beta v_{ln}^\alpha f_{nl} }{ \hbar\bar\omega_1 -\epsilon_{ln} }
-\frac{ f_{lm} v_{ml}^\alpha v_{ln}^\beta }{ \hbar\bar\omega_1 -\epsilon_{ml} }
\bigg)
\nonumber\\ &
%
%
-\frac{ v_{nm}^\phi \, \bar\delta_{ml} \bar\delta_{ln} }{ 
\hbar\bar\omega_s -\epsilon_{mn} }
\bigg[
 \frac{ v_{ml}^\lambda / \epsilon_{ml} }{
 \hbar(\bar\omega_2 +\bar\omega_1 ) -\epsilon_{ln} } \bigg(
\frac{ v_{ln}^\alpha f_{nl} /\epsilon_{ln} }{ \hbar\bar\omega_1 -\epsilon_{ln} }
\bigg)_{;\beta}
%
%
-\bigg(
\frac{ f_{lm} v_{ml}^\alpha /\epsilon_{ml} }{ \hbar\bar\omega_1 -\epsilon_{ml} }
  \bigg)_{;\beta}
 \frac{ v_{ln}^\lambda / \epsilon_{ln} }{
 \hbar(\bar\omega_2 +\bar\omega_1) -\epsilon_{ml} }
\bigg]
\nonumber \\ &-
%
%
%
\frac{1}{\hbar\bar\omega_1}
\frac{ \bar\delta_{ml} \bar\delta_{ln} \, v_{nm}^\phi 
}{ \hbar\bar\omega_s -\epsilon_{mn} }
\bigg[
 \frac{ v_{ml}^\lambda v_{ln}^\beta / \epsilon_{ml}\epsilon_{ln}
}{ \hbar(\bar\omega_2 +\bar\omega_1) -\epsilon_{ln} }
\frac{ \partial f_{nl} }{ \partial k_\alpha }
-\frac{ v_{ml}^\beta v_{ln}^\lambda / \epsilon_{ml}\epsilon_{ln}
}{ \hbar( \bar\omega_2 +\bar\omega_1) -\epsilon_{ml} }
\frac{ \partial f_{lm} }{ \partial k_\alpha }
\bigg] \bigg\} \, ,
\\
%
%
\label{eq:s3:iie}
\sigma_{\phi\lambda\beta\alpha}^{\{iie\}}( \omega_3, \omega_2, \omega_1 ) &=
\frac{ -g \sigma_3 }{ i\Omega } \frac{ \hbar^2 \gamma_0^2 }{ a_0^2 }
\sum_{mn} \sum_\mathbf{k}
\bigg[
\bigg( 
\frac{ v_{nm}^\phi }{ \hbar\bar\omega_s -\epsilon_{mn} }
 \bigg)_{;\lambda}
\frac{ \bar\delta_{mn} }{ \hbar(\bar\omega_2 +\bar\omega_1) -\epsilon_{mn} } 
\bigg(
\frac{ v_{mn}^\alpha f_{nm}/\epsilon_{mn} }{ \hbar\bar\omega_1 -\epsilon_{mn} }
\bigg)_{;\beta}
\nonumber \\ &
%
%
%
+\bigg( \frac{ \bar\delta_{mn} \,  v_{nm}^\phi }{
\hbar\bar\omega_s -\epsilon_{mn} } \bigg)_{;\lambda}
\frac{ v_{mn}^\beta/( \epsilon_{mn} \hbar\bar\omega_1 )
}{ \hbar(\bar\omega_2 +\bar\omega_1) -\epsilon_{mn} }
\frac{ \partial f_{nm} }{ \partial k_\alpha }
%
%
%
-\frac{ 1 }{ \hbar(\bar\omega_2 +\bar\omega_1)\hbar\bar\omega_1 }
\frac{ \bar\delta_{mn} \, v_{nm}^\phi }{ \hbar\bar\omega_s -\epsilon_{mn} }
\frac{ v_{mn}^\lambda }{ \epsilon_{mn} }
\frac{ \partial^2 f_{nm} }{ \partial k_\beta \partial k_\alpha }
\bigg] \, ,
\\
%
%
\label{eq:s3:iii}
\sigma_{\phi\lambda\beta\alpha}^{iii}( \omega_3, \omega_2, \omega_1 ) &=
\frac{ g \sigma_3 }{ i\Omega } \frac{ \hbar \gamma_0^2/a_0^2 }{
\hbar\bar\omega_s \hbar(\bar\omega_2+\omega_1) \hbar\bar\omega_1 }
\sum_n\sum_\mathbf{k}
\frac{ \partial v_{nn}^\phi }{ \partial k_\lambda }
\frac{ \partial^2 f_n }{\partial k_\beta \partial k_\alpha } \, ,
\end{align}
\end{subequations}
where the third-order nonlinear 2D conductivity unit reads $ \sigma_3\tequiv 
e^4 a_0^2 / 8 \gamma_0^2 \hbar $, and we make use of the contracted notation 
$\{iee\} = iee+eie+eei$, $\{iie\} = iie+iei+eii$.
\end{widetext}
It is important to highlight that we made use of several permutations of dummy 
indices and applied integration by parts for periodic functions
to derive the results shown in \Eqref{eq:s3}.

As identified in \Rcite{Aversa1995}, \Eqsref{eq:s2}, and \eqref{eq:s3} contain 
spurious divergences that can affect the evaluation of the nonlinear optical 
response of cold insulators.
For example, the $ee$ term, \Eqref{eq:s3:eee}, exhibits spurious divergences 
for all $m = n$ or $m = l$.
In addition to the spurious divergences, the results shown \Eqsref{eq:s1}, 
\eqref{eq:s2}, and \eqref{eq:s3}, include the expectable Drude-like terms 
associated with intraband motion of electrons.
As discussed in the main text, Sec.~\ref{sec:effTen}, the divergence appearing 
in the linear response, \Eqref{eq:s1}, represents intraband motion of the 
electronic system that is responsible for the Drude peak.
Beyond linear order, the problem becomes more complex and many unphysical 
terms can plague the calculation of the nonlinear response.
Below, we introduce a straightforward, although lengthy, procedure that 
disentangles the physical Drude-like terms from the regular part of the 
nonlinear response and removes the spurious divergences.

\subsection{Outline of the procedure}

By making use of the effective nonlinear rank-2 tensor defined in the main 
text, \Eqref{eq:sEff}, we generalize the procedure introduced in 
\Rcite{Hipolito2018} to regularize the nonlinear response.
We consider the case that all frequencies are taken to the DC limit 
simultaneously. In this limit, the analytically continued frequencies 
$\bar\omega \equiv \omega +i\eta$ can be mapped to a unique frequency 
$\bar\omega_n = x_n \varpi $, where $x_n$ are real numbers.
This transformation allows us to consider only one limit, $\varpi \to 0$, and 
thus handle the divergences occurring in the DC limit.
The analytic continuation of the frequencies ensures that the transformation 
holds, even for cases such as the nonlinear OR, where $\bar\omega_2 
+\bar\omega_1= \pm( \omega_1-\omega_1 ) +2i\eta $.
Moreover, it can be shown all divergences that arise from the combination of 
positive and negative frequency components are always canceled by a symmetric 
contribution that stems from the complementary frequency combination.

The procedure follows naturally from the definition of the effective tensor and 
can be summarized as follows:
group tensor elements according to the number of inter- and 
intraband transitions;
isolate all divergences in $\varpi$ by means of partial fraction 
decomposition with the identities
\begin{align*}
\frac{ 1 }{ \hbar\varpi(\epsilon \mp \hbar\varpi) } = 
\frac{ 1 }{ \hbar\varpi \,\epsilon} \pm \frac{ 1 }{ \epsilon(\epsilon \mp 
\hbar\varpi) }
\, ;
\end{align*}
for each divergence occurring at frequency component $\omega_s = 
\omega_n+\dots+\omega_1$, add all terms associated with the permutations of the 
spatial indices, \ie add all combinations for each value $\nu$.

\subsection{Isolating the divergences}

The procedure leads to a redefinition of the conductivity tensors in terms of a 
power series in $1/\varpi$.
The effective rank-2 tensors for the second-order process read
\begin{subequations}
\label{eq:s2:eff}
\begin{align}
\bar\sigma_{\phi\nu}^{ee} (\omega_s) &\equiv
\sump_{\omega_2\omega_1} \sump_{\lambda\alpha} \big(
  A_{\phi\lambda\alpha}^{ee}
+ a_{\phi\lambda\alpha}^{ee}
+ b_{\phi\lambda\alpha}^{ee} / \varpi \big) \, ,
\\
\bar\sigma_{\phi\nu}^{ei} (\omega_s) &\equiv
\sump_{\omega_2\omega_1} \sump_{\lambda\alpha} \big(
  A_{\phi\lambda\alpha}^{ei}
+ B_{\phi\lambda\alpha}^{ei} / \varpi \big) \, ,
\\
\bar\sigma_{\phi\nu}^{ie} (\omega_s) &\equiv
\sump_{\omega_2\omega_1} \sump_{\lambda\alpha}
A_{\phi\lambda\alpha}^{ie} \, ,
\\
\bar\sigma_{\phi\nu}^{ii} (\omega_s) &\equiv
\sump_{\omega_2\omega_1} \sump_{\lambda\alpha}
C_{\phi\lambda\alpha}^{ii} / \varpi^2 \, .
\end{align}
\end{subequations}
To facilitate the identification of the regular and divergent terms, we adopt 
the following convention: regular terms are represented by $A$ and $a$; linear 
divergences by $B$ and $b$; quadratic by $C$ and $c$; cubic by $D$.
In all cases, upper-case latin letters represent terms that involve the same 
number of interband transitions as the original tensor, whereas lower-case 
represent terms with one less interband transition.
The elements that can be trivially shown to vanish are not displayed, \eg 
$B_{\phi\lambda\beta\alpha}^{eee} = 0$.
In addition, the primed summation over the frequencies is evaluated with 
the restriction $\omega_s = \omega_n+\ldots+\omega_1$, while the primed 
summation over spatial indices $\lambda\ldots\alpha$ respects the combinations 
for effective tensor index $\nu$ listed in \Tbref{tab:nu}.

Note that the definitions of $A_{\phi\beta\alpha}^{ie}$ and 
$C_{\phi\beta\alpha}^{ii}$ are unchanged with respect to the original 
definitions in \Eqref{eq:s2}, as the former contains no divergences and the 
latter only contains terms that diverge quadratically with the photon frequency 
in the DC limit.
By the same token, the effective tensor for the third-order is cast as
\begin{subequations}
\label{eq:s3:eff}
\begin{align}
\bar\sigma_{\phi\nu}^{eee} (\omega_s) &\equiv
\sump_{\omega_3\omega_2\omega_1} \sump_{\lambda\beta\alpha} \big(
  A_{\phi\lambda\beta\alpha}^{eee}
+ a_{\phi\lambda\beta\alpha}^{eee}
+ b_{\phi\lambda\beta\alpha}^{eee} / \varpi \big) \, ,
\\
\bar\sigma_{\phi\nu}^{\{iee\}} (\omega_s) &\equiv
\sump_{\omega_3\omega_2\omega_1} \sump_{\lambda\beta\alpha} \big(
  A_{\phi\lambda\beta\alpha}^{\{iee\}}
+ a_{\phi\lambda\beta\alpha}^{\{iee\}}
+ B_{\phi\lambda\beta\alpha}^{\{iee\}} / \varpi
\nonumber\\ &
+ b_{\phi\lambda\beta\alpha}^{\{iee\}} / \varpi
+ c_{\phi\lambda\beta\alpha}^{\{iee\}} / \varpi^2 \big) \, ,
\, \\
\bar\sigma_{\phi\nu}^{\{iie\}} (\omega_s) &\equiv
\sump_{\omega_3\omega_2\omega_1} \sump_{\lambda\beta\alpha} \big(
  A_{\phi\lambda\beta\alpha}^{\{iie\}}
+ B_{\phi\lambda\beta\alpha}^{\{iie\}} / \varpi
\nonumber\\ &   
+ C_{\phi\lambda\beta\alpha}^{\{iie\}} / \varpi^2 \big) \, .
\, \\
\bar\sigma_{\phi\nu}^{iii} (\omega_s) &\equiv
\sump_{\omega_3\omega_2\omega_1} \sump_{\lambda\beta\alpha}
D_{\phi\lambda\beta\alpha}^{iii} / \varpi^3 \, ,
\end{align}
\end{subequations}
with $D_{\phi\lambda\beta\alpha}^{iii}$ inheriting the original definition used 
in \Eqref{eq:s3:iii}.
We begin by identifying the explicit form of all regular terms and address the 
divergent terms in the following subsections, where we remove the spurious 
divergences and identify all nonlinear Drude-like terms.
Note that throughout this process we relabeled several summation indices.
The regular terms at second-order read
\begin{widetext}
\begin{subequations}
\label{eq:reg:s2}
\begin{align}
\label{eq:s2:divs:A:ee}
A_{\phi\lambda\alpha}^{ee} &=
\hbar^2 C_2
\sum_{lmn} \sum_\mathbf{k}
\frac{ v_{nm}^\phi \, \bar\delta_{nm} }{ 
\hbar(\bar\omega_2 +\bar\omega_1) -\epsilon_{mn} }
\frac{ \bar\delta_{ml}\bar\delta_{ln} }{ \epsilon_{ml}\epsilon_{ln} }
\bigg( 
\frac{ v_{ml}^\lambda v_{ln}^\alpha f_{nl} }{ \hbar\bar\omega_1 -\epsilon_{ln} }
-\frac{ f_{lm} v_{ml}^\alpha v_{ln}^\lambda }{ \hbar\bar\omega_1 -\epsilon_{ml} 
}
\bigg) \, ,
\\ 
\label{eq:s2:divs:a:ee}
a_{\phi\lambda\alpha}^{ee} &=
-\hbar^2 C_2
\frac{ \bar\omega_1 }{ \bar\omega_2 +\bar\omega_1 }
\sum_{mn} \sum_\mathbf{k}
\frac{ v_{nn}^\phi \, \bar\delta_{mn} \, f_{mn} }{ \epsilon_{mn}^3 }
\bigg( 
 \frac{ v_{nm}^\lambda v_{mn}^\alpha }{ \hbar\bar\omega_1 -\epsilon_{mn} }
-\frac{ v_{nm}^\alpha v_{mn}^\lambda }{ \hbar\bar\omega_1 +\epsilon_{mn} }
\bigg) \, ,
\\ 
\label{eq:s2:divs:A:ei}
A_{\phi\lambda\alpha}^{ei} &=
-\hbar C_2
\frac{ \bar\omega_2 +\bar\omega_1 }{ \bar\omega_1 }
\sum_{mn} \sum_\mathbf{k}
\frac{ v_{nm}^\phi v_{mn}^\lambda }{ 
\hbar( \bar\omega_2 +\bar\omega_1 ) -\epsilon_{mn} }
\frac{ \bar\delta_{mn} }{ \epsilon_{mn} }
\frac{ \partial f_{nm} }{ \partial k_\alpha }
\, ,
\\ 
\label{eq:s2:divs:A:ie}
A_{\phi\lambda\alpha}^{ie} &=
-\hbar C_2 \sum_{mn} \sum_\mathbf{k}
\frac{ v_{nm}^\phi \bar\delta_{nm} }{
\hbar( \bar\omega_2 +\bar\omega_1 ) -\epsilon_{mn} }
\bigg( \frac{ v_{mn}^\alpha f_{nm} /\epsilon_{mn} }{ 
\hbar\bar\omega_1 -\epsilon_{mn} }
\bigg)_{;\lambda}
\, ,
\end{align}
\end{subequations}
with $C_2 = (  g\sigma_2/\Omega )( \hbar\gamma_0/a_0 )$. The regularized 
expressions for third-order processes read
\begin{subequations}
\label{eq:reg:s3}
\begin{align}
\label{eq:s3:divs:A:eee}
A_{\phi\lambda\beta\alpha}^{eee} &=
-\hbar^4 C_3
\sum_{plmn} \sum_\mathbf{k}
\frac{ \bar\delta_{pm} 
\bar\delta_{ml} \bar\delta_{lp} \bar\delta_{pn}
\bar\delta_{nm} }{ \epsilon_{ml} \epsilon_{lp} \epsilon_{pn} }
\bigg[
\frac{ v_{nm}^\phi }{ \hbar\bar\omega_s -\epsilon_{mn} }
\bigg(
 \frac{ v_{ml}^\beta v_{lp}^\alpha f_{pl} }{ \hbar\bar\omega_1 -\epsilon_{lp} }
-\frac{ f_{lm} v_{ml}^\alpha v_{lp}^\beta }{ \hbar\bar\omega_1 -\epsilon_{ml} } 
\bigg)
\frac{ v_{pn}^\lambda }{ \hbar(\bar\omega_2 +\bar\omega_1) -\epsilon_{mp} }
\nonumber\\ &+
\frac{ v_{np}^\lambda }{
\hbar(\bar\omega_2 +\bar\omega_1 ) +\epsilon_{mp} } \bigg(
 \frac{ v_{pl}^\beta v_{lm}^\alpha f_{ml} }{ \hbar\bar\omega_1 -\epsilon_{lm} }
-\frac{ f_{lp} v_{pl}^\alpha v_{lm}^\beta }{ \hbar\bar\omega_1 -\epsilon_{pl} } 
\bigg)
\frac{ v_{mn}^\phi }{ \hbar\bar\omega_s +\epsilon_{mn} }
\bigg]
\, ,
\\ 
\label{eq:s3:divs:A:iee}
A_{\phi\lambda\beta\alpha}^{\{iee\}} &=
\hbar^3 C_3
\sum_{lmn} \sum_\mathbf{k} \bar\delta_{nm} \bar\delta_{ml} \bar\delta_{ln}
\bigg\{
\bigg( 
\frac{ v_{nm}^\phi }{ \hbar\bar\omega_s 
-\epsilon_{mn} }
\bigg)_{;\lambda}
\frac{ 1 / \epsilon_{ml} \epsilon_{ln}
}{ \hbar(\bar\omega_2 +\bar\omega_1) -\epsilon_{mn} }
\bigg(
 \frac{ v_{ml}^\beta v_{ln}^\alpha f_{nl} }{ \hbar\bar\omega_1 -\epsilon_{ln} }
-\frac{ f_{lm} v_{ml}^\alpha v_{ln}^\beta }{ \hbar\bar\omega_1 -\epsilon_{ml} }
\bigg)
\nonumber \\ &
-\frac{v_{nm}^\phi }{ \hbar\bar\omega_s -\epsilon_{mn} } \bigg[ 
  \frac{ v_{ml}^\lambda /\epsilon_{ml} }{
\hbar(\bar\omega_2+\bar\omega_1) -\epsilon_{ln} }
 \bigg( \frac{ v_{ln}^\alpha f_{nl} /\epsilon_{ln}
}{ \hbar\bar\omega_1 -\epsilon_{ln} }
\bigg)_{;\beta}
-\bigg( \frac{ f_{lm} v_{ml}^\alpha /\epsilon_{ml}
}{ \hbar\bar\omega_1 -\epsilon_{ml} }
\bigg)_{;\beta}
\frac{ v_{ln}^\lambda /\epsilon_{ln} }{
\hbar(\bar\omega_2+\bar\omega_1) -\epsilon_{ml} }
\bigg]
\nonumber\\ &
-\frac{ v_{nm}^\phi }{ \epsilon_{mn} \epsilon_{ml} \epsilon_{ln} }
\bigg[
\frac{ \bar\omega_s /\bar\omega_1 }{
\hbar\bar\omega_s -\epsilon_{mn} }
\bigg(
\frac{ v_{ml}^\lambda v_{ln}^\beta }{ 
\hbar(\bar\omega_2+\bar\omega_1) -\epsilon_{ln} }
 \frac{ \partial f_{nl} }{ \partial k_\alpha }
-\frac{ \partial f_{lm} }{ \partial k_\alpha }
\frac{ v_{ml}^\beta v_{ln}^\lambda }{ 
\hbar(\bar\omega_2+\bar\omega_1) -\epsilon_{ml} }
\bigg)
\nonumber \\&
-\frac{ \bar\omega_2 +\bar\omega_1 }{ \bar\omega_1 }
\bigg(
\frac{ v_{ml}^\lambda v_{ln}^\beta / \epsilon_{ln} 
}{ \hbar(\bar\omega_2+\bar\omega_1) -\epsilon_{ln} }
 \frac{ \partial f_{nl} }{ \partial k_\alpha }
-\frac{ \partial f_{lm} }{ \partial k_\alpha }
\frac{ v_{ml}^\beta v_{ln}^\lambda / \epsilon_{ml}
}{ \hbar(\bar\omega_2+\bar\omega_1) -\epsilon_{ml} }
\bigg) \bigg] \bigg\} \, ,
\\ 
\label{eq:s3:divs:a:iie}
A_{\phi\lambda\beta\alpha}^{\{iie\}} &=
-\hbar^2 C_3
\sum_{mn} \sum_\mathbf{k}
 \bar\delta_{mn}
\Bigg[ 
\bigg( 
\frac{ v_{nm}^\phi }{ \hbar\bar\omega_s -\epsilon_{mn} } \bigg)_{;\lambda}
\frac{ 1 }{ \hbar(\bar\omega_2 +\bar\omega_1) -\epsilon_{mn} } 
\bigg(
\frac{ v_{mn}^\alpha f_{nm}/\epsilon_{mn} }{ \hbar\bar\omega_1 -\epsilon_{mn} }
\bigg)_{;\beta}
%
%
%
-\frac{ \bar\omega_s^2/\bar\omega_1  }{ 
\bar\omega_1 +\bar\omega_2 }
\frac{ v_{nm}^\phi v_{mn}^\lambda / \epsilon_{mn}^3 }{
\hbar\bar\omega_s -\epsilon_{mn} }
\frac{ \partial^2 f_{nm} }{ \partial k_\beta   \partial k_\alpha }
\nonumber \\ &
+\bigg[ \bigg( 
\frac{ v_{nm}^\phi }{ \hbar\bar\omega_s -\epsilon_{mn} } \bigg)_{;\lambda}
\frac{ (\bar\omega_2 +\bar\omega_1)/\bar\omega_1 }{
\hbar(\bar\omega_2 +\bar\omega_1) -\epsilon_{mn} }
-\frac{ \bar\omega_s }{ \bar \omega_1 }
\bigg( \frac{ v_{nm}^\phi /\epsilon_{mn}
}{ \hbar\bar\omega_s -\epsilon_{mn} } \bigg)_{;\lambda}
\bigg]
\frac{ v_{mn}^\beta }{ \epsilon_{mn}^2 }
\frac{ \partial f_{nm} }{ \partial k_\alpha }
\Bigg] \, ,
\\
\label{eq:s3:divs:a:eee}
a_{\phi\lambda\beta\alpha}^{eee} &=
-\hbar^4 C_3
\sum_{lmn} \sum_\mathbf{k} \Bigg(
\frac{ \bar\delta_{nm} \bar\delta_{ml} \bar\delta_{ln}
}{ \epsilon_{mn} \epsilon_{ml} \epsilon_{ln} } \bigg\{
\nonumber\\&
\frac{ \bar\omega_2 +\bar\omega_1 }{ \bar\omega_s }
\bigg[
 \bigg(
   \frac{ v_{nl}^\beta v_{lm}^\alpha f_{lm} 
    }{ \hbar\bar\omega_1 +\epsilon_{ml} }
  -\frac{ f_{nl} v_{nl}^\alpha v_{lm}^\beta
    }{ \hbar\bar\omega_1 +\epsilon_{ln} }
 \bigg)
 \frac{ v_{mn}^\lambda v_{nn}^\phi /\epsilon_{mn} 
    }{ \hbar\bar\omega_s +\epsilon_{mn} }
+\frac{ v_{nn}^\phi v_{nm}^\lambda /\epsilon_{mn} 
    }{ \hbar\bar\omega_s -\epsilon_{mn} }
 \bigg(
   \frac{ v_{nl}^\beta v_{lm}^\alpha f_{nl}
    }{ \hbar\bar\omega_1 -\epsilon_{ln} }
  -\frac{ f_{lm} v_{nl}^\alpha v_{lm}^\beta 
    }{ \hbar\bar\omega_1 +\epsilon_{ml} }
 \bigg)
\bigg]
\nonumber\\&
+\frac{ \bar\omega_1 }{ \bar\omega_s }
 \frac{ v_{nn}^\phi }{ \epsilon_{mn} }
\bigg[
\bigg( 
  \frac{ v_{nl}^\beta v_{lm}^\alpha f_{lm}/\epsilon_{ml}
  }{ \hbar\bar\omega_1 +\epsilon_{ml} }
 -\frac{ f_{nl} v_{nl}^\alpha v_{lm}^\beta /\epsilon_{ln}
  }{ \hbar\bar\omega_1 +\epsilon_{ln} }
\bigg) v_{mn}^\lambda
-v_{nm}^\lambda \bigg( 
  \frac{ v_{ml}^\beta v_{ln}^\alpha f_{nl}/\epsilon_{ln}
  }{ \hbar\bar\omega_1 -\epsilon_{ln} }
 -\frac{ f_{lm} v_{ml}^\alpha v_{ln}^\beta /\epsilon_{ml}
  }{ \hbar\bar\omega_1 -\epsilon_{ml} }
\bigg)
\bigg] \bigg\}
\nonumber\\&
-\frac{ \bar\delta_{ml} \bar\delta_{mn}
}{ \epsilon_{ml}^2 \epsilon_{mn} } \bigg[
\frac{ \bar\omega_1 }{ \bar\omega_2 +\bar\omega_1 } \bigg(
  \frac{ v_{nm}^\phi v_{mn}^\lambda }{ \hbar\bar\omega_s -\epsilon_{mn} }
 +\frac{ v_{nm}^\lambda v_{mn}^\phi }{ \hbar\bar\omega_s +\epsilon_{mn} }
\bigg)
\frac{ f_{lm} }{ \epsilon_{ml} }
\bigg(
  \frac{ v_{ml}^\beta v_{lm}^\alpha }{ \hbar\bar\omega_1 +\epsilon_{ml} }
 -\frac{ v_{ml}^\alpha v_{lm}^\beta }{ \hbar\bar\omega_1 -\epsilon_{ml} }
\bigg)
\nonumber\\&
-\frac{ \bar\omega_2 +\bar\omega_1 }{ \bar\omega_s } \bigg(
  \frac{ v_{nm}^\phi v_{mn}^\lambda }{ \hbar\bar\omega_s -\epsilon_{mn} }
 +\frac{ v_{nm}^\lambda v_{mn}^\phi }{ \hbar\bar\omega_s +\epsilon_{mn} }
\bigg)
\frac{ f_{lm} }{ \epsilon_{mn} }
\frac{ v_{ml}^\beta v_{lm}^\alpha -v_{ml}^\alpha v_{lm}^\beta
}{ \epsilon_{ml} }
\bigg]
\Bigg) \, ,
\\ 
\label{eq:s3:divs:a:iee}
a_{\phi\lambda\beta\alpha}^{\{iee\}} &=
\hbar^3 C_3
\sum_{mn} \sum_\mathbf{k} \bar\delta_{mn}
\Bigg(
-\frac{ \bar\omega_1^2 \, \big( v_{nn}^\phi \big)_{;\lambda}
}{ \bar\omega_s( \bar\omega_2 +\bar\omega_1 ) }
\frac{ f_{nm} }{ \epsilon_{mn}^4 }
\bigg(
 \frac{ v_{nm}^\beta  v_{mn}^\alpha }{ \hbar\bar\omega_1 -\epsilon_{mn} }
-\frac{ v_{nm}^\alpha v_{mn}^\beta  }{ \hbar\bar\omega_1 +\epsilon_{mn} }
\bigg)
\nonumber \\ &
+
\frac{ \bar\omega_1  }{ \bar\omega_s }
\frac{ v_{nn}^\phi }{ \epsilon_{mn} } \bigg[
\frac{ v_{nm}^\lambda }{ \hbar(\bar\omega_1 +\bar\omega_1) -\epsilon_{mn} }
\bigg(
\frac{ v_{mn}^\alpha f_{nm} /\epsilon_{mn}^2 
}{ \hbar\bar\omega_1-\epsilon_{mn} } \bigg)_{;\beta}
-
\bigg( \frac{ v_{nm}^\alpha f_{nm}/\epsilon_{mn}^2 
}{ \hbar\bar\omega_1 +\epsilon_{mn} } \bigg)_{;\beta}
\frac{ v_{mn}^\lambda }{ \hbar(\bar\omega_1 +\bar\omega_1) +\epsilon_{mn} }
\bigg]
\nonumber \\&-
\frac{ ( \bar\omega_2 +\bar\omega_1 ) }{ \bar\omega_s }
\frac{ v_{nn}^\phi }{ \epsilon_{mn}^2 }
\bigg[ 
\frac{ v_{nm}^\lambda }{ \hbar( \bar\omega_2 +\bar\omega_1 ) -\epsilon_{mn} }
\bigg( \frac{ v_{mn}^\alpha f_{nm}
}{ \epsilon_{mn}^2 } \bigg)_{;\beta}
+
\bigg( \frac{ v_{nm}^\alpha f_{nm}
}{ \epsilon_{mn}^2 } \bigg)_{;\beta}
\frac{ v_{mn}^\lambda }{ \hbar( \bar\omega_2 +\bar\omega_1 ) -\epsilon_{mn} }
\bigg]
\nonumber \\ &+
\frac{ ( \bar\omega_2 +\bar\omega_1 )^2 }{
\bar\omega_s \, \bar\omega_1 }
\frac{ v_{nn}^\phi }{ \epsilon_{mn}^4 }
\frac{ \partial f_{nm} }{ \partial k_\alpha }
\bigg(
 \frac{ v_{nm}^\lambda v_{mn}^\beta }{ 
 \hbar(\bar\omega_2+\bar\omega_1) -\epsilon_{mn} }
+\frac{ v_{nm}^\beta v_{mn}^\lambda }{ 
 \hbar(\bar\omega_2+\bar\omega_1) +\epsilon_{mn} }
\bigg) \Bigg)\, .
\end{align}
\end{subequations}
where $C_3 = ( g\sigma_3/ i\Omega )( \gamma_0^2/a_0^2 )$.
These expressions are regular at all frequencies and include all terms 
necessary to characterize the nonlinear response of cold insulators up to 
third-order.

\subsection{Divergences: spurious and Drude-like}
\label{sec:reg}
Having identified all divergent terms, we explicitly remove the spurious 
terms by showing that these terms vanish in the calculation of the effective 
tensor and also identify the nonlinear contributions to the Drude-like terms.
We start by addressing the second-order response and then proceed to the 
third-order.
\subsubsection{Quadratic response}
At second-order, the current density response contains divergences in three 
terms, namely in the $ee$, $ei$, and $ii$ terms.
The divergences in the purely interband term, $b_{\phi\lambda\alpha}^{ee}$, 
are immediately shown to vanish for all cases with $\lambda = \alpha$, \ie $\nu 
= \{1,2,3\}$.
For the remaining cases, $\nu = \{4,5,6\}$, it is sufficient to consider the 
following combination
\begin{align}
 \frac{ b_{\phi\lambda\alpha}^{ee} }{ \varpi } 
+\frac{ b_{\phi\alpha\lambda}^{ee} }{ \varpi } &=
\frac{ \hbar^2 C_2 }{ \hbar ( \bar\omega_2 +\bar\omega_1 ) }
\sum_{mn} \sum_\mathbf{k}
\frac{ v_{nn}^\phi \, \bar\delta_{mn} \, f_{mn} }{ \epsilon_{mn} }
\frac{ 
 v_{nm}^\lambda v_{mn}^\alpha -v_{nm}^\alpha v_{mn}^\lambda
-v_{nm}^\alpha v_{mn}^\lambda +v_{nm}^\lambda v_{mn}^\alpha 
}{ \epsilon_{mn}^2 } = 0 \, ,
\end{align}
thus showing that all divergences in the purely interband term are vanish.
Upon decomposition of the regular and divergent parts of the $ei$ term, we 
verify that 
\begin{align}
\label{eq:s2:ei}
\frac{ B_{\phi\lambda\alpha}^{ei} }{ \varpi } &=
\frac{ \hbar C_2 }{ \hbar\bar\omega_1 }
\sum_{mn} \sum_\mathbf{k}
\frac{ \bar\delta_{mn} v_{nm}^\phi v_{mn}^\lambda }{ \epsilon_{mn}^2 }
\frac{ \partial f_{nm} }{ \partial k_\alpha }
\end{align}
is a natural second-order Drude-like term that contributes only to the 
response of metallic systems or doped insulators.
The purely intraband term $ii$ does not contain any regular parts and defines 
the quadratic Drude-like peak
\begin{align}
\label{eq:s2:ii}
\frac{ C_{\phi\lambda\alpha}^{ii} }{ \varpi^2 } =
\frac{ C_2 }{ \hbar( \bar\omega_2 +\bar\omega_1 )\hbar\bar\omega_1 }
\sum_n \sum_\mathbf{k}
v_{nn}^\phi
\frac{ \partial^2 f_n }{ \partial k_\lambda \partial k_\alpha }
\, . 
\end{align}

\subsubsection{3rd-order purely interband and purely intraband}
At third-order, the separation of the natural contributions to the nonlinear 
Drude peak from the spurious divergences is not trivial, particularly in 
processes involving inter- and intraband transitions.
We start by addressing the spurious divergences in the purely interband 
contribution
\begin{subequations}
\begin{align}
 \frac{ b _{\phi\nu}^{eee} }{ \varpi } &= 
-\hbar^4 C_3 \sum_{lmn} \sump_{\lambda\beta\alpha} \sum_\mathbf{k}
\bigg\{
\frac{ \bar\delta_{nm} \bar\delta_{ml} \bar\delta_{ln} 
}{ \epsilon_{mn}^2 \epsilon_{ml}^2 \epsilon_{ln}^2 } 
\frac{ v_{nn}^\phi }{ \hbar \bar\omega_s }
\Big[
 f_{lm} \epsilon_{ln} \big( 
 V_{nlm}^{\beta\alpha\lambda} +V_{nml}^{\lambda\alpha\beta} \big)
-f_{nl} \epsilon_{ml} \big( 
 V_{nlm}^{\alpha\beta\lambda} +V_{nml}^{\lambda\alpha\beta} \big)
 \Big]
\nonumber\\ &
+
\frac{ \bar\delta_{ml} \bar\delta_{ln} }{ \epsilon_{ml}^2 \epsilon_{mn}^2 } 
f_{lm} \frac{ 
v_{nm}^\phi v_{mn}^\lambda -v_{nm}^\lambda v_{mn}^\phi }{ \hbar \bar\omega_1 }
\frac{ 
v_{ml}^\beta v_{lm}^\alpha -v_{ml}^\alpha v_{lm}^\beta }{ \epsilon_{ml} }
\bigg\}
\, ,
\end{align}
where $V_{nml}^{\lambda\alpha\beta} \equiv v_{nm}^\lambda v_{ml}^\beta 
v_{ln}^\alpha$.
The latter arises from the cases where $m = l$ in \Eqref{eq:s3:eee} (note that 
we replace $p\to m$ to recover the $l$ index) and vanishes for all $\{lmn\}$ by 
considering all combinations to the effective tensor.
To prove that the former [associated with $m=n$ terms in \Eqref{eq:s3:eee}]
vanishes, we begin by expanding the Fermi energy differences, add a second copy 
with interchanged indices $m,l$ and then make use of the combinations that 
define the effective tensor
\begin{align}
 \frac{ b _{\phi\nu}^{eee} }{ \varpi } &= 
-\frac{ \hbar^4 C_3 }2 \sum_{lmn} \sum_\mathbf{k}
\frac{ \bar\delta_{nm} \bar\delta_{ml} \bar\delta_{ln} 
}{ \epsilon_{mn}^2 \epsilon_{ml}^2 \epsilon_{ln}^2 } 
\frac{ v_{nn}^\phi }{ \hbar \bar\omega_s }
\sump_{\lambda\beta\alpha} \Big[
 f_l \epsilon_{mn} \big(
  V_{nlm}^{\beta\alpha\lambda} +V_{nml}^{\lambda\alpha\beta}
 -V_{nml}^{\beta\alpha\lambda} -V_{nlm}^{\lambda\alpha\beta} \big)
\nonumber\\ &
+( f_m \epsilon_{ln} -f_n \epsilon_{mn} ) 
 \big( V_{nlm}^{\beta\alpha\lambda} +V_{nml}^{\lambda\alpha\beta}
      -V_{nml}^{\beta\alpha\lambda} -V_{nlm}^{\lambda\alpha\beta} \big) 
\Big]
= 0\, ,
\end{align}
\end{subequations}
that vanishes for all effective tensors.
There are no quadratic divergences and the purely intraband contribution has no 
spurious divergences, it is physical and defines the cubic Drude-like term for 
third-order response
\begin{align}
\label{eq:s3:divs:D:iii}
\frac{ D_{\phi\lambda\beta\alpha}^{iii} }{ \varpi^3 } =
\frac{ \hbar C_3 }{
\hbar \bar\omega_s \hbar( \bar\omega_2 +\bar\omega_1 )\hbar\bar\omega_1 }
\sum_n \sum_\mathbf{k}
\frac{ \partial v_{nn}^\phi }{ \partial k_\lambda  }
\frac{ \partial^2 f_n }{ \partial k_\beta \partial k_\alpha }
\, .
\end{align}

\subsubsection{3rd-order mixed \{iee\} --- two inter- and one intraband}

The analysis of the divergent terms involving three bands can be facilitated, 
provided that we consider not the individual tensor, but rather 
$B_{\phi\lambda\beta\alpha}^{\{iee\}} \to \big(
B_{\phi\lambda\beta\alpha}^{\{iee\}}
+B_{\phi\beta\lambda\alpha}^{\{iee\}} \big)/2$.
The permutation of indices $\{\lambda,\beta\}$ is consistent with all
effective tensor elements.
Using this transformation, the divergence involving three band processes 
defines a contribution for a linear Drude-like term in the third-order 
response. 
\begin{align}
 \label{eq:s3:divs:B:iee}
\frac{ B_{\phi\lambda\beta\alpha}^{\{iee\}} 
      +B_{\phi\beta\lambda\alpha}^{\{iee\}} }{ 2\varpi } &=
-\frac{ \hbar^3 C_3 }{ 2\hbar \bar\omega_1 }
\sum_{lmn} \sum_\mathbf{k}
\frac{ v_{nm}^\phi \bar\delta_{nm} \bar\delta_{ml} \bar\delta_{ln}
}{ \epsilon_{mn} \epsilon_{ml} \epsilon_{ln} }
\frac{ v_{ml}^\lambda v_{ln}^\beta +v_{ml}^\beta v_{ln}^\lambda 
}{ \epsilon_{ml} \epsilon_{ln} }
\bigg(
 \frac{ \partial f_{n} }{ \partial k_\alpha } \epsilon_{ml}
+\frac{ \partial f_{m} }{ \partial k_\alpha } \epsilon_{ln}
-\frac{ \partial f_{l} }{ \partial k_\alpha } \epsilon_{mn}
\bigg) \, .
\end{align}
The origin of this can be traced to the $eei$ branch of the density 
matrix, that is usually discarded from the onset in the calculation of the 
response of cold insulators \cite{Aversa1995}.

Proceeding to the divergence present in the processes involving only two bands, 
we find
\begin{subequations}
\begin{align}
\label{eq:s3:divs:b}
\frac{ b_{\phi\lambda\beta\alpha}^{\{iee\}} }{ \varpi } &=
\hbar^3 C_3
\sum_{mn} \sum_\mathbf{k}
\frac{ \bar\delta_{mn} }{ \hbar \bar\omega_s }
\bigg\{
\frac{1}{x_2}
\frac{ \partial v_{nn}^\phi }{ \partial k_\lambda }
\frac{ v_{nm}^\beta v_{mn}^\alpha +v_{nm}^\alpha v_{mn}^\beta 
}{ \epsilon_{mn}^4 }
f_{nm}
+\frac{ v_{nn}^\phi }{ \epsilon_{mn}^2 } \bigg[
  v_{nm}^\lambda \bigg( 
  \frac{ v_{mn}^\alpha f_{nm} }{ \epsilon_{mn}^2 } \bigg)_{;\beta}
 +\bigg( \frac{ v_{nm}^\alpha f_{nm} }{ \epsilon_{mn}^2 } \bigg)_{;\beta}
  v_{mn}^\lambda \bigg]
\nonumber\\ &
+x_2
\frac{ v_{nn}^\phi }{ \epsilon_{mn}^2 }
\frac{ v_{nm}^\lambda v_{mn}^\beta +v_{nm}^\beta v_{mn}^\lambda
}{ \epsilon_{mn}^2 }
\frac{ \partial f_{nm} }{ \partial \beta }
\bigg\} \, .
\end{align}
The explicit dependence of the response on 
$1/x_2 \equiv \bar\omega_1/(\bar\omega_2 +\bar\omega_1)$ and 
$x_2   \equiv (\bar\omega_2 +\bar\omega_1)/\bar\omega_1$ indicates that we must 
consider with care the permutations of frequencies whenever $\omega_2 \neq 
\omega_1$
\begin{align}
\label{eq:s3:divs:b2}
\frac{ b_{\phi\lambda\beta\alpha}^{\{iee\}} }{ \varpi } &\equiv
\frac{ 
 b_{\phi\lambda\beta\alpha}^{\{iee\}}( \omega_3, \omega_2, \omega_1 )
+b_{\phi\lambda\beta\alpha}^{\{iee\}}( \omega_3, \omega_1, \omega_2 )
}{ 2\varpi }
\nonumber\\& =
\frac{ \hbar^3 C_3 }{ 2 }
\sum_{mn} \sum_\mathbf{k}
\frac{ \bar\delta_{mn} }{ \hbar \bar\omega_s }
\bigg\{
\frac{ \partial v_{nn}^\phi }{ \partial k_\lambda }
\frac{ v_{nm}^\beta v_{mn}^\alpha +v_{nm}^\alpha v_{mn}^\beta 
}{ \epsilon_{mn}^4 }
f_{nm}
+2\frac{ v_{nn}^\phi }{ \epsilon_{mn}^2 } \bigg[
  v_{nm}^\lambda \bigg( 
  \frac{ v_{mn}^\alpha f_{nm} }{ \epsilon_{mn}^2 } \bigg)_{;\beta}
 +\bigg( \frac{ v_{nm}^\alpha f_{nm} }{ \epsilon_{mn}^2 } \bigg)_{;\beta}
  v_{mn}^\lambda \bigg]
\nonumber\\ &
+\frac{ ( \bar\omega_1 +\bar\omega_2 )^2 }{ \bar\omega_1 \bar\omega_2 }
\frac{ v_{nn}^\phi }{ \epsilon_{mn}^2 }
\frac{ v_{nm}^\lambda v_{mn}^\beta +v_{nm}^\beta v_{mn}^\lambda
}{ \epsilon_{mn}^2 }
\frac{ \partial f_{nm} }{ \partial \beta }
\bigg\} \, .
\end{align}
As in the three-band process, this term represents a physical contribution to 
the response of metallic systems and doped semiconductors that results in an 
additional linear Drude-like term for the cubic response function.
\end{subequations}

The remaining term in this class of processes
\begin{align}
\label{eq:s3:divs:c:iee}
\frac{ c_{\phi\lambda\beta\alpha}^{\{iee\}} }{ \varpi^2 } &= 
-\frac{ \hbar^3 C_3  }{ \hbar \bar\omega_s  \hbar \bar \omega_1 }
\sum_{mn} \sum_\mathbf{k}
v_{nn}^\phi \bar\delta_{mn}
\Bigg[
\bigg( f_{nm}
\frac{ v_{nm}^\beta v_{mn}^\alpha -v_{nm}^\alpha v_{mn}^\beta
}{ \epsilon_{mn}^3 } \bigg)_{;\lambda} 
+\frac{ \partial f_{nm} }{ \partial k_\alpha }
 \frac{ v_{nm}^\lambda v_{mn}^\beta -v_{nm}^\beta v_{mn}^\lambda
 }{ \epsilon_{mn}^3 } \Bigg] = 0 \, 
\end{align}
is shown to vanish for all effective tensor, as the numerators in the form 
$v_{nm}^\beta v_{mn}^\alpha -v_{nm}^\alpha v_{mn}^\beta$ cancel for all 
$\nu$.

\subsubsection{3rd-order mixed \{iie\} --- one inter- and two intraband}
The last divergences involve a single interband transition and two 
intraband processes.
Such terms are never considered in \Rcite{Aversa1995}, as these divergences 
stem from the $iei$ and $eii$ branches of the density matrix, that are 
discarded from the onset in the response of cold insulators. 
The linear divergences represent a physical contribution to the Drude-peak
\begin{align}
\label{eq:s3:divs:B:iie}
 \frac{ B_{\phi\nu}^{\{iie\}} }{ \varpi }&=
-\frac{ \hbar^2 C_3  }{ \hbar \bar\omega_1 }
\sum_{mn} \sum_\mathbf{k} \sump_{\lambda\beta\alpha}
\bar\delta_{mn} \bigg[
\bigg( \frac{ v_{nm}^\phi }{ \epsilon_{mn} } \bigg)_{;\lambda}
\frac{ v_{mn}^\beta }{ \epsilon_{mn}^2 }
\frac{ \partial f_{nm} }{ \partial k_\alpha }
+\frac{ \bar\omega_s }{ \bar\omega_2 +\bar\omega_1 }
\frac{v_{nm}^\phi v_{nm}^\lambda }{ \epsilon_{mn}^3 }
\frac{ \partial^2 f_{nm} }{ \partial k_\beta \partial k_\alpha }
\bigg]\, .
\end{align}
In contrast, the quadratic divergence is spurious and shown to vanish, upon 
integration, for all effective tensors upon summation of pairs of bands
\begin{align}
\label{eq:s3:divs:C:iie}
 \frac{ C_{\phi\nu}^{\{iie\}} }{ \varpi^2 }&=
-\frac{ \hbar^2 C_3  }{ \hbar(\bar\omega_2+\bar\omega_1)\hbar \bar\omega_1 }
\sum_{mn} \sum_\mathbf{k} \sump_{\lambda\beta\alpha}
\bar\delta_{mn}
\frac{v_{nm}^\phi v_{mn}^\lambda }{ \epsilon_{mn}^2 }
\frac{ \partial^2 f_{nm} }{ \partial k_\beta \partial k_\alpha }
\nonumber\\&=
-\frac{ \hbar^2 C_3/2  }{ \hbar(\bar\omega_2+\bar\omega_1)\hbar \bar\omega_1 }
\sum_{mn} \sum_\mathbf{k} \sump_{\lambda\beta\alpha}
\bar\delta_{mn}
\frac{ v_{nm}^\phi v_{mn}^\lambda -v_{mn}^\phi v_{nm}^\lambda
}{ \epsilon_{mn}^2 }
\frac{ \partial^2 f_{nm} }{ \partial k_\beta \partial k_\alpha } = 0
\, .
\end{align}
These results isolate and identify all nonlinear contributions to the Drude
peak and by removing all spurious divergences show that only odd powers of the 
frequency can contribute to the nonlinear Drude-like response.
\end{widetext}

\bibliography{3rdGraphDoped}

\begin{thebibliography}{109}%
\makeatletter
\providecommand \@ifxundefined [1]{%
 \@ifx{#1\undefined}
}%
\providecommand \@ifnum [1]{%
 \ifnum #1\expandafter \@firstoftwo
 \else \expandafter \@secondoftwo
 \fi
}%
\providecommand \@ifx [1]{%
 \ifx #1\expandafter \@firstoftwo
 \else \expandafter \@secondoftwo
 \fi
}%
\providecommand \natexlab [1]{#1}%
\providecommand \enquote  [1]{``#1''}%
\providecommand \bibnamefont  [1]{#1}%
\providecommand \bibfnamefont [1]{#1}%
\providecommand \citenamefont [1]{#1}%
\providecommand \href@noop [0]{\@secondoftwo}%
\providecommand \href [0]{\begingroup \@sanitize@url \@href}%
\providecommand \@href[1]{\@@startlink{#1}\@@href}%
\providecommand \@@href[1]{\endgroup#1\@@endlink}%
\providecommand \@sanitize@url [0]{\catcode `\\12\catcode `\$12\catcode
  `\&12\catcode `\#12\catcode `\^12\catcode `\_12\catcode `\%12\relax}%
\providecommand \@@startlink[1]{}%
\providecommand \@@endlink[0]{}%
\providecommand \url  [0]{\begingroup\@sanitize@url \@url }%
\providecommand \@url [1]{\endgroup\@href {#1}{\urlprefix }}%
\providecommand \urlprefix  [0]{URL }%
\providecommand \Eprint [0]{\href }%
\providecommand \doibase [0]{http://dx.doi.org/}%
\providecommand \selectlanguage [0]{\@gobble}%
\providecommand \bibinfo  [0]{\@secondoftwo}%
\providecommand \bibfield  [0]{\@secondoftwo}%
\providecommand \translation [1]{[#1]}%
\providecommand \BibitemOpen [0]{}%
\providecommand \bibitemStop [0]{}%
\providecommand \bibitemNoStop [0]{.\EOS\space}%
\providecommand \EOS [0]{\spacefactor3000\relax}%
\providecommand \BibitemShut  [1]{\csname bibitem#1\endcsname}%
\let\auto@bib@innerbib\@empty
\bibitem [{\citenamefont {Shen}(2002)}]{Shen2002}%
  \BibitemOpen
  \bibfield  {author} {\bibinfo {author} {\bibfnamefont {Y.~R.}\ \bibnamefont
  {Shen}},\ }\href@noop {} {\emph {\bibinfo {title} {The Principles of
  Nonlinear Optics}}}\ (\bibinfo  {publisher} {Wiley-Interscience},\ \bibinfo
  {year} {2002})\BibitemShut {NoStop}%
\bibitem [{\citenamefont {Boyd}(2008)}]{Boyd2008}%
  \BibitemOpen
  \bibfield  {author} {\bibinfo {author} {\bibfnamefont {R.~W.}\ \bibnamefont
  {Boyd}},\ }\href@noop {} {\emph {\bibinfo {title} {Nonlinear Optics}}},\
  \bibinfo {edition} {3rd}\ ed.\ (\bibinfo  {publisher} {Elsevier Science
  Publishing Co Inc},\ \bibinfo {year} {2008})\BibitemShut {NoStop}%
\bibitem [{\citenamefont {Gu}\ \emph {et~al.}(2012)\citenamefont {Gu},
  \citenamefont {Petrone}, \citenamefont {McMillan}, \citenamefont {van~der
  Zande}, \citenamefont {Yu}, \citenamefont {Lo}, \citenamefont {Kwong},
  \citenamefont {Hone},\ and\ \citenamefont {Wong}}]{Gu2012}%
  \BibitemOpen
  \bibfield  {author} {\bibinfo {author} {\bibfnamefont {T.}~\bibnamefont
  {Gu}}, \bibinfo {author} {\bibfnamefont {N.}~\bibnamefont {Petrone}},
  \bibinfo {author} {\bibfnamefont {J.~F.}\ \bibnamefont {McMillan}}, \bibinfo
  {author} {\bibfnamefont {A.}~\bibnamefont {van~der Zande}}, \bibinfo {author}
  {\bibfnamefont {M.}~\bibnamefont {Yu}}, \bibinfo {author} {\bibfnamefont
  {G.~Q.}\ \bibnamefont {Lo}}, \bibinfo {author} {\bibfnamefont {D.~L.}\
  \bibnamefont {Kwong}}, \bibinfo {author} {\bibfnamefont {J.}~\bibnamefont
  {Hone}}, \ and\ \bibinfo {author} {\bibfnamefont {C.~W.}\ \bibnamefont
  {Wong}},\ }\href {\doibase 10.1038/nphoton.2012.147} {\bibfield  {journal}
  {\bibinfo  {journal} {Nat. Photon.}\ }\textbf {\bibinfo {volume} {6}},\
  \bibinfo {pages} {554} (\bibinfo {year} {2012})}\BibitemShut {NoStop}%
\bibitem [{\citenamefont {Hendry}\ \emph {et~al.}(2010)\citenamefont {Hendry},
  \citenamefont {Hale}, \citenamefont {Moger}, \citenamefont {Savchenko},\ and\
  \citenamefont {Mikhailov}}]{Hendry2010}%
  \BibitemOpen
  \bibfield  {author} {\bibinfo {author} {\bibfnamefont {E.}~\bibnamefont
  {Hendry}}, \bibinfo {author} {\bibfnamefont {P.~J.}\ \bibnamefont {Hale}},
  \bibinfo {author} {\bibfnamefont {J.}~\bibnamefont {Moger}}, \bibinfo
  {author} {\bibfnamefont {A.~K.}\ \bibnamefont {Savchenko}}, \ and\ \bibinfo
  {author} {\bibfnamefont {S.~A.}\ \bibnamefont {Mikhailov}},\ }\href {\doibase
  10.1103/PhysRevLett.105.097401} {\bibfield  {journal} {\bibinfo  {journal}
  {Phys. Rev. Lett.}\ }\textbf {\bibinfo {volume} {105}},\ \bibinfo {pages}
  {097401} (\bibinfo {year} {2010})}\BibitemShut {NoStop}%
\bibitem [{\citenamefont {Mikhailov}(2007)}]{Mikhailov2007}%
  \BibitemOpen
  \bibfield  {author} {\bibinfo {author} {\bibfnamefont {S.~A.}\ \bibnamefont
  {Mikhailov}},\ }\href {\doibase 10.1209/0295-5075/79/27002} {\bibfield
  {journal} {\bibinfo  {journal} {Europhysics Letters (EPL)}\ }\textbf
  {\bibinfo {volume} {79}},\ \bibinfo {pages} {27002} (\bibinfo {year}
  {2007})}\BibitemShut {NoStop}%
\bibitem [{\citenamefont {Ishikawa}(2010)}]{Ishikawa2010}%
  \BibitemOpen
  \bibfield  {author} {\bibinfo {author} {\bibfnamefont {K.~L.}\ \bibnamefont
  {Ishikawa}},\ }\href {\doibase 10.1103/PhysRevB.82.201402} {\bibfield
  {journal} {\bibinfo  {journal} {Phys. Rev. B}\ }\textbf {\bibinfo {volume}
  {82}},\ \bibinfo {pages} {201402} (\bibinfo {year} {2010})}\BibitemShut
  {NoStop}%
\bibitem [{\citenamefont {Avetissian}\ \emph {et~al.}(2012)\citenamefont
  {Avetissian}, \citenamefont {Avetissian}, \citenamefont {Mkrtchian},\ and\
  \citenamefont {Sedrakian}}]{Avetissian2012}%
  \BibitemOpen
  \bibfield  {author} {\bibinfo {author} {\bibfnamefont {H.~K.}\ \bibnamefont
  {Avetissian}}, \bibinfo {author} {\bibfnamefont {A.~K.}\ \bibnamefont
  {Avetissian}}, \bibinfo {author} {\bibfnamefont {G.~F.}\ \bibnamefont
  {Mkrtchian}}, \ and\ \bibinfo {author} {\bibfnamefont {K.~V.}\ \bibnamefont
  {Sedrakian}},\ }\href {\doibase 10.1103/PhysRevB.85.115443} {\bibfield
  {journal} {\bibinfo  {journal} {Phys. Rev. B}\ }\textbf {\bibinfo {volume}
  {85}},\ \bibinfo {pages} {115443} (\bibinfo {year} {2012})}\BibitemShut
  {NoStop}%
\bibitem [{\citenamefont {Wu}\ \emph {et~al.}(2015)\citenamefont {Wu},
  \citenamefont {Buckley}, \citenamefont {Schaibley}, \citenamefont {Feng},
  \citenamefont {Yan}, \citenamefont {Mandrus}, \citenamefont {Hatami},
  \citenamefont {Yao}, \citenamefont {Vu{\v{c}}kovi{\'{c}}}, \citenamefont
  {Majumdar},\ and\ \citenamefont {Xu}}]{Wu2015}%
  \BibitemOpen
  \bibfield  {author} {\bibinfo {author} {\bibfnamefont {S.}~\bibnamefont
  {Wu}}, \bibinfo {author} {\bibfnamefont {S.}~\bibnamefont {Buckley}},
  \bibinfo {author} {\bibfnamefont {J.~R.}\ \bibnamefont {Schaibley}}, \bibinfo
  {author} {\bibfnamefont {L.}~\bibnamefont {Feng}}, \bibinfo {author}
  {\bibfnamefont {J.}~\bibnamefont {Yan}}, \bibinfo {author} {\bibfnamefont
  {D.~G.}\ \bibnamefont {Mandrus}}, \bibinfo {author} {\bibfnamefont
  {F.}~\bibnamefont {Hatami}}, \bibinfo {author} {\bibfnamefont
  {W.}~\bibnamefont {Yao}}, \bibinfo {author} {\bibfnamefont {J.}~\bibnamefont
  {Vu{\v{c}}kovi{\'{c}}}}, \bibinfo {author} {\bibfnamefont {A.}~\bibnamefont
  {Majumdar}}, \ and\ \bibinfo {author} {\bibfnamefont {X.}~\bibnamefont
  {Xu}},\ }\href {\doibase 10.1038/nature14290} {\bibfield  {journal} {\bibinfo
   {journal} {Nat.}\ }\textbf {\bibinfo {volume} {520}},\ \bibinfo {pages} {69}
  (\bibinfo {year} {2015})}\BibitemShut {NoStop}%
\bibitem [{\citenamefont {Chizhova}\ \emph {et~al.}(2016)\citenamefont
  {Chizhova}, \citenamefont {Libisch},\ and\ \citenamefont
  {Burgd{\"{o}}rfer}}]{Chizhova2016}%
  \BibitemOpen
  \bibfield  {author} {\bibinfo {author} {\bibfnamefont {L.~A.}\ \bibnamefont
  {Chizhova}}, \bibinfo {author} {\bibfnamefont {F.}~\bibnamefont {Libisch}}, \
  and\ \bibinfo {author} {\bibfnamefont {J.}~\bibnamefont {Burgd{\"{o}}rfer}},\
  }\href {\doibase 10.1103/PhysRevB.94.075412} {\bibfield  {journal} {\bibinfo
  {journal} {Phys. Rev. B}\ }\textbf {\bibinfo {volume} {94}},\ \bibinfo
  {pages} {075412} (\bibinfo {year} {2016})}\BibitemShut {NoStop}%
\bibitem [{\citenamefont {Mikhailov}\ and\ \citenamefont
  {Ziegler}(2008)}]{Mikhailov2008}%
  \BibitemOpen
  \bibfield  {author} {\bibinfo {author} {\bibfnamefont {S.~A.}\ \bibnamefont
  {Mikhailov}}\ and\ \bibinfo {author} {\bibfnamefont {K.}~\bibnamefont
  {Ziegler}},\ }\href {\doibase 10.1088/0953-8984/20/38/384204} {\bibfield
  {journal} {\bibinfo  {journal} {J. Phys. Condens. Matter}\ }\textbf {\bibinfo
  {volume} {20}},\ \bibinfo {pages} {384204} (\bibinfo {year}
  {2008})}\BibitemShut {NoStop}%
\bibitem [{\citenamefont {Hong}\ \emph {et~al.}(2013)\citenamefont {Hong},
  \citenamefont {Dadap}, \citenamefont {Petrone}, \citenamefont {Yeh},
  \citenamefont {Hone},\ and\ \citenamefont {Osgood}}]{Hong2013}%
  \BibitemOpen
  \bibfield  {author} {\bibinfo {author} {\bibfnamefont {S.-Y.}\ \bibnamefont
  {Hong}}, \bibinfo {author} {\bibfnamefont {J.~I.}\ \bibnamefont {Dadap}},
  \bibinfo {author} {\bibfnamefont {N.}~\bibnamefont {Petrone}}, \bibinfo
  {author} {\bibfnamefont {P.-C.}\ \bibnamefont {Yeh}}, \bibinfo {author}
  {\bibfnamefont {J.}~\bibnamefont {Hone}}, \ and\ \bibinfo {author}
  {\bibfnamefont {R.~M.}\ \bibnamefont {Osgood}},\ }\href {\doibase
  10.1103/PhysRevX.3.021014} {\bibfield  {journal} {\bibinfo  {journal} {Phys.
  Rev. X}\ }\textbf {\bibinfo {volume} {3}},\ \bibinfo {pages} {021014}
  (\bibinfo {year} {2013})}\BibitemShut {NoStop}%
\bibitem [{\citenamefont {Kumar}\ \emph {et~al.}(2013)\citenamefont {Kumar},
  \citenamefont {Kumar}, \citenamefont {Gerstenkorn}, \citenamefont {Wang},
  \citenamefont {Chiu}, \citenamefont {Smirl},\ and\ \citenamefont
  {Zhao}}]{Kumar2013}%
  \BibitemOpen
  \bibfield  {author} {\bibinfo {author} {\bibfnamefont {N.}~\bibnamefont
  {Kumar}}, \bibinfo {author} {\bibfnamefont {J.}~\bibnamefont {Kumar}},
  \bibinfo {author} {\bibfnamefont {C.}~\bibnamefont {Gerstenkorn}}, \bibinfo
  {author} {\bibfnamefont {R.}~\bibnamefont {Wang}}, \bibinfo {author}
  {\bibfnamefont {H.-Y.}\ \bibnamefont {Chiu}}, \bibinfo {author}
  {\bibfnamefont {A.~L.}\ \bibnamefont {Smirl}}, \ and\ \bibinfo {author}
  {\bibfnamefont {H.}~\bibnamefont {Zhao}},\ }\href {\doibase
  10.1103/PhysRevB.87.121406} {\bibfield  {journal} {\bibinfo  {journal} {Phys.
  Rev. B}\ }\textbf {\bibinfo {volume} {87}},\ \bibinfo {pages} {121406}
  (\bibinfo {year} {2013})}\BibitemShut {NoStop}%
\bibitem [{\citenamefont {Li}\ \emph {et~al.}(2013)\citenamefont {Li},
  \citenamefont {Rao}, \citenamefont {Mak}, \citenamefont {You}, \citenamefont
  {Wang}, \citenamefont {Dean},\ and\ \citenamefont {Heinz}}]{Li2013a}%
  \BibitemOpen
  \bibfield  {author} {\bibinfo {author} {\bibfnamefont {Y.}~\bibnamefont
  {Li}}, \bibinfo {author} {\bibfnamefont {Y.}~\bibnamefont {Rao}}, \bibinfo
  {author} {\bibfnamefont {K.~F.}\ \bibnamefont {Mak}}, \bibinfo {author}
  {\bibfnamefont {Y.}~\bibnamefont {You}}, \bibinfo {author} {\bibfnamefont
  {S.}~\bibnamefont {Wang}}, \bibinfo {author} {\bibfnamefont {C.~R.}\
  \bibnamefont {Dean}}, \ and\ \bibinfo {author} {\bibfnamefont {T.~F.}\
  \bibnamefont {Heinz}},\ }\href {\doibase 10.1021/nl401561r} {\bibfield
  {journal} {\bibinfo  {journal} {Nano Lett.}\ }\textbf {\bibinfo {volume}
  {13}},\ \bibinfo {pages} {3329} (\bibinfo {year} {2013})}\BibitemShut
  {NoStop}%
\bibitem [{\citenamefont {Zeng}\ \emph {et~al.}(2013)\citenamefont {Zeng},
  \citenamefont {Liu}, \citenamefont {Dai}, \citenamefont {Yan}, \citenamefont
  {Zhu}, \citenamefont {He}, \citenamefont {Xie}, \citenamefont {Xu},
  \citenamefont {Chen}, \citenamefont {Yao},\ and\ \citenamefont
  {Cui}}]{Zeng2013}%
  \BibitemOpen
  \bibfield  {author} {\bibinfo {author} {\bibfnamefont {H.}~\bibnamefont
  {Zeng}}, \bibinfo {author} {\bibfnamefont {G.-B.}\ \bibnamefont {Liu}},
  \bibinfo {author} {\bibfnamefont {J.}~\bibnamefont {Dai}}, \bibinfo {author}
  {\bibfnamefont {Y.}~\bibnamefont {Yan}}, \bibinfo {author} {\bibfnamefont
  {B.}~\bibnamefont {Zhu}}, \bibinfo {author} {\bibfnamefont {R.}~\bibnamefont
  {He}}, \bibinfo {author} {\bibfnamefont {L.}~\bibnamefont {Xie}}, \bibinfo
  {author} {\bibfnamefont {S.}~\bibnamefont {Xu}}, \bibinfo {author}
  {\bibfnamefont {X.}~\bibnamefont {Chen}}, \bibinfo {author} {\bibfnamefont
  {W.}~\bibnamefont {Yao}}, \ and\ \bibinfo {author} {\bibfnamefont
  {X.}~\bibnamefont {Cui}},\ }\href {\doibase 10.1038/srep01608} {\bibfield
  {journal} {\bibinfo  {journal} {Sci. Rep.}\ }\textbf {\bibinfo {volume}
  {3}},\ \bibinfo {pages} {1608} (\bibinfo {year} {2013})}\BibitemShut
  {NoStop}%
\bibitem [{\citenamefont {Avetissian}\ \emph {et~al.}(2013)\citenamefont
  {Avetissian}, \citenamefont {Mkrtchian}, \citenamefont {Batrakov},
  \citenamefont {Maksimenko},\ and\ \citenamefont {Hoffmann}}]{Avetissian2013}%
  \BibitemOpen
  \bibfield  {author} {\bibinfo {author} {\bibfnamefont {H.~K.}\ \bibnamefont
  {Avetissian}}, \bibinfo {author} {\bibfnamefont {G.~F.}\ \bibnamefont
  {Mkrtchian}}, \bibinfo {author} {\bibfnamefont {K.~G.}\ \bibnamefont
  {Batrakov}}, \bibinfo {author} {\bibfnamefont {S.~A.}\ \bibnamefont
  {Maksimenko}}, \ and\ \bibinfo {author} {\bibfnamefont {A.}~\bibnamefont
  {Hoffmann}},\ }\href {\doibase 10.1103/PhysRevB.88.165411} {\bibfield
  {journal} {\bibinfo  {journal} {Phys. Rev. B}\ }\textbf {\bibinfo {volume}
  {88}},\ \bibinfo {pages} {165411} (\bibinfo {year} {2013})}\BibitemShut
  {NoStop}%
\bibitem [{\citenamefont {Janisch}\ \emph {et~al.}(2015)\citenamefont
  {Janisch}, \citenamefont {Wang}, \citenamefont {Ma}, \citenamefont {Mehta},
  \citenamefont {El{\'{i}}as}, \citenamefont {Perea-L{\'{o}}pez}, \citenamefont
  {Terrones}, \citenamefont {Crespi},\ and\ \citenamefont {Liu}}]{Janisch2014}%
  \BibitemOpen
  \bibfield  {author} {\bibinfo {author} {\bibfnamefont {C.}~\bibnamefont
  {Janisch}}, \bibinfo {author} {\bibfnamefont {Y.}~\bibnamefont {Wang}},
  \bibinfo {author} {\bibfnamefont {D.}~\bibnamefont {Ma}}, \bibinfo {author}
  {\bibfnamefont {N.}~\bibnamefont {Mehta}}, \bibinfo {author} {\bibfnamefont
  {A.~L.}\ \bibnamefont {El{\'{i}}as}}, \bibinfo {author} {\bibfnamefont
  {N.}~\bibnamefont {Perea-L{\'{o}}pez}}, \bibinfo {author} {\bibfnamefont
  {M.}~\bibnamefont {Terrones}}, \bibinfo {author} {\bibfnamefont
  {V.}~\bibnamefont {Crespi}}, \ and\ \bibinfo {author} {\bibfnamefont
  {Z.}~\bibnamefont {Liu}},\ }\href {\doibase 10.1038/srep05530} {\bibfield
  {journal} {\bibinfo  {journal} {Sci. Rep.}\ }\textbf {\bibinfo {volume}
  {4}},\ \bibinfo {pages} {5530} (\bibinfo {year} {2015})}\BibitemShut
  {NoStop}%
\bibitem [{\citenamefont {Wang}\ \emph
  {et~al.}(2015{\natexlab{a}})\citenamefont {Wang}, \citenamefont {Marie},
  \citenamefont {Gerber}, \citenamefont {Amand}, \citenamefont {Lagarde},
  \citenamefont {Bouet}, \citenamefont {Vidal}, \citenamefont {Balocchi},\ and\
  \citenamefont {Urbaszek}}]{Wang2015a}%
  \BibitemOpen
  \bibfield  {author} {\bibinfo {author} {\bibfnamefont {G.}~\bibnamefont
  {Wang}}, \bibinfo {author} {\bibfnamefont {X.}~\bibnamefont {Marie}},
  \bibinfo {author} {\bibfnamefont {I.}~\bibnamefont {Gerber}}, \bibinfo
  {author} {\bibfnamefont {T.}~\bibnamefont {Amand}}, \bibinfo {author}
  {\bibfnamefont {D.}~\bibnamefont {Lagarde}}, \bibinfo {author} {\bibfnamefont
  {L.}~\bibnamefont {Bouet}}, \bibinfo {author} {\bibfnamefont
  {M.}~\bibnamefont {Vidal}}, \bibinfo {author} {\bibfnamefont
  {A.}~\bibnamefont {Balocchi}}, \ and\ \bibinfo {author} {\bibfnamefont
  {B.}~\bibnamefont {Urbaszek}},\ }\href {\doibase
  10.1103/PhysRevLett.114.097403} {\bibfield  {journal} {\bibinfo  {journal}
  {Phys. Rev. Lett.}\ }\textbf {\bibinfo {volume} {114}},\ \bibinfo {pages}
  {097403} (\bibinfo {year} {2015}{\natexlab{a}})}\BibitemShut {NoStop}%
\bibitem [{\citenamefont {Yin}\ \emph {et~al.}(2014)\citenamefont {Yin},
  \citenamefont {Ye}, \citenamefont {Chenet}, \citenamefont {Ye}, \citenamefont
  {O'Brien}, \citenamefont {Hone},\ and\ \citenamefont {Zhang}}]{Yin2014a}%
  \BibitemOpen
  \bibfield  {author} {\bibinfo {author} {\bibfnamefont {X.}~\bibnamefont
  {Yin}}, \bibinfo {author} {\bibfnamefont {Z.}~\bibnamefont {Ye}}, \bibinfo
  {author} {\bibfnamefont {D.~A.}\ \bibnamefont {Chenet}}, \bibinfo {author}
  {\bibfnamefont {Y.}~\bibnamefont {Ye}}, \bibinfo {author} {\bibfnamefont
  {K.}~\bibnamefont {O'Brien}}, \bibinfo {author} {\bibfnamefont {J.~C.}\
  \bibnamefont {Hone}}, \ and\ \bibinfo {author} {\bibfnamefont
  {X.}~\bibnamefont {Zhang}},\ }\href {\doibase 10.1126/science.1250564}
  {\bibfield  {journal} {\bibinfo  {journal} {Sci.}\ }\textbf {\bibinfo
  {volume} {344}},\ \bibinfo {pages} {488} (\bibinfo {year}
  {2014})}\BibitemShut {NoStop}%
\bibitem [{\citenamefont {Clark}\ \emph {et~al.}(2015)\citenamefont {Clark},
  \citenamefont {Senthilkumar}, \citenamefont {Le}, \citenamefont {Weerawarne},
  \citenamefont {Shim}, \citenamefont {Jang}, \citenamefont {Shim},
  \citenamefont {Cho}, \citenamefont {Sim}, \citenamefont {Seong},
  \citenamefont {Rhim}, \citenamefont {Freeman}, \citenamefont {Chung},\ and\
  \citenamefont {Kim}}]{Clark2014}%
  \BibitemOpen
  \bibfield  {author} {\bibinfo {author} {\bibfnamefont {D.~J.}\ \bibnamefont
  {Clark}}, \bibinfo {author} {\bibfnamefont {V.}~\bibnamefont {Senthilkumar}},
  \bibinfo {author} {\bibfnamefont {C.~T.}\ \bibnamefont {Le}}, \bibinfo
  {author} {\bibfnamefont {D.~L.}\ \bibnamefont {Weerawarne}}, \bibinfo
  {author} {\bibfnamefont {B.}~\bibnamefont {Shim}}, \bibinfo {author}
  {\bibfnamefont {J.~I.}\ \bibnamefont {Jang}}, \bibinfo {author}
  {\bibfnamefont {J.~H.}\ \bibnamefont {Shim}}, \bibinfo {author}
  {\bibfnamefont {J.}~\bibnamefont {Cho}}, \bibinfo {author} {\bibfnamefont
  {Y.}~\bibnamefont {Sim}}, \bibinfo {author} {\bibfnamefont {M.-J.}\
  \bibnamefont {Seong}}, \bibinfo {author} {\bibfnamefont {S.~H.}\ \bibnamefont
  {Rhim}}, \bibinfo {author} {\bibfnamefont {A.~J.}\ \bibnamefont {Freeman}},
  \bibinfo {author} {\bibfnamefont {K.-H.}\ \bibnamefont {Chung}}, \ and\
  \bibinfo {author} {\bibfnamefont {Y.~S.}\ \bibnamefont {Kim}},\ }\href
  {\doibase 10.1103/PhysRevB.90.121409} {\bibfield  {journal} {\bibinfo
  {journal} {Phys. Rev. B}\ }\textbf {\bibinfo {volume} {92}},\ \bibinfo
  {pages} {121409} (\bibinfo {year} {2015})}\BibitemShut {NoStop}%
\bibitem [{\citenamefont {Al-Naib}\ \emph {et~al.}(2015)\citenamefont
  {Al-Naib}, \citenamefont {Sipe},\ and\ \citenamefont {Dignam}}]{Al-Naib2015}%
  \BibitemOpen
  \bibfield  {author} {\bibinfo {author} {\bibfnamefont {I.}~\bibnamefont
  {Al-Naib}}, \bibinfo {author} {\bibfnamefont {J.~E.}\ \bibnamefont {Sipe}}, \
  and\ \bibinfo {author} {\bibfnamefont {M.~M.}\ \bibnamefont {Dignam}},\
  }\href {\doibase 10.1088/1367-2630/17/11/113018} {\bibfield  {journal}
  {\bibinfo  {journal} {New J. Phys.}\ }\textbf {\bibinfo {volume} {17}},\
  \bibinfo {pages} {113018} (\bibinfo {year} {2015})}\BibitemShut {NoStop}%
\bibitem [{\citenamefont {Chizhova}\ \emph {et~al.}(2017)\citenamefont
  {Chizhova}, \citenamefont {Libisch},\ and\ \citenamefont
  {Burgd{\"{o}}rfer}}]{Chizhova2017}%
  \BibitemOpen
  \bibfield  {author} {\bibinfo {author} {\bibfnamefont {L.~A.}\ \bibnamefont
  {Chizhova}}, \bibinfo {author} {\bibfnamefont {F.}~\bibnamefont {Libisch}}, \
  and\ \bibinfo {author} {\bibfnamefont {J.}~\bibnamefont {Burgd{\"{o}}rfer}},\
  }\href {\doibase 10.1103/PhysRevB.95.085436} {\bibfield  {journal} {\bibinfo
  {journal} {Phys. Rev. B}\ }\textbf {\bibinfo {volume} {95}},\ \bibinfo
  {pages} {085436} (\bibinfo {year} {2017})}\BibitemShut {NoStop}%
\bibitem [{\citenamefont {Hipolito}\ and\ \citenamefont
  {Pereira}(2017)}]{Hipolito2017}%
  \BibitemOpen
  \bibfield  {author} {\bibinfo {author} {\bibfnamefont {F.}~\bibnamefont
  {Hipolito}}\ and\ \bibinfo {author} {\bibfnamefont {V.~M.}\ \bibnamefont
  {Pereira}},\ }\href {\doibase 10.1088/2053-1583/aa6f4d} {\bibfield  {journal}
  {\bibinfo  {journal} {2D Mater.}\ }\textbf {\bibinfo {volume} {4}},\ \bibinfo
  {pages} {021027} (\bibinfo {year} {2017})}\BibitemShut {NoStop}%
\bibitem [{\citenamefont {McIver}\ \emph {et~al.}(2011)\citenamefont {McIver},
  \citenamefont {Hsieh}, \citenamefont {Steinberg}, \citenamefont
  {Jarillo-Herrero},\ and\ \citenamefont {Gedik}}]{McIver2012}%
  \BibitemOpen
  \bibfield  {author} {\bibinfo {author} {\bibfnamefont {J.~W.}\ \bibnamefont
  {McIver}}, \bibinfo {author} {\bibfnamefont {D.}~\bibnamefont {Hsieh}},
  \bibinfo {author} {\bibfnamefont {H.}~\bibnamefont {Steinberg}}, \bibinfo
  {author} {\bibfnamefont {P.}~\bibnamefont {Jarillo-Herrero}}, \ and\ \bibinfo
  {author} {\bibfnamefont {N.}~\bibnamefont {Gedik}},\ }\href {\doibase
  10.1038/nnano.2011.214} {\bibfield  {journal} {\bibinfo  {journal} {Nat.
  Nanotechnol.}\ }\textbf {\bibinfo {volume} {7}},\ \bibinfo {pages} {96}
  (\bibinfo {year} {2011})}\BibitemShut {NoStop}%
\bibitem [{\citenamefont {Taghizadeh}\ \emph {et~al.}(2017)\citenamefont
  {Taghizadeh}, \citenamefont {Hipolito},\ and\ \citenamefont
  {Pedersen}}]{Taghizadeh2017a}%
  \BibitemOpen
  \bibfield  {author} {\bibinfo {author} {\bibfnamefont {A.}~\bibnamefont
  {Taghizadeh}}, \bibinfo {author} {\bibfnamefont {F.}~\bibnamefont
  {Hipolito}}, \ and\ \bibinfo {author} {\bibfnamefont {T.~G.}\ \bibnamefont
  {Pedersen}},\ }\href {\doibase 10.1103/PhysRevB.96.195413} {\bibfield
  {journal} {\bibinfo  {journal} {Phys. Rev. B}\ }\textbf {\bibinfo {volume}
  {96}},\ \bibinfo {pages} {195413} (\bibinfo {year} {2017})}\BibitemShut
  {NoStop}%
\bibitem [{\citenamefont {Taghizadeh}\ and\ \citenamefont
  {Pedersen}(2018)}]{Taghizadeh2018}%
  \BibitemOpen
  \bibfield  {author} {\bibinfo {author} {\bibfnamefont {A.}~\bibnamefont
  {Taghizadeh}}\ and\ \bibinfo {author} {\bibfnamefont {T.~G.}\ \bibnamefont
  {Pedersen}},\ }\href {\doibase 10.1103/PhysRevB.97.205432} {\bibfield
  {journal} {\bibinfo  {journal} {Phys. Rev. B}\ }\textbf {\bibinfo {volume}
  {97}},\ \bibinfo {pages} {205432} (\bibinfo {year} {2018})}\BibitemShut
  {NoStop}%
\bibitem [{\citenamefont {Aversa}\ and\ \citenamefont
  {Sipe}(1995)}]{Aversa1995}%
  \BibitemOpen
  \bibfield  {author} {\bibinfo {author} {\bibfnamefont {C.}~\bibnamefont
  {Aversa}}\ and\ \bibinfo {author} {\bibfnamefont {J.~E.}\ \bibnamefont
  {Sipe}},\ }\href {\doibase 10.1103/PhysRevB.52.14636} {\bibfield  {journal}
  {\bibinfo  {journal} {Phys. Rev. B}\ }\textbf {\bibinfo {volume} {52}},\
  \bibinfo {pages} {14636} (\bibinfo {year} {1995})}\BibitemShut {NoStop}%
\bibitem [{\citenamefont {Hipolito}\ and\ \citenamefont
  {Pedersen}(2018)}]{Hipolito2018}%
  \BibitemOpen
  \bibfield  {author} {\bibinfo {author} {\bibfnamefont {F.}~\bibnamefont
  {Hipolito}}\ and\ \bibinfo {author} {\bibfnamefont {T.~G.}\ \bibnamefont
  {Pedersen}},\ }\href {\doibase 10.1103/PhysRevB.97.035431} {\bibfield
  {journal} {\bibinfo  {journal} {Phys. Rev. B}\ }\textbf {\bibinfo {volume}
  {97}},\ \bibinfo {pages} {035431} (\bibinfo {year} {2018})}\BibitemShut
  {NoStop}%
\bibitem [{\citenamefont {Cheng}\ \emph {et~al.}(2014)\citenamefont {Cheng},
  \citenamefont {Vermeulen},\ and\ \citenamefont {Sipe}}]{Cheng2014}%
  \BibitemOpen
  \bibfield  {author} {\bibinfo {author} {\bibfnamefont {J.~L.}\ \bibnamefont
  {Cheng}}, \bibinfo {author} {\bibfnamefont {N.}~\bibnamefont {Vermeulen}}, \
  and\ \bibinfo {author} {\bibfnamefont {J.~E.}\ \bibnamefont {Sipe}},\ }\href
  {\doibase 10.1364/OE.22.015868} {\bibfield  {journal} {\bibinfo  {journal}
  {Opt. Express}\ }\textbf {\bibinfo {volume} {22}},\ \bibinfo {pages} {15868}
  (\bibinfo {year} {2014})}\BibitemShut {NoStop}%
\bibitem [{\citenamefont {Mikhailov}(2014)}]{Mikhailov2014}%
  \BibitemOpen
  \bibfield  {author} {\bibinfo {author} {\bibfnamefont {S.~A.}\ \bibnamefont
  {Mikhailov}},\ }\href {\doibase 10.1103/PhysRevB.90.241301} {\bibfield
  {journal} {\bibinfo  {journal} {Phys. Rev. B}\ }\textbf {\bibinfo {volume}
  {90}},\ \bibinfo {pages} {241301} (\bibinfo {year} {2014})}\BibitemShut
  {NoStop}%
\bibitem [{\citenamefont {Soavi}\ \emph {et~al.}(2018)\citenamefont {Soavi},
  \citenamefont {Wang}, \citenamefont {Rostami}, \citenamefont {Purdie},
  \citenamefont {{De Fazio}}, \citenamefont {Ma}, \citenamefont {Luo},
  \citenamefont {Wang}, \citenamefont {Ott}, \citenamefont {Yoon},
  \citenamefont {Bourelle}, \citenamefont {Muench}, \citenamefont {Goykhman},
  \citenamefont {{Dal Conte}}, \citenamefont {Celebrano}, \citenamefont
  {Tomadin}, \citenamefont {Polini}, \citenamefont {Cerullo},\ and\
  \citenamefont {Ferrari}}]{Soavi2017}%
  \BibitemOpen
  \bibfield  {author} {\bibinfo {author} {\bibfnamefont {G.}~\bibnamefont
  {Soavi}}, \bibinfo {author} {\bibfnamefont {G.}~\bibnamefont {Wang}},
  \bibinfo {author} {\bibfnamefont {H.}~\bibnamefont {Rostami}}, \bibinfo
  {author} {\bibfnamefont {D.~G.}\ \bibnamefont {Purdie}}, \bibinfo {author}
  {\bibfnamefont {D.}~\bibnamefont {{De Fazio}}}, \bibinfo {author}
  {\bibfnamefont {T.}~\bibnamefont {Ma}}, \bibinfo {author} {\bibfnamefont
  {B.}~\bibnamefont {Luo}}, \bibinfo {author} {\bibfnamefont {J.}~\bibnamefont
  {Wang}}, \bibinfo {author} {\bibfnamefont {A.~K.}\ \bibnamefont {Ott}},
  \bibinfo {author} {\bibfnamefont {D.}~\bibnamefont {Yoon}}, \bibinfo {author}
  {\bibfnamefont {S.~A.}\ \bibnamefont {Bourelle}}, \bibinfo {author}
  {\bibfnamefont {J.~E.}\ \bibnamefont {Muench}}, \bibinfo {author}
  {\bibfnamefont {I.}~\bibnamefont {Goykhman}}, \bibinfo {author}
  {\bibfnamefont {S.}~\bibnamefont {{Dal Conte}}}, \bibinfo {author}
  {\bibfnamefont {M.}~\bibnamefont {Celebrano}}, \bibinfo {author}
  {\bibfnamefont {A.}~\bibnamefont {Tomadin}}, \bibinfo {author} {\bibfnamefont
  {M.}~\bibnamefont {Polini}}, \bibinfo {author} {\bibfnamefont
  {G.}~\bibnamefont {Cerullo}}, \ and\ \bibinfo {author} {\bibfnamefont
  {A.~C.}\ \bibnamefont {Ferrari}},\ }\href {\doibase
  10.1038/s41565-018-0145-8} {\bibfield  {journal} {\bibinfo  {journal} {Nat.
  Nanotechnol.}\ }\textbf {\bibinfo {volume} {13}},\ \bibinfo {pages} {583}
  (\bibinfo {year} {2018})}\BibitemShut {NoStop}%
\bibitem [{\citenamefont {Pedersen}\ and\ \citenamefont
  {Pedersen}(2009)}]{Pedersen2009}%
  \BibitemOpen
  \bibfield  {author} {\bibinfo {author} {\bibfnamefont {T.~G.}\ \bibnamefont
  {Pedersen}}\ and\ \bibinfo {author} {\bibfnamefont {K.}~\bibnamefont
  {Pedersen}},\ }\href {\doibase 10.1103/PhysRevB.79.035422} {\bibfield
  {journal} {\bibinfo  {journal} {Phys. Rev. B}\ }\textbf {\bibinfo {volume}
  {79}},\ \bibinfo {pages} {035422} (\bibinfo {year} {2009})}\BibitemShut
  {NoStop}%
\bibitem [{\citenamefont {Zhou}\ \emph {et~al.}(2007)\citenamefont {Zhou},
  \citenamefont {Gweon}, \citenamefont {Fedorov}, \citenamefont {First},
  \citenamefont {de~Heer}, \citenamefont {Lee}, \citenamefont {Guinea},
  \citenamefont {{Castro Neto}},\ and\ \citenamefont {Lanzara}}]{Zhou2007}%
  \BibitemOpen
  \bibfield  {author} {\bibinfo {author} {\bibfnamefont {S.~Y.}\ \bibnamefont
  {Zhou}}, \bibinfo {author} {\bibfnamefont {G.-H.}\ \bibnamefont {Gweon}},
  \bibinfo {author} {\bibfnamefont {a.~V.}\ \bibnamefont {Fedorov}}, \bibinfo
  {author} {\bibfnamefont {P.~N.}\ \bibnamefont {First}}, \bibinfo {author}
  {\bibfnamefont {W.~A.}\ \bibnamefont {de~Heer}}, \bibinfo {author}
  {\bibfnamefont {D.-H.}\ \bibnamefont {Lee}}, \bibinfo {author} {\bibfnamefont
  {F.}~\bibnamefont {Guinea}}, \bibinfo {author} {\bibfnamefont {A.~H.}\
  \bibnamefont {{Castro Neto}}}, \ and\ \bibinfo {author} {\bibfnamefont
  {A.}~\bibnamefont {Lanzara}},\ }\href {\doibase 10.1038/nmat2056} {\bibfield
  {journal} {\bibinfo  {journal} {Nat. Mater.}\ }\textbf {\bibinfo {volume}
  {6}},\ \bibinfo {pages} {916} (\bibinfo {year} {2007})}\BibitemShut {NoStop}%
\bibitem [{\citenamefont {Riedl}\ \emph {et~al.}(2009)\citenamefont {Riedl},
  \citenamefont {Coletti}, \citenamefont {Iwasaki}, \citenamefont {Zakharov},\
  and\ \citenamefont {Starke}}]{Riedl2009}%
  \BibitemOpen
  \bibfield  {author} {\bibinfo {author} {\bibfnamefont {C.}~\bibnamefont
  {Riedl}}, \bibinfo {author} {\bibfnamefont {C.}~\bibnamefont {Coletti}},
  \bibinfo {author} {\bibfnamefont {T.}~\bibnamefont {Iwasaki}}, \bibinfo
  {author} {\bibfnamefont {A.~A.}\ \bibnamefont {Zakharov}}, \ and\ \bibinfo
  {author} {\bibfnamefont {U.}~\bibnamefont {Starke}},\ }\href {\doibase
  10.1103/PhysRevLett.103.246804} {\bibfield  {journal} {\bibinfo  {journal}
  {Phys. Rev. Lett.}\ }\textbf {\bibinfo {volume} {103}},\ \bibinfo {pages}
  {246804} (\bibinfo {year} {2009})}\BibitemShut {NoStop}%
\bibitem [{\citenamefont {Woods}\ \emph {et~al.}(2014)\citenamefont {Woods},
  \citenamefont {Britnell}, \citenamefont {Eckmann}, \citenamefont {Ma},
  \citenamefont {Lu}, \citenamefont {Guo}, \citenamefont {Lin}, \citenamefont
  {Yu}, \citenamefont {Cao}, \citenamefont {Gorbachev}, \citenamefont
  {Kretinin}, \citenamefont {Park}, \citenamefont {Ponomarenko}, \citenamefont
  {Katsnelson}, \citenamefont {Gornostyrev}, \citenamefont {Watanabe},
  \citenamefont {Taniguchi}, \citenamefont {Casiraghi}, \citenamefont {Gao},
  \citenamefont {Geim},\ and\ \citenamefont {Novoselov}}]{Woods2014}%
  \BibitemOpen
  \bibfield  {author} {\bibinfo {author} {\bibfnamefont {C.~R.}\ \bibnamefont
  {Woods}}, \bibinfo {author} {\bibfnamefont {L.}~\bibnamefont {Britnell}},
  \bibinfo {author} {\bibfnamefont {A.}~\bibnamefont {Eckmann}}, \bibinfo
  {author} {\bibfnamefont {R.~S.}\ \bibnamefont {Ma}}, \bibinfo {author}
  {\bibfnamefont {J.~C.}\ \bibnamefont {Lu}}, \bibinfo {author} {\bibfnamefont
  {H.~M.}\ \bibnamefont {Guo}}, \bibinfo {author} {\bibfnamefont
  {X.}~\bibnamefont {Lin}}, \bibinfo {author} {\bibfnamefont {G.~L.}\
  \bibnamefont {Yu}}, \bibinfo {author} {\bibfnamefont {Y.}~\bibnamefont
  {Cao}}, \bibinfo {author} {\bibfnamefont {R.~V.}\ \bibnamefont {Gorbachev}},
  \bibinfo {author} {\bibfnamefont {A.~V.}\ \bibnamefont {Kretinin}}, \bibinfo
  {author} {\bibfnamefont {J.}~\bibnamefont {Park}}, \bibinfo {author}
  {\bibfnamefont {L.~A.}\ \bibnamefont {Ponomarenko}}, \bibinfo {author}
  {\bibfnamefont {M.~I.}\ \bibnamefont {Katsnelson}}, \bibinfo {author}
  {\bibfnamefont {Y.~N.}\ \bibnamefont {Gornostyrev}}, \bibinfo {author}
  {\bibfnamefont {K.}~\bibnamefont {Watanabe}}, \bibinfo {author}
  {\bibfnamefont {T.}~\bibnamefont {Taniguchi}}, \bibinfo {author}
  {\bibfnamefont {C.}~\bibnamefont {Casiraghi}}, \bibinfo {author}
  {\bibfnamefont {H.-j.~J.}\ \bibnamefont {Gao}}, \bibinfo {author}
  {\bibfnamefont {A.~K.}\ \bibnamefont {Geim}}, \ and\ \bibinfo {author}
  {\bibfnamefont {K.~S.}\ \bibnamefont {Novoselov}},\ }\href {\doibase
  10.1038/nphys2954} {\bibfield  {journal} {\bibinfo  {journal} {Nat. Phys.}\
  }\textbf {\bibinfo {volume} {10}},\ \bibinfo {pages} {451} (\bibinfo {year}
  {2014})}\BibitemShut {NoStop}%
\bibitem [{\citenamefont {Brun}\ and\ \citenamefont
  {Pedersen}(2015)}]{Brun2015}%
  \BibitemOpen
  \bibfield  {author} {\bibinfo {author} {\bibfnamefont {S.~J.}\ \bibnamefont
  {Brun}}\ and\ \bibinfo {author} {\bibfnamefont {T.~G.}\ \bibnamefont
  {Pedersen}},\ }\href {\doibase 10.1103/PhysRevB.91.205405} {\bibfield
  {journal} {\bibinfo  {journal} {Phys. Rev. B}\ }\textbf {\bibinfo {volume}
  {91}},\ \bibinfo {pages} {205405} (\bibinfo {year} {2015})}\BibitemShut
  {NoStop}%
\bibitem [{\citenamefont {Elias}\ \emph {et~al.}(2009)\citenamefont {Elias},
  \citenamefont {Nair}, \citenamefont {Mohiuddin}, \citenamefont {Morozov},
  \citenamefont {Blake}, \citenamefont {Halsall}, \citenamefont {Ferrari},
  \citenamefont {Boukhvalov}, \citenamefont {Katsnelson}, \citenamefont
  {Geim},\ and\ \citenamefont {Novoselov}}]{Elias2009}%
  \BibitemOpen
  \bibfield  {author} {\bibinfo {author} {\bibfnamefont {D.~C.}\ \bibnamefont
  {Elias}}, \bibinfo {author} {\bibfnamefont {R.~R.}\ \bibnamefont {Nair}},
  \bibinfo {author} {\bibfnamefont {T.~M.~G.}\ \bibnamefont {Mohiuddin}},
  \bibinfo {author} {\bibfnamefont {S.~V.}\ \bibnamefont {Morozov}}, \bibinfo
  {author} {\bibfnamefont {P.}~\bibnamefont {Blake}}, \bibinfo {author}
  {\bibfnamefont {M.~P.}\ \bibnamefont {Halsall}}, \bibinfo {author}
  {\bibfnamefont {A.~C.}\ \bibnamefont {Ferrari}}, \bibinfo {author}
  {\bibfnamefont {D.~W.}\ \bibnamefont {Boukhvalov}}, \bibinfo {author}
  {\bibfnamefont {M.~I.}\ \bibnamefont {Katsnelson}}, \bibinfo {author}
  {\bibfnamefont {A.~K.}\ \bibnamefont {Geim}}, \ and\ \bibinfo {author}
  {\bibfnamefont {K.~S.}\ \bibnamefont {Novoselov}},\ }\href {\doibase
  10.1126/science.1167130} {\bibfield  {journal} {\bibinfo  {journal} {Sci.}\
  }\textbf {\bibinfo {volume} {323}},\ \bibinfo {pages} {610} (\bibinfo {year}
  {2009})}\BibitemShut {NoStop}%
\bibitem [{\citenamefont {Balog}\ \emph {et~al.}(2010)\citenamefont {Balog},
  \citenamefont {J{\o}rgensen}, \citenamefont {Nilsson}, \citenamefont
  {Andersen}, \citenamefont {Rienks}, \citenamefont {Bianchi}, \citenamefont
  {Fanetti}, \citenamefont {L{\ae}gsgaard}, \citenamefont {Baraldi},
  \citenamefont {Lizzit}, \citenamefont {Sljivancanin}, \citenamefont
  {Besenbacher}, \citenamefont {Hammer}, \citenamefont {Pedersen},
  \citenamefont {Hofmann},\ and\ \citenamefont {Hornek{\ae}r}}]{Balog2010}%
  \BibitemOpen
  \bibfield  {author} {\bibinfo {author} {\bibfnamefont {R.}~\bibnamefont
  {Balog}}, \bibinfo {author} {\bibfnamefont {B.}~\bibnamefont {J{\o}rgensen}},
  \bibinfo {author} {\bibfnamefont {L.}~\bibnamefont {Nilsson}}, \bibinfo
  {author} {\bibfnamefont {M.}~\bibnamefont {Andersen}}, \bibinfo {author}
  {\bibfnamefont {E.}~\bibnamefont {Rienks}}, \bibinfo {author} {\bibfnamefont
  {M.}~\bibnamefont {Bianchi}}, \bibinfo {author} {\bibfnamefont
  {M.}~\bibnamefont {Fanetti}}, \bibinfo {author} {\bibfnamefont
  {E.}~\bibnamefont {L{\ae}gsgaard}}, \bibinfo {author} {\bibfnamefont
  {A.}~\bibnamefont {Baraldi}}, \bibinfo {author} {\bibfnamefont
  {S.}~\bibnamefont {Lizzit}}, \bibinfo {author} {\bibfnamefont
  {Z.}~\bibnamefont {Sljivancanin}}, \bibinfo {author} {\bibfnamefont
  {F.}~\bibnamefont {Besenbacher}}, \bibinfo {author} {\bibfnamefont
  {B.}~\bibnamefont {Hammer}}, \bibinfo {author} {\bibfnamefont {T.~G.}\
  \bibnamefont {Pedersen}}, \bibinfo {author} {\bibfnamefont {P.}~\bibnamefont
  {Hofmann}}, \ and\ \bibinfo {author} {\bibfnamefont {L.}~\bibnamefont
  {Hornek{\ae}r}},\ }\href {\doibase 10.1038/nmat2710} {\bibfield  {journal}
  {\bibinfo  {journal} {Nat. Mater.}\ }\textbf {\bibinfo {volume} {9}},\
  \bibinfo {pages} {315} (\bibinfo {year} {2010})}\BibitemShut {NoStop}%
\bibitem [{\citenamefont {Son}\ \emph {et~al.}(2016)\citenamefont {Son},
  \citenamefont {Lee}, \citenamefont {Kim}, \citenamefont {Park}, \citenamefont
  {Lee}, \citenamefont {Kim},\ and\ \citenamefont {Kim}}]{Son2016}%
  \BibitemOpen
  \bibfield  {author} {\bibinfo {author} {\bibfnamefont {J.}~\bibnamefont
  {Son}}, \bibinfo {author} {\bibfnamefont {S.}~\bibnamefont {Lee}}, \bibinfo
  {author} {\bibfnamefont {S.~J.}\ \bibnamefont {Kim}}, \bibinfo {author}
  {\bibfnamefont {B.~C.}\ \bibnamefont {Park}}, \bibinfo {author}
  {\bibfnamefont {H.-k.}\ \bibnamefont {Lee}}, \bibinfo {author} {\bibfnamefont
  {S.~J.}\ \bibnamefont {Kim}}, \ and\ \bibinfo {author} {\bibfnamefont
  {J.~H.}\ \bibnamefont {Kim}},\ }\href {\doibase 10.1038/ncomms13261}
  {\bibfield  {journal} {\bibinfo  {journal} {Nat. Commun.}\ }\textbf {\bibinfo
  {volume} {7}},\ \bibinfo {pages} {13261} (\bibinfo {year}
  {2016})}\BibitemShut {NoStop}%
\bibitem [{\citenamefont {Fetter}\ and\ \citenamefont
  {Walecka}(1971)}]{Fetter1971}%
  \BibitemOpen
  \bibfield  {author} {\bibinfo {author} {\bibfnamefont {A.~L.}\ \bibnamefont
  {Fetter}}\ and\ \bibinfo {author} {\bibfnamefont {J.~D.}\ \bibnamefont
  {Walecka}},\ }\href@noop {} {\emph {\bibinfo {title} {Quantum Theory of
  Many-Particle Systems}}},\ \bibinfo {edition} {dover}\ ed.\ (\bibinfo
  {publisher} {Dover Publication, Inc.},\ \bibinfo {address} {Mineola, New
  York},\ \bibinfo {year} {1971})\BibitemShut {NoStop}%
\bibitem [{\citenamefont {Wirtz}\ \emph {et~al.}(2006)\citenamefont {Wirtz},
  \citenamefont {Marini},\ and\ \citenamefont {Rubio}}]{Wirtz2006}%
  \BibitemOpen
  \bibfield  {author} {\bibinfo {author} {\bibfnamefont {L.}~\bibnamefont
  {Wirtz}}, \bibinfo {author} {\bibfnamefont {A.}~\bibnamefont {Marini}}, \
  and\ \bibinfo {author} {\bibfnamefont {A.}~\bibnamefont {Rubio}},\ }\href
  {\doibase 10.1103/PhysRevLett.96.126104} {\bibfield  {journal} {\bibinfo
  {journal} {Phys. Rev. Lett.}\ }\textbf {\bibinfo {volume} {96}},\ \bibinfo
  {pages} {126104} (\bibinfo {year} {2006})}\BibitemShut {NoStop}%
\bibitem [{\citenamefont {Gr{\"{u}}ning}\ and\ \citenamefont
  {Attaccalite}(2014)}]{Gruning2014}%
  \BibitemOpen
  \bibfield  {author} {\bibinfo {author} {\bibfnamefont {M.}~\bibnamefont
  {Gr{\"{u}}ning}}\ and\ \bibinfo {author} {\bibfnamefont {C.}~\bibnamefont
  {Attaccalite}},\ }\href {\doibase 10.1103/PhysRevB.89.081102} {\bibfield
  {journal} {\bibinfo  {journal} {Phys. Rev. B}\ }\textbf {\bibinfo {volume}
  {89}},\ \bibinfo {pages} {081102} (\bibinfo {year} {2014})}\BibitemShut
  {NoStop}%
\bibitem [{\citenamefont {Pedersen}(2015)}]{Pedersen2015}%
  \BibitemOpen
  \bibfield  {author} {\bibinfo {author} {\bibfnamefont {T.~G.}\ \bibnamefont
  {Pedersen}},\ }\href {\doibase 10.1103/PhysRevB.92.235432} {\bibfield
  {journal} {\bibinfo  {journal} {Phys. Rev. B}\ }\textbf {\bibinfo {volume}
  {92}},\ \bibinfo {pages} {235432} (\bibinfo {year} {2015})}\BibitemShut
  {NoStop}%
\bibitem [{\citenamefont {Wagoner}\ \emph {et~al.}(1998)\citenamefont
  {Wagoner}, \citenamefont {Persans}, \citenamefont {{Van Wagenen}},\ and\
  \citenamefont {Korenowski}}]{Wagoner1998}%
  \BibitemOpen
  \bibfield  {author} {\bibinfo {author} {\bibfnamefont {G.~A.}\ \bibnamefont
  {Wagoner}}, \bibinfo {author} {\bibfnamefont {P.~D.}\ \bibnamefont
  {Persans}}, \bibinfo {author} {\bibfnamefont {E.~A.}\ \bibnamefont {{Van
  Wagenen}}}, \ and\ \bibinfo {author} {\bibfnamefont {G.~M.}\ \bibnamefont
  {Korenowski}},\ }\href {\doibase 10.1364/JOSAB.15.001017} {\bibfield
  {journal} {\bibinfo  {journal} {J. Opt. Soc. Am. B}\ }\textbf {\bibinfo
  {volume} {15}},\ \bibinfo {pages} {1017} (\bibinfo {year}
  {1998})}\BibitemShut {NoStop}%
\bibitem [{\citenamefont {Splendiani}\ \emph {et~al.}(2010)\citenamefont
  {Splendiani}, \citenamefont {Sun}, \citenamefont {Zhang}, \citenamefont {Li},
  \citenamefont {Kim}, \citenamefont {Chim}, \citenamefont {Galli},\ and\
  \citenamefont {Wang}}]{Splendiani2010}%
  \BibitemOpen
  \bibfield  {author} {\bibinfo {author} {\bibfnamefont {A.}~\bibnamefont
  {Splendiani}}, \bibinfo {author} {\bibfnamefont {L.}~\bibnamefont {Sun}},
  \bibinfo {author} {\bibfnamefont {Y.}~\bibnamefont {Zhang}}, \bibinfo
  {author} {\bibfnamefont {T.}~\bibnamefont {Li}}, \bibinfo {author}
  {\bibfnamefont {J.}~\bibnamefont {Kim}}, \bibinfo {author} {\bibfnamefont
  {C.-Y.}\ \bibnamefont {Chim}}, \bibinfo {author} {\bibfnamefont
  {G.}~\bibnamefont {Galli}}, \ and\ \bibinfo {author} {\bibfnamefont
  {F.}~\bibnamefont {Wang}},\ }\href {\doibase 10.1021/nl903868w} {\bibfield
  {journal} {\bibinfo  {journal} {Nano Lett.}\ }\textbf {\bibinfo {volume}
  {10}},\ \bibinfo {pages} {1271} (\bibinfo {year} {2010})}\BibitemShut
  {NoStop}%
\bibitem [{\citenamefont {Zeng}\ \emph {et~al.}(2012)\citenamefont {Zeng},
  \citenamefont {Dai}, \citenamefont {Yao}, \citenamefont {Xiao},\ and\
  \citenamefont {Cui}}]{Zeng2012}%
  \BibitemOpen
  \bibfield  {author} {\bibinfo {author} {\bibfnamefont {H.}~\bibnamefont
  {Zeng}}, \bibinfo {author} {\bibfnamefont {J.}~\bibnamefont {Dai}}, \bibinfo
  {author} {\bibfnamefont {W.}~\bibnamefont {Yao}}, \bibinfo {author}
  {\bibfnamefont {D.}~\bibnamefont {Xiao}}, \ and\ \bibinfo {author}
  {\bibfnamefont {X.}~\bibnamefont {Cui}},\ }\href {\doibase
  10.1038/nnano.2012.95} {\bibfield  {journal} {\bibinfo  {journal} {Nat.
  Nanotechnol.}\ }\textbf {\bibinfo {volume} {7}},\ \bibinfo {pages} {490}
  (\bibinfo {year} {2012})}\BibitemShut {NoStop}%
\bibitem [{\citenamefont {Jiang}\ \emph {et~al.}(2014)\citenamefont {Jiang},
  \citenamefont {Liu}, \citenamefont {Huang}, \citenamefont {Zhang},
  \citenamefont {Li}, \citenamefont {Gong}, \citenamefont {Shen}, \citenamefont
  {Liu},\ and\ \citenamefont {Wu}}]{Jiang2014}%
  \BibitemOpen
  \bibfield  {author} {\bibinfo {author} {\bibfnamefont {T.}~\bibnamefont
  {Jiang}}, \bibinfo {author} {\bibfnamefont {H.}~\bibnamefont {Liu}}, \bibinfo
  {author} {\bibfnamefont {D.}~\bibnamefont {Huang}}, \bibinfo {author}
  {\bibfnamefont {S.}~\bibnamefont {Zhang}}, \bibinfo {author} {\bibfnamefont
  {Y.}~\bibnamefont {Li}}, \bibinfo {author} {\bibfnamefont {X.}~\bibnamefont
  {Gong}}, \bibinfo {author} {\bibfnamefont {Y.-R.}\ \bibnamefont {Shen}},
  \bibinfo {author} {\bibfnamefont {W.-T.}\ \bibnamefont {Liu}}, \ and\
  \bibinfo {author} {\bibfnamefont {S.}~\bibnamefont {Wu}},\ }\href {\doibase
  10.1038/nnano.2014.176} {\bibfield  {journal} {\bibinfo  {journal} {Nat.
  Nanotechnol.}\ }\textbf {\bibinfo {volume} {9}},\ \bibinfo {pages} {825}
  (\bibinfo {year} {2014})}\BibitemShut {NoStop}%
\bibitem [{\citenamefont {Wang}\ \emph
  {et~al.}(2015{\natexlab{b}})\citenamefont {Wang}, \citenamefont {Jones},
  \citenamefont {Seyler}, \citenamefont {Tran}, \citenamefont {Jia},
  \citenamefont {Zhao}, \citenamefont {Wang}, \citenamefont {Yang},
  \citenamefont {Xu},\ and\ \citenamefont {Xia}}]{Wang2015c}%
  \BibitemOpen
  \bibfield  {author} {\bibinfo {author} {\bibfnamefont {X.}~\bibnamefont
  {Wang}}, \bibinfo {author} {\bibfnamefont {A.~M.}\ \bibnamefont {Jones}},
  \bibinfo {author} {\bibfnamefont {K.~L.}\ \bibnamefont {Seyler}}, \bibinfo
  {author} {\bibfnamefont {V.}~\bibnamefont {Tran}}, \bibinfo {author}
  {\bibfnamefont {Y.}~\bibnamefont {Jia}}, \bibinfo {author} {\bibfnamefont
  {H.}~\bibnamefont {Zhao}}, \bibinfo {author} {\bibfnamefont {H.}~\bibnamefont
  {Wang}}, \bibinfo {author} {\bibfnamefont {L.}~\bibnamefont {Yang}}, \bibinfo
  {author} {\bibfnamefont {X.}~\bibnamefont {Xu}}, \ and\ \bibinfo {author}
  {\bibfnamefont {F.}~\bibnamefont {Xia}},\ }\href {\doibase
  10.1038/nnano.2015.71} {\bibfield  {journal} {\bibinfo  {journal} {Nat.
  Nanotechnol.}\ }\textbf {\bibinfo {volume} {10}},\ \bibinfo {pages} {517}
  (\bibinfo {year} {2015}{\natexlab{b}})}\BibitemShut {NoStop}%
\bibitem [{\citenamefont {Trolle}\ \emph {et~al.}(2015)\citenamefont {Trolle},
  \citenamefont {Tsao}, \citenamefont {Pedersen},\ and\ \citenamefont
  {Pedersen}}]{Trolle2015}%
  \BibitemOpen
  \bibfield  {author} {\bibinfo {author} {\bibfnamefont {M.~L.}\ \bibnamefont
  {Trolle}}, \bibinfo {author} {\bibfnamefont {Y.-C.}\ \bibnamefont {Tsao}},
  \bibinfo {author} {\bibfnamefont {K.}~\bibnamefont {Pedersen}}, \ and\
  \bibinfo {author} {\bibfnamefont {T.~G.}\ \bibnamefont {Pedersen}},\ }\href
  {\doibase 10.1103/PhysRevB.92.161409} {\bibfield  {journal} {\bibinfo
  {journal} {Phys. Rev. B}\ }\textbf {\bibinfo {volume} {92}},\ \bibinfo
  {pages} {161409} (\bibinfo {year} {2015})}\BibitemShut {NoStop}%
\bibitem [{\citenamefont {Li}\ \emph {et~al.}(2016)\citenamefont {Li},
  \citenamefont {Kim}, \citenamefont {Jin}, \citenamefont {Ye}, \citenamefont
  {Qiu}, \citenamefont {da~Jornada}, \citenamefont {Shi}, \citenamefont {Chen},
  \citenamefont {Zhang}, \citenamefont {Yang}, \citenamefont {Watanabe},
  \citenamefont {Taniguchi}, \citenamefont {Ren}, \citenamefont {Louie},
  \citenamefont {Chen}, \citenamefont {Zhang},\ and\ \citenamefont
  {Wang}}]{Li2016a}%
  \BibitemOpen
  \bibfield  {author} {\bibinfo {author} {\bibfnamefont {L.}~\bibnamefont
  {Li}}, \bibinfo {author} {\bibfnamefont {J.}~\bibnamefont {Kim}}, \bibinfo
  {author} {\bibfnamefont {C.}~\bibnamefont {Jin}}, \bibinfo {author}
  {\bibfnamefont {G.~J.}\ \bibnamefont {Ye}}, \bibinfo {author} {\bibfnamefont
  {D.~Y.}\ \bibnamefont {Qiu}}, \bibinfo {author} {\bibfnamefont {F.~H.}\
  \bibnamefont {da~Jornada}}, \bibinfo {author} {\bibfnamefont
  {Z.}~\bibnamefont {Shi}}, \bibinfo {author} {\bibfnamefont {L.}~\bibnamefont
  {Chen}}, \bibinfo {author} {\bibfnamefont {Z.}~\bibnamefont {Zhang}},
  \bibinfo {author} {\bibfnamefont {F.}~\bibnamefont {Yang}}, \bibinfo {author}
  {\bibfnamefont {K.}~\bibnamefont {Watanabe}}, \bibinfo {author}
  {\bibfnamefont {T.}~\bibnamefont {Taniguchi}}, \bibinfo {author}
  {\bibfnamefont {W.}~\bibnamefont {Ren}}, \bibinfo {author} {\bibfnamefont
  {S.~G.}\ \bibnamefont {Louie}}, \bibinfo {author} {\bibfnamefont {X.~H.}\
  \bibnamefont {Chen}}, \bibinfo {author} {\bibfnamefont {Y.}~\bibnamefont
  {Zhang}}, \ and\ \bibinfo {author} {\bibfnamefont {F.}~\bibnamefont {Wang}},\
  }\href {\doibase 10.1038/nnano.2016.171} {\bibfield  {journal} {\bibinfo
  {journal} {Nat. Nanotechnol.}\ }\textbf {\bibinfo {volume} {12}},\ \bibinfo
  {pages} {21} (\bibinfo {year} {2016})}\BibitemShut {NoStop}%
\bibitem [{\citenamefont {Zhang}\ \emph {et~al.}(2017)\citenamefont {Zhang},
  \citenamefont {Huang}, \citenamefont {Chaves}, \citenamefont {Song},
  \citenamefont {{\"{O}}z{\c{c}}elik}, \citenamefont {Low},\ and\ \citenamefont
  {Yan}}]{Zhang2017}%
  \BibitemOpen
  \bibfield  {author} {\bibinfo {author} {\bibfnamefont {G.}~\bibnamefont
  {Zhang}}, \bibinfo {author} {\bibfnamefont {S.}~\bibnamefont {Huang}},
  \bibinfo {author} {\bibfnamefont {A.}~\bibnamefont {Chaves}}, \bibinfo
  {author} {\bibfnamefont {C.}~\bibnamefont {Song}}, \bibinfo {author}
  {\bibfnamefont {V.~O.}\ \bibnamefont {{\"{O}}z{\c{c}}elik}}, \bibinfo
  {author} {\bibfnamefont {T.}~\bibnamefont {Low}}, \ and\ \bibinfo {author}
  {\bibfnamefont {H.}~\bibnamefont {Yan}},\ }\href {\doibase
  10.1038/ncomms14071} {\bibfield  {journal} {\bibinfo  {journal} {Nat.
  Commun.}\ }\textbf {\bibinfo {volume} {8}},\ \bibinfo {pages} {14071}
  (\bibinfo {year} {2017})}\BibitemShut {NoStop}%
\bibitem [{\citenamefont {Elias}\ \emph {et~al.}(2011)\citenamefont {Elias},
  \citenamefont {Gorbachev}, \citenamefont {Mayorov}, \citenamefont {Morozov},
  \citenamefont {Zhukov}, \citenamefont {Blake}, \citenamefont {Ponomarenko},
  \citenamefont {Grigorieva}, \citenamefont {Novoselov}, \citenamefont
  {Guinea},\ and\ \citenamefont {Geim}}]{Elias2011}%
  \BibitemOpen
  \bibfield  {author} {\bibinfo {author} {\bibfnamefont {D.~C.}\ \bibnamefont
  {Elias}}, \bibinfo {author} {\bibfnamefont {R.~V.}\ \bibnamefont
  {Gorbachev}}, \bibinfo {author} {\bibfnamefont {A.~S.}\ \bibnamefont
  {Mayorov}}, \bibinfo {author} {\bibfnamefont {S.~V.}\ \bibnamefont
  {Morozov}}, \bibinfo {author} {\bibfnamefont {A.~A.}\ \bibnamefont {Zhukov}},
  \bibinfo {author} {\bibfnamefont {P.}~\bibnamefont {Blake}}, \bibinfo
  {author} {\bibfnamefont {L.~A.}\ \bibnamefont {Ponomarenko}}, \bibinfo
  {author} {\bibfnamefont {I.~V.}\ \bibnamefont {Grigorieva}}, \bibinfo
  {author} {\bibfnamefont {K.~S.}\ \bibnamefont {Novoselov}}, \bibinfo {author}
  {\bibfnamefont {F.}~\bibnamefont {Guinea}}, \ and\ \bibinfo {author}
  {\bibfnamefont {A.~K.}\ \bibnamefont {Geim}},\ }\href {\doibase
  10.1038/nphys2049} {\bibfield  {journal} {\bibinfo  {journal} {Nat. Phys.}\
  }\textbf {\bibinfo {volume} {7}},\ \bibinfo {pages} {701} (\bibinfo {year}
  {2011})}\BibitemShut {NoStop}%
\bibitem [{\citenamefont {Park}\ \emph {et~al.}(2009)\citenamefont {Park},
  \citenamefont {Giustino}, \citenamefont {Spataru}, \citenamefont {Cohen},\
  and\ \citenamefont {Louie}}]{Park2009}%
  \BibitemOpen
  \bibfield  {author} {\bibinfo {author} {\bibfnamefont {C.-H.}\ \bibnamefont
  {Park}}, \bibinfo {author} {\bibfnamefont {F.}~\bibnamefont {Giustino}},
  \bibinfo {author} {\bibfnamefont {C.~D.}\ \bibnamefont {Spataru}}, \bibinfo
  {author} {\bibfnamefont {M.~L.}\ \bibnamefont {Cohen}}, \ and\ \bibinfo
  {author} {\bibfnamefont {S.~G.}\ \bibnamefont {Louie}},\ }\href {\doibase
  10.1021/nl902448v} {\bibfield  {journal} {\bibinfo  {journal} {Nano Lett.}\
  }\textbf {\bibinfo {volume} {9}},\ \bibinfo {pages} {4234} (\bibinfo {year}
  {2009})}\BibitemShut {NoStop}%
\bibitem [{\citenamefont {Hwang}\ \emph {et~al.}(2011)\citenamefont {Hwang},
  \citenamefont {Park}, \citenamefont {Siegel}, \citenamefont {Fedorov},
  \citenamefont {Louie},\ and\ \citenamefont {Lanzara}}]{Hwang2011}%
  \BibitemOpen
  \bibfield  {author} {\bibinfo {author} {\bibfnamefont {C.}~\bibnamefont
  {Hwang}}, \bibinfo {author} {\bibfnamefont {C.-H.}\ \bibnamefont {Park}},
  \bibinfo {author} {\bibfnamefont {D.~A.}\ \bibnamefont {Siegel}}, \bibinfo
  {author} {\bibfnamefont {A.~V.}\ \bibnamefont {Fedorov}}, \bibinfo {author}
  {\bibfnamefont {S.~G.}\ \bibnamefont {Louie}}, \ and\ \bibinfo {author}
  {\bibfnamefont {A.}~\bibnamefont {Lanzara}},\ }\href {\doibase
  10.1103/PhysRevB.84.125422} {\bibfield  {journal} {\bibinfo  {journal} {Phys.
  Rev. B}\ }\textbf {\bibinfo {volume} {84}},\ \bibinfo {pages} {125422}
  (\bibinfo {year} {2011})}\BibitemShut {NoStop}%
\bibitem [{\citenamefont {Grushin}\ \emph {et~al.}(2009)\citenamefont
  {Grushin}, \citenamefont {Valenzuela},\ and\ \citenamefont
  {Vozmediano}}]{Grushin2009}%
  \BibitemOpen
  \bibfield  {author} {\bibinfo {author} {\bibfnamefont {A.~G.}\ \bibnamefont
  {Grushin}}, \bibinfo {author} {\bibfnamefont {B.}~\bibnamefont {Valenzuela}},
  \ and\ \bibinfo {author} {\bibfnamefont {M.~A.~H.}\ \bibnamefont
  {Vozmediano}},\ }\href {\doibase 10.1103/PhysRevB.80.155417} {\bibfield
  {journal} {\bibinfo  {journal} {Phys. Rev. B}\ }\textbf {\bibinfo {volume}
  {80}},\ \bibinfo {pages} {155417} (\bibinfo {year} {2009})}\BibitemShut
  {NoStop}%
\bibitem [{\citenamefont {Peres}\ \emph {et~al.}(2010)\citenamefont {Peres},
  \citenamefont {Ribeiro},\ and\ \citenamefont {{Castro Neto}}}]{Peres2010b}%
  \BibitemOpen
  \bibfield  {author} {\bibinfo {author} {\bibfnamefont {N.~M.~R.}\
  \bibnamefont {Peres}}, \bibinfo {author} {\bibfnamefont {R.~M.}\ \bibnamefont
  {Ribeiro}}, \ and\ \bibinfo {author} {\bibfnamefont {A.~H.}\ \bibnamefont
  {{Castro Neto}}},\ }\href {\doibase 10.1103/PhysRevLett.105.055501}
  {\bibfield  {journal} {\bibinfo  {journal} {Phys. Rev. Lett.}\ }\textbf
  {\bibinfo {volume} {105}},\ \bibinfo {pages} {055501} (\bibinfo {year}
  {2010})}\BibitemShut {NoStop}%
\bibitem [{\citenamefont {Stroucken}\ \emph {et~al.}(2011)\citenamefont
  {Stroucken}, \citenamefont {Gr{\"{o}}nqvist},\ and\ \citenamefont
  {Koch}}]{Stroucken2011}%
  \BibitemOpen
  \bibfield  {author} {\bibinfo {author} {\bibfnamefont {T.}~\bibnamefont
  {Stroucken}}, \bibinfo {author} {\bibfnamefont {J.~H.}\ \bibnamefont
  {Gr{\"{o}}nqvist}}, \ and\ \bibinfo {author} {\bibfnamefont {S.~W.}\
  \bibnamefont {Koch}},\ }\href {\doibase 10.1103/PhysRevB.84.205445}
  {\bibfield  {journal} {\bibinfo  {journal} {Phys. Rev. B}\ }\textbf {\bibinfo
  {volume} {84}},\ \bibinfo {pages} {205445} (\bibinfo {year}
  {2011})}\BibitemShut {NoStop}%
\bibitem [{\citenamefont {Breusing}\ \emph {et~al.}(2011)\citenamefont
  {Breusing}, \citenamefont {Kuehn}, \citenamefont {Winzer}, \citenamefont
  {Mali{\'{c}}}, \citenamefont {Milde}, \citenamefont {Severin}, \citenamefont
  {Rabe}, \citenamefont {Ropers}, \citenamefont {Knorr},\ and\ \citenamefont
  {Elsaesser}}]{Breusing2011}%
  \BibitemOpen
  \bibfield  {author} {\bibinfo {author} {\bibfnamefont {M.}~\bibnamefont
  {Breusing}}, \bibinfo {author} {\bibfnamefont {S.}~\bibnamefont {Kuehn}},
  \bibinfo {author} {\bibfnamefont {T.}~\bibnamefont {Winzer}}, \bibinfo
  {author} {\bibfnamefont {E.}~\bibnamefont {Mali{\'{c}}}}, \bibinfo {author}
  {\bibfnamefont {F.}~\bibnamefont {Milde}}, \bibinfo {author} {\bibfnamefont
  {N.}~\bibnamefont {Severin}}, \bibinfo {author} {\bibfnamefont {J.~P.}\
  \bibnamefont {Rabe}}, \bibinfo {author} {\bibfnamefont {C.}~\bibnamefont
  {Ropers}}, \bibinfo {author} {\bibfnamefont {A.}~\bibnamefont {Knorr}}, \
  and\ \bibinfo {author} {\bibfnamefont {T.}~\bibnamefont {Elsaesser}},\ }\href
  {\doibase 10.1103/PhysRevB.83.153410} {\bibfield  {journal} {\bibinfo
  {journal} {Phys. Rev. B}\ }\textbf {\bibinfo {volume} {83}},\ \bibinfo
  {pages} {153410} (\bibinfo {year} {2011})}\BibitemShut {NoStop}%
\bibitem [{\citenamefont {Sun}\ \emph {et~al.}(2012{\natexlab{a}})\citenamefont
  {Sun}, \citenamefont {Aivazian}, \citenamefont {Jones}, \citenamefont {Ross},
  \citenamefont {Yao}, \citenamefont {Cobden},\ and\ \citenamefont
  {Xu}}]{Sun2012}%
  \BibitemOpen
  \bibfield  {author} {\bibinfo {author} {\bibfnamefont {D.}~\bibnamefont
  {Sun}}, \bibinfo {author} {\bibfnamefont {G.}~\bibnamefont {Aivazian}},
  \bibinfo {author} {\bibfnamefont {A.~M.}\ \bibnamefont {Jones}}, \bibinfo
  {author} {\bibfnamefont {J.~S.}\ \bibnamefont {Ross}}, \bibinfo {author}
  {\bibfnamefont {W.}~\bibnamefont {Yao}}, \bibinfo {author} {\bibfnamefont
  {D.}~\bibnamefont {Cobden}}, \ and\ \bibinfo {author} {\bibfnamefont
  {X.}~\bibnamefont {Xu}},\ }\href {\doibase 10.1038/nnano.2011.243} {\bibfield
   {journal} {\bibinfo  {journal} {Nat. Nanotechnol.}\ }\textbf {\bibinfo
  {volume} {7}},\ \bibinfo {pages} {114} (\bibinfo {year}
  {2012}{\natexlab{a}})}\BibitemShut {NoStop}%
\bibitem [{\citenamefont {Malic}\ \emph {et~al.}(2011)\citenamefont {Malic},
  \citenamefont {Winzer}, \citenamefont {Bobkin},\ and\ \citenamefont
  {Knorr}}]{Malic2011}%
  \BibitemOpen
  \bibfield  {author} {\bibinfo {author} {\bibfnamefont {E.}~\bibnamefont
  {Malic}}, \bibinfo {author} {\bibfnamefont {T.}~\bibnamefont {Winzer}},
  \bibinfo {author} {\bibfnamefont {E.}~\bibnamefont {Bobkin}}, \ and\ \bibinfo
  {author} {\bibfnamefont {A.}~\bibnamefont {Knorr}},\ }\href {\doibase
  10.1103/PhysRevB.84.205406} {\bibfield  {journal} {\bibinfo  {journal} {Phys.
  Rev. B}\ }\textbf {\bibinfo {volume} {84}},\ \bibinfo {pages} {205406}
  (\bibinfo {year} {2011})}\BibitemShut {NoStop}%
\bibitem [{\citenamefont {Sule}\ \emph {et~al.}(2014)\citenamefont {Sule},
  \citenamefont {Willis}, \citenamefont {Hagness},\ and\ \citenamefont
  {Knezevic}}]{Sule2014}%
  \BibitemOpen
  \bibfield  {author} {\bibinfo {author} {\bibfnamefont {N.}~\bibnamefont
  {Sule}}, \bibinfo {author} {\bibfnamefont {K.~J.}\ \bibnamefont {Willis}},
  \bibinfo {author} {\bibfnamefont {S.~C.}\ \bibnamefont {Hagness}}, \ and\
  \bibinfo {author} {\bibfnamefont {I.}~\bibnamefont {Knezevic}},\ }\href
  {\doibase 10.1103/PhysRevB.90.045431} {\bibfield  {journal} {\bibinfo
  {journal} {Phys. Rev. B}\ }\textbf {\bibinfo {volume} {90}},\ \bibinfo
  {pages} {045431} (\bibinfo {year} {2014})}\BibitemShut {NoStop}%
\bibitem [{\citenamefont {Vasko}\ \emph {et~al.}(2012)\citenamefont {Vasko},
  \citenamefont {Mitin}, \citenamefont {Ryzhii},\ and\ \citenamefont
  {Otsuji}}]{Vasko2012}%
  \BibitemOpen
  \bibfield  {author} {\bibinfo {author} {\bibfnamefont {F.~T.}\ \bibnamefont
  {Vasko}}, \bibinfo {author} {\bibfnamefont {V.~V.}\ \bibnamefont {Mitin}},
  \bibinfo {author} {\bibfnamefont {V.}~\bibnamefont {Ryzhii}}, \ and\ \bibinfo
  {author} {\bibfnamefont {T.}~\bibnamefont {Otsuji}},\ }\href {\doibase
  10.1103/PhysRevB.86.235424} {\bibfield  {journal} {\bibinfo  {journal} {Phys.
  Rev. B}\ }\textbf {\bibinfo {volume} {86}},\ \bibinfo {pages} {235424}
  (\bibinfo {year} {2012})}\BibitemShut {NoStop}%
\bibitem [{\citenamefont {Avetissian}\ and\ \citenamefont
  {Mkrtchian}(2018)}]{Avetissian2018}%
  \BibitemOpen
  \bibfield  {author} {\bibinfo {author} {\bibfnamefont {H.~K.}\ \bibnamefont
  {Avetissian}}\ and\ \bibinfo {author} {\bibfnamefont {G.~F.}\ \bibnamefont
  {Mkrtchian}},\ }\href {\doibase 10.1103/PhysRevB.97.115454} {\bibfield
  {journal} {\bibinfo  {journal} {Phys. Rev. B}\ }\textbf {\bibinfo {volume}
  {97}},\ \bibinfo {pages} {115454} (\bibinfo {year} {2018})}\BibitemShut
  {NoStop}%
\bibitem [{\citenamefont {Avetissian}\ \emph {et~al.}(2018)\citenamefont
  {Avetissian}, \citenamefont {Avetissian}, \citenamefont {Avchyan},\ and\
  \citenamefont {Mkrtchian}}]{Avetissian2018a}%
  \BibitemOpen
  \bibfield  {author} {\bibinfo {author} {\bibfnamefont {H.~K.}\ \bibnamefont
  {Avetissian}}, \bibinfo {author} {\bibfnamefont {A.~K.}\ \bibnamefont
  {Avetissian}}, \bibinfo {author} {\bibfnamefont {B.~R.}\ \bibnamefont
  {Avchyan}}, \ and\ \bibinfo {author} {\bibfnamefont {G.~F.}\ \bibnamefont
  {Mkrtchian}},\ }\href {\doibase 10.1088/1361-648X/aab989} {\bibfield
  {journal} {\bibinfo  {journal} {J. Physics: Condensed Matter}\ }\textbf
  {\bibinfo {volume} {30}},\ \bibinfo {pages} {185302} (\bibinfo {year}
  {2018})}\BibitemShut {NoStop}%
\bibitem [{\citenamefont {Yu}\ \emph {et~al.}(2013)\citenamefont {Yu},
  \citenamefont {Jalil}, \citenamefont {Belle}, \citenamefont {Mayorov},
  \citenamefont {Blake}, \citenamefont {Schedin}, \citenamefont {Morozov},
  \citenamefont {Ponomarenko}, \citenamefont {Chiappini}, \citenamefont
  {Wiedmann}, \citenamefont {Zeitler}, \citenamefont {Katsnelson},
  \citenamefont {Geim}, \citenamefont {Novoselov},\ and\ \citenamefont
  {Elias}}]{Yu2013}%
  \BibitemOpen
  \bibfield  {author} {\bibinfo {author} {\bibfnamefont {G.~L.}\ \bibnamefont
  {Yu}}, \bibinfo {author} {\bibfnamefont {R.}~\bibnamefont {Jalil}}, \bibinfo
  {author} {\bibfnamefont {B.}~\bibnamefont {Belle}}, \bibinfo {author}
  {\bibfnamefont {A.~S.}\ \bibnamefont {Mayorov}}, \bibinfo {author}
  {\bibfnamefont {P.}~\bibnamefont {Blake}}, \bibinfo {author} {\bibfnamefont
  {F.}~\bibnamefont {Schedin}}, \bibinfo {author} {\bibfnamefont {S.~V.}\
  \bibnamefont {Morozov}}, \bibinfo {author} {\bibfnamefont {L.~A.}\
  \bibnamefont {Ponomarenko}}, \bibinfo {author} {\bibfnamefont
  {F.}~\bibnamefont {Chiappini}}, \bibinfo {author} {\bibfnamefont
  {S.}~\bibnamefont {Wiedmann}}, \bibinfo {author} {\bibfnamefont
  {U.}~\bibnamefont {Zeitler}}, \bibinfo {author} {\bibfnamefont {M.~I.}\
  \bibnamefont {Katsnelson}}, \bibinfo {author} {\bibfnamefont {A.~K.}\
  \bibnamefont {Geim}}, \bibinfo {author} {\bibfnamefont {K.~S.}\ \bibnamefont
  {Novoselov}}, \ and\ \bibinfo {author} {\bibfnamefont {D.~C.}\ \bibnamefont
  {Elias}},\ }\href {\doibase 10.1073/pnas.1300599110} {\bibfield  {journal}
  {\bibinfo  {journal} {PNAS}\ }\textbf {\bibinfo {volume} {110}},\ \bibinfo
  {pages} {3282} (\bibinfo {year} {2013})}\BibitemShut {NoStop}%
\bibitem [{\citenamefont {Stauber}\ \emph {et~al.}(2017)\citenamefont
  {Stauber}, \citenamefont {Parida}, \citenamefont {Trushin}, \citenamefont
  {Ulybyshev}, \citenamefont {Boyda},\ and\ \citenamefont
  {Schliemann}}]{Stauber2017}%
  \BibitemOpen
  \bibfield  {author} {\bibinfo {author} {\bibfnamefont {T.}~\bibnamefont
  {Stauber}}, \bibinfo {author} {\bibfnamefont {P.}~\bibnamefont {Parida}},
  \bibinfo {author} {\bibfnamefont {M.}~\bibnamefont {Trushin}}, \bibinfo
  {author} {\bibfnamefont {M.~V.}\ \bibnamefont {Ulybyshev}}, \bibinfo {author}
  {\bibfnamefont {D.~L.}\ \bibnamefont {Boyda}}, \ and\ \bibinfo {author}
  {\bibfnamefont {J.}~\bibnamefont {Schliemann}},\ }\href {\doibase
  10.1103/PhysRevLett.118.266801} {\bibfield  {journal} {\bibinfo  {journal}
  {Phys. Rev. Lett.}\ }\textbf {\bibinfo {volume} {118}},\ \bibinfo {pages}
  {266801} (\bibinfo {year} {2017})}\BibitemShut {NoStop}%
\bibitem [{\citenamefont {Sun}\ \emph {et~al.}(2012{\natexlab{b}})\citenamefont
  {Sun}, \citenamefont {Divin}, \citenamefont {Mihnev}, \citenamefont {Winzer},
  \citenamefont {Malic}, \citenamefont {Knorr}, \citenamefont {Sipe},
  \citenamefont {Berger}, \citenamefont {de~Heer}, \citenamefont {First},\ and\
  \citenamefont {Norris}}]{Sun2012a}%
  \BibitemOpen
  \bibfield  {author} {\bibinfo {author} {\bibfnamefont {D.}~\bibnamefont
  {Sun}}, \bibinfo {author} {\bibfnamefont {C.}~\bibnamefont {Divin}}, \bibinfo
  {author} {\bibfnamefont {M.}~\bibnamefont {Mihnev}}, \bibinfo {author}
  {\bibfnamefont {T.}~\bibnamefont {Winzer}}, \bibinfo {author} {\bibfnamefont
  {E.}~\bibnamefont {Malic}}, \bibinfo {author} {\bibfnamefont
  {A.}~\bibnamefont {Knorr}}, \bibinfo {author} {\bibfnamefont {J.~E.}\
  \bibnamefont {Sipe}}, \bibinfo {author} {\bibfnamefont {C.}~\bibnamefont
  {Berger}}, \bibinfo {author} {\bibfnamefont {W.~A.}\ \bibnamefont {de~Heer}},
  \bibinfo {author} {\bibfnamefont {P.~N.}\ \bibnamefont {First}}, \ and\
  \bibinfo {author} {\bibfnamefont {T.~B.}\ \bibnamefont {Norris}},\ }\href
  {\doibase 10.1088/1367-2630/14/10/105012} {\bibfield  {journal} {\bibinfo
  {journal} {New J. Phys.}\ }\textbf {\bibinfo {volume} {14}},\ \bibinfo
  {pages} {105012} (\bibinfo {year} {2012}{\natexlab{b}})}\BibitemShut
  {NoStop}%
\bibitem [{\citenamefont {Peres}(2010)}]{Peres2010}%
  \BibitemOpen
  \bibfield  {author} {\bibinfo {author} {\bibfnamefont {N.~M.~R.}\
  \bibnamefont {Peres}},\ }\href {\doibase 10.1103/RevModPhys.82.2673}
  {\bibfield  {journal} {\bibinfo  {journal} {Rev. Mod. Phys.}\ }\textbf
  {\bibinfo {volume} {82}},\ \bibinfo {pages} {2673} (\bibinfo {year}
  {2010})}\BibitemShut {NoStop}%
\bibitem [{\citenamefont {Park}\ and\ \citenamefont {Louie}(2010)}]{Park2010}%
  \BibitemOpen
  \bibfield  {author} {\bibinfo {author} {\bibfnamefont {C.~H.}\ \bibnamefont
  {Park}}\ and\ \bibinfo {author} {\bibfnamefont {S.~G.}\ \bibnamefont
  {Louie}},\ }\href {\doibase 10.1021/nl902932k} {\bibfield  {journal}
  {\bibinfo  {journal} {Nano Lett.}\ }\textbf {\bibinfo {volume} {10}},\
  \bibinfo {pages} {426} (\bibinfo {year} {2010})}\BibitemShut {NoStop}%
\bibitem [{\citenamefont {Ju}\ \emph {et~al.}(2017)\citenamefont {Ju},
  \citenamefont {Wang}, \citenamefont {Cao}, \citenamefont {Taniguchi},
  \citenamefont {Watanabe}, \citenamefont {Louie}, \citenamefont {Rana},
  \citenamefont {Park}, \citenamefont {Hone}, \citenamefont {Wang},\ and\
  \citenamefont {McEuen}}]{Ju2017}%
  \BibitemOpen
  \bibfield  {author} {\bibinfo {author} {\bibfnamefont {L.}~\bibnamefont
  {Ju}}, \bibinfo {author} {\bibfnamefont {L.}~\bibnamefont {Wang}}, \bibinfo
  {author} {\bibfnamefont {T.}~\bibnamefont {Cao}}, \bibinfo {author}
  {\bibfnamefont {T.}~\bibnamefont {Taniguchi}}, \bibinfo {author}
  {\bibfnamefont {K.}~\bibnamefont {Watanabe}}, \bibinfo {author}
  {\bibfnamefont {S.~G.}\ \bibnamefont {Louie}}, \bibinfo {author}
  {\bibfnamefont {F.}~\bibnamefont {Rana}}, \bibinfo {author} {\bibfnamefont
  {J.}~\bibnamefont {Park}}, \bibinfo {author} {\bibfnamefont {J.}~\bibnamefont
  {Hone}}, \bibinfo {author} {\bibfnamefont {F.}~\bibnamefont {Wang}}, \ and\
  \bibinfo {author} {\bibfnamefont {P.~L.}\ \bibnamefont {McEuen}},\ }\href
  {\doibase 10.1126/science.aam9175} {\bibfield  {journal} {\bibinfo  {journal}
  {Sci.}\ }\textbf {\bibinfo {volume} {358}},\ \bibinfo {pages} {907} (\bibinfo
  {year} {2017})}\BibitemShut {NoStop}%
\bibitem [{\citenamefont {Horng}\ \emph {et~al.}(2011)\citenamefont {Horng},
  \citenamefont {Chen}, \citenamefont {Geng}, \citenamefont {Girit},
  \citenamefont {Zhang}, \citenamefont {Hao}, \citenamefont {Bechtel},
  \citenamefont {Martin}, \citenamefont {Zettl}, \citenamefont {Crommie},
  \citenamefont {Shen},\ and\ \citenamefont {Wang}}]{Horng2011}%
  \BibitemOpen
  \bibfield  {author} {\bibinfo {author} {\bibfnamefont {J.}~\bibnamefont
  {Horng}}, \bibinfo {author} {\bibfnamefont {C.-F.}\ \bibnamefont {Chen}},
  \bibinfo {author} {\bibfnamefont {B.}~\bibnamefont {Geng}}, \bibinfo {author}
  {\bibfnamefont {C.}~\bibnamefont {Girit}}, \bibinfo {author} {\bibfnamefont
  {Y.}~\bibnamefont {Zhang}}, \bibinfo {author} {\bibfnamefont
  {Z.}~\bibnamefont {Hao}}, \bibinfo {author} {\bibfnamefont {H.~A.}\
  \bibnamefont {Bechtel}}, \bibinfo {author} {\bibfnamefont {M.~C.}\
  \bibnamefont {Martin}}, \bibinfo {author} {\bibfnamefont {A.}~\bibnamefont
  {Zettl}}, \bibinfo {author} {\bibfnamefont {M.~F.}\ \bibnamefont {Crommie}},
  \bibinfo {author} {\bibfnamefont {Y.~R.}\ \bibnamefont {Shen}}, \ and\
  \bibinfo {author} {\bibfnamefont {F.}~\bibnamefont {Wang}},\ }\href {\doibase
  10.1103/PhysRevB.83.165113} {\bibfield  {journal} {\bibinfo  {journal} {Phys.
  Rev. B}\ }\textbf {\bibinfo {volume} {83}},\ \bibinfo {pages} {165113}
  (\bibinfo {year} {2011})}\BibitemShut {NoStop}%
\bibitem [{\citenamefont {Stauber}\ \emph
  {et~al.}(2008{\natexlab{a}})\citenamefont {Stauber}, \citenamefont {Peres},\
  and\ \citenamefont {Geim}}]{Stauber2008}%
  \BibitemOpen
  \bibfield  {author} {\bibinfo {author} {\bibfnamefont {T.}~\bibnamefont
  {Stauber}}, \bibinfo {author} {\bibfnamefont {N.~M.~R.}\ \bibnamefont
  {Peres}}, \ and\ \bibinfo {author} {\bibfnamefont {A.~K.}\ \bibnamefont
  {Geim}},\ }\href {\doibase 10.1103/PhysRevB.78.085432} {\bibfield  {journal}
  {\bibinfo  {journal} {Phys. Rev. B}\ }\textbf {\bibinfo {volume} {78}},\
  \bibinfo {pages} {085432} (\bibinfo {year} {2008}{\natexlab{a}})}\BibitemShut
  {NoStop}%
\bibitem [{\citenamefont {Stauber}\ and\ \citenamefont
  {Peres}(2008)}]{Stauber2008a}%
  \BibitemOpen
  \bibfield  {author} {\bibinfo {author} {\bibfnamefont {T.}~\bibnamefont
  {Stauber}}\ and\ \bibinfo {author} {\bibfnamefont {N.~M.~R.}\ \bibnamefont
  {Peres}},\ }\href {\doibase 10.1088/0953-8984/20/5/055002} {\bibfield
  {journal} {\bibinfo  {journal} {J. Phys. Condens. Matter}\ }\textbf {\bibinfo
  {volume} {20}},\ \bibinfo {pages} {055002} (\bibinfo {year}
  {2008})}\BibitemShut {NoStop}%
\bibitem [{\citenamefont {Stauber}\ \emph
  {et~al.}(2008{\natexlab{b}})\citenamefont {Stauber}, \citenamefont {Peres},\
  and\ \citenamefont {{Castro Neto}}}]{Stauber2008b}%
  \BibitemOpen
  \bibfield  {author} {\bibinfo {author} {\bibfnamefont {T.}~\bibnamefont
  {Stauber}}, \bibinfo {author} {\bibfnamefont {N.~M.~R.}\ \bibnamefont
  {Peres}}, \ and\ \bibinfo {author} {\bibfnamefont {A.~H.}\ \bibnamefont
  {{Castro Neto}}},\ }\href {\doibase 10.1103/PhysRevB.78.085418} {\bibfield
  {journal} {\bibinfo  {journal} {Phys. Rev. B}\ }\textbf {\bibinfo {volume}
  {78}},\ \bibinfo {pages} {085418} (\bibinfo {year}
  {2008}{\natexlab{b}})}\BibitemShut {NoStop}%
\bibitem [{\citenamefont {Peres}\ \emph {et~al.}(2008)\citenamefont {Peres},
  \citenamefont {Stauber},\ and\ \citenamefont {{Castro Neto}}}]{Peres2008a}%
  \BibitemOpen
  \bibfield  {author} {\bibinfo {author} {\bibfnamefont {N.~M.~R.}\
  \bibnamefont {Peres}}, \bibinfo {author} {\bibfnamefont {T.}~\bibnamefont
  {Stauber}}, \ and\ \bibinfo {author} {\bibfnamefont {A.~H.}\ \bibnamefont
  {{Castro Neto}}},\ }\href {\doibase 10.1209/0295-5075/84/38002} {\bibfield
  {journal} {\bibinfo  {journal} {EPL}\ }\textbf {\bibinfo {volume} {84}},\
  \bibinfo {pages} {38002} (\bibinfo {year} {2008})}\BibitemShut {NoStop}%
\bibitem [{\citenamefont {Zhang}\ \emph {et~al.}(2008)\citenamefont {Zhang},
  \citenamefont {Li}, \citenamefont {Basov}, \citenamefont {Fogler},
  \citenamefont {Hao},\ and\ \citenamefont {Martin}}]{Zhang2008}%
  \BibitemOpen
  \bibfield  {author} {\bibinfo {author} {\bibfnamefont {L.~M.}\ \bibnamefont
  {Zhang}}, \bibinfo {author} {\bibfnamefont {Z.~Q.}\ \bibnamefont {Li}},
  \bibinfo {author} {\bibfnamefont {D.~N.}\ \bibnamefont {Basov}}, \bibinfo
  {author} {\bibfnamefont {M.~M.}\ \bibnamefont {Fogler}}, \bibinfo {author}
  {\bibfnamefont {Z.}~\bibnamefont {Hao}}, \ and\ \bibinfo {author}
  {\bibfnamefont {M.~C.}\ \bibnamefont {Martin}},\ }\href {\doibase
  10.1103/PhysRevB.78.235408} {\bibfield  {journal} {\bibinfo  {journal} {Phys.
  Rev. B}\ }\textbf {\bibinfo {volume} {78}},\ \bibinfo {pages} {235408}
  (\bibinfo {year} {2008})}\BibitemShut {NoStop}%
\bibitem [{\citenamefont {Hipolito}\ \emph {et~al.}(2016)\citenamefont
  {Hipolito}, \citenamefont {Pedersen},\ and\ \citenamefont
  {Pereira}}]{Hipolito2016}%
  \BibitemOpen
  \bibfield  {author} {\bibinfo {author} {\bibfnamefont {F.}~\bibnamefont
  {Hipolito}}, \bibinfo {author} {\bibfnamefont {T.~G.}\ \bibnamefont
  {Pedersen}}, \ and\ \bibinfo {author} {\bibfnamefont {V.~M.}\ \bibnamefont
  {Pereira}},\ }\href {\doibase 10.1103/PhysRevB.94.045434} {\bibfield
  {journal} {\bibinfo  {journal} {Phys. Rev. B}\ }\textbf {\bibinfo {volume}
  {94}},\ \bibinfo {pages} {045434} (\bibinfo {year} {2016})}\BibitemShut
  {NoStop}%
\bibitem [{\citenamefont {Cheng}\ \emph {et~al.}(2015)\citenamefont {Cheng},
  \citenamefont {Vermeulen},\ and\ \citenamefont {Sipe}}]{Cheng2015b}%
  \BibitemOpen
  \bibfield  {author} {\bibinfo {author} {\bibfnamefont {J.~L.}\ \bibnamefont
  {Cheng}}, \bibinfo {author} {\bibfnamefont {N.}~\bibnamefont {Vermeulen}}, \
  and\ \bibinfo {author} {\bibfnamefont {J.~E.}\ \bibnamefont {Sipe}},\ }\href
  {\doibase 10.1103/PhysRevB.91.235320} {\bibfield  {journal} {\bibinfo
  {journal} {Phys. Rev. B}\ }\textbf {\bibinfo {volume} {91}},\ \bibinfo
  {pages} {235320} (\bibinfo {year} {2015})}\BibitemShut {NoStop}%
\bibitem [{\citenamefont {Mikhailov}(2016)}]{Mikhailov2016}%
  \BibitemOpen
  \bibfield  {author} {\bibinfo {author} {\bibfnamefont {S.~A.}\ \bibnamefont
  {Mikhailov}},\ }\href {\doibase 10.1103/PhysRevB.93.085403} {\bibfield
  {journal} {\bibinfo  {journal} {Phys. Rev. B}\ }\textbf {\bibinfo {volume}
  {93}},\ \bibinfo {pages} {085403} (\bibinfo {year} {2016})}\BibitemShut
  {NoStop}%
\bibitem [{\citenamefont {Mikhailov}(2017)}]{Mikhailov2017}%
  \BibitemOpen
  \bibfield  {author} {\bibinfo {author} {\bibfnamefont {S.~A.}\ \bibnamefont
  {Mikhailov}},\ }\href {\doibase 10.1103/PhysRevB.95.085432} {\bibfield
  {journal} {\bibinfo  {journal} {Phys. Rev. B}\ }\textbf {\bibinfo {volume}
  {95}},\ \bibinfo {pages} {085432} (\bibinfo {year} {2017})}\BibitemShut
  {NoStop}%
\bibitem [{\citenamefont {Ventura}\ \emph {et~al.}(2017)\citenamefont
  {Ventura}, \citenamefont {Passos}, \citenamefont {Lopes~dos Santos},
  \citenamefont {Lopes Viana~Parente},\ and\ \citenamefont
  {Peres}}]{Ventura2017}%
  \BibitemOpen
  \bibfield  {author} {\bibinfo {author} {\bibfnamefont {G.~B.}\ \bibnamefont
  {Ventura}}, \bibinfo {author} {\bibfnamefont {D.~J.}\ \bibnamefont {Passos}},
  \bibinfo {author} {\bibfnamefont {J.~M.~B.}\ \bibnamefont {Lopes~dos
  Santos}}, \bibinfo {author} {\bibfnamefont {J.~M.}\ \bibnamefont {Lopes
  Viana~Parente}}, \ and\ \bibinfo {author} {\bibfnamefont {N.~M.~R.}\
  \bibnamefont {Peres}},\ }\href {\doibase 10.1103/PhysRevB.96.035431}
  {\bibfield  {journal} {\bibinfo  {journal} {Phys. Rev. B}\ }\textbf {\bibinfo
  {volume} {96}},\ \bibinfo {pages} {035431} (\bibinfo {year}
  {2017})}\BibitemShut {NoStop}%
\bibitem [{\citenamefont {Dimitrovski}\ \emph {et~al.}(2017)\citenamefont
  {Dimitrovski}, \citenamefont {Madsen},\ and\ \citenamefont
  {Pedersen}}]{Dimitrovski2017a}%
  \BibitemOpen
  \bibfield  {author} {\bibinfo {author} {\bibfnamefont {D.}~\bibnamefont
  {Dimitrovski}}, \bibinfo {author} {\bibfnamefont {L.~B.}\ \bibnamefont
  {Madsen}}, \ and\ \bibinfo {author} {\bibfnamefont {T.~G.}\ \bibnamefont
  {Pedersen}},\ }\href {\doibase 10.1103/PhysRevB.95.035405} {\bibfield
  {journal} {\bibinfo  {journal} {Phys. Rev. B}\ }\textbf {\bibinfo {volume}
  {95}},\ \bibinfo {pages} {035405} (\bibinfo {year} {2017})}\BibitemShut
  {NoStop}%
\bibitem [{\citenamefont {McGouran}\ and\ \citenamefont
  {Dignam}(2017)}]{McGouran2017}%
  \BibitemOpen
  \bibfield  {author} {\bibinfo {author} {\bibfnamefont {R.}~\bibnamefont
  {McGouran}}\ and\ \bibinfo {author} {\bibfnamefont {M.~M.}\ \bibnamefont
  {Dignam}},\ }\href {\doibase 10.1103/PhysRevB.96.045439} {\bibfield
  {journal} {\bibinfo  {journal} {Phys. Rev. B}\ }\textbf {\bibinfo {volume}
  {96}},\ \bibinfo {pages} {045439} (\bibinfo {year} {2017})}\BibitemShut
  {NoStop}%
\bibitem [{\citenamefont {Mucha-Kruczy{\'{n}}ski}\ \emph
  {et~al.}(2008)\citenamefont {Mucha-Kruczy{\'{n}}ski}, \citenamefont
  {Tsyplyatyev}, \citenamefont {Grishin}, \citenamefont {McCann}, \citenamefont
  {Fal'ko}, \citenamefont {Bostwick},\ and\ \citenamefont
  {Rotenberg}}]{Mucha-Kruczynski2008}%
  \BibitemOpen
  \bibfield  {author} {\bibinfo {author} {\bibfnamefont {M.}~\bibnamefont
  {Mucha-Kruczy{\'{n}}ski}}, \bibinfo {author} {\bibfnamefont {O.}~\bibnamefont
  {Tsyplyatyev}}, \bibinfo {author} {\bibfnamefont {A.}~\bibnamefont
  {Grishin}}, \bibinfo {author} {\bibfnamefont {E.}~\bibnamefont {McCann}},
  \bibinfo {author} {\bibfnamefont {V.~I.}\ \bibnamefont {Fal'ko}}, \bibinfo
  {author} {\bibfnamefont {A.}~\bibnamefont {Bostwick}}, \ and\ \bibinfo
  {author} {\bibfnamefont {E.}~\bibnamefont {Rotenberg}},\ }\href {\doibase
  10.1103/PhysRevB.77.195403} {\bibfield  {journal} {\bibinfo  {journal} {Phys.
  Rev. B}\ }\textbf {\bibinfo {volume} {77}},\ \bibinfo {pages} {195403}
  (\bibinfo {year} {2008})}\BibitemShut {NoStop}%
\bibitem [{\citenamefont {Gr{\"{u}}neis}\ \emph {et~al.}(2008)\citenamefont
  {Gr{\"{u}}neis}, \citenamefont {Attaccalite}, \citenamefont {Wirtz},
  \citenamefont {Shiozawa}, \citenamefont {Saito}, \citenamefont {Pichler},\
  and\ \citenamefont {Rubio}}]{Gruneis2008}%
  \BibitemOpen
  \bibfield  {author} {\bibinfo {author} {\bibfnamefont {A.}~\bibnamefont
  {Gr{\"{u}}neis}}, \bibinfo {author} {\bibfnamefont {C.}~\bibnamefont
  {Attaccalite}}, \bibinfo {author} {\bibfnamefont {L.}~\bibnamefont {Wirtz}},
  \bibinfo {author} {\bibfnamefont {H.}~\bibnamefont {Shiozawa}}, \bibinfo
  {author} {\bibfnamefont {R.}~\bibnamefont {Saito}}, \bibinfo {author}
  {\bibfnamefont {T.}~\bibnamefont {Pichler}}, \ and\ \bibinfo {author}
  {\bibfnamefont {A.}~\bibnamefont {Rubio}},\ }\href {\doibase
  10.1103/PhysRevB.78.205425} {\bibfield  {journal} {\bibinfo  {journal} {Phys.
  Rev. B}\ }\textbf {\bibinfo {volume} {78}},\ \bibinfo {pages} {205425}
  (\bibinfo {year} {2008})}\BibitemShut {NoStop}%
\bibitem [{\citenamefont {Kuzmenko}\ \emph
  {et~al.}(2009{\natexlab{a}})\citenamefont {Kuzmenko}, \citenamefont
  {Crassee}, \citenamefont {van~der Marel}, \citenamefont {Blake},\ and\
  \citenamefont {Novoselov}}]{Kuzmenko2009a}%
  \BibitemOpen
  \bibfield  {author} {\bibinfo {author} {\bibfnamefont {A.~B.}\ \bibnamefont
  {Kuzmenko}}, \bibinfo {author} {\bibfnamefont {I.}~\bibnamefont {Crassee}},
  \bibinfo {author} {\bibfnamefont {D.}~\bibnamefont {van~der Marel}}, \bibinfo
  {author} {\bibfnamefont {P.}~\bibnamefont {Blake}}, \ and\ \bibinfo {author}
  {\bibfnamefont {K.~S.}\ \bibnamefont {Novoselov}},\ }\href {\doibase
  10.1103/PhysRevB.80.165406} {\bibfield  {journal} {\bibinfo  {journal} {Phys.
  Rev. B}\ }\textbf {\bibinfo {volume} {80}},\ \bibinfo {pages} {165406}
  (\bibinfo {year} {2009}{\natexlab{a}})}\BibitemShut {NoStop}%
\bibitem [{\citenamefont {Partoens}\ and\ \citenamefont
  {Peeters}(2006)}]{Partoens2006}%
  \BibitemOpen
  \bibfield  {author} {\bibinfo {author} {\bibfnamefont {B.}~\bibnamefont
  {Partoens}}\ and\ \bibinfo {author} {\bibfnamefont {F.~M.}\ \bibnamefont
  {Peeters}},\ }\href {\doibase 10.1103/PhysRevB.74.075404} {\bibfield
  {journal} {\bibinfo  {journal} {Phys. Rev. B}\ }\textbf {\bibinfo {volume}
  {74}},\ \bibinfo {pages} {075404} (\bibinfo {year} {2006})}\BibitemShut
  {NoStop}%
\bibitem [{\citenamefont {McCann}\ and\ \citenamefont
  {Fal'ko}(2006)}]{McCann2006}%
  \BibitemOpen
  \bibfield  {author} {\bibinfo {author} {\bibfnamefont {E.}~\bibnamefont
  {McCann}}\ and\ \bibinfo {author} {\bibfnamefont {V.~I.}\ \bibnamefont
  {Fal'ko}},\ }\href {\doibase 10.1103/PhysRevLett.96.086805} {\bibfield
  {journal} {\bibinfo  {journal} {Phys. Rev. Lett.}\ }\textbf {\bibinfo
  {volume} {96}},\ \bibinfo {pages} {086805} (\bibinfo {year}
  {2006})}\BibitemShut {NoStop}%
\bibitem [{\citenamefont {Malard}\ \emph {et~al.}(2007)\citenamefont {Malard},
  \citenamefont {Nilsson}, \citenamefont {Elias}, \citenamefont {Brant},
  \citenamefont {Plentz}, \citenamefont {Alves}, \citenamefont {{Castro
  Neto}},\ and\ \citenamefont {Pimenta}}]{Malard2007}%
  \BibitemOpen
  \bibfield  {author} {\bibinfo {author} {\bibfnamefont {L.~M.}\ \bibnamefont
  {Malard}}, \bibinfo {author} {\bibfnamefont {J.}~\bibnamefont {Nilsson}},
  \bibinfo {author} {\bibfnamefont {D.~C.}\ \bibnamefont {Elias}}, \bibinfo
  {author} {\bibfnamefont {J.~C.}\ \bibnamefont {Brant}}, \bibinfo {author}
  {\bibfnamefont {F.}~\bibnamefont {Plentz}}, \bibinfo {author} {\bibfnamefont
  {E.~S.}\ \bibnamefont {Alves}}, \bibinfo {author} {\bibfnamefont {A.~H.}\
  \bibnamefont {{Castro Neto}}}, \ and\ \bibinfo {author} {\bibfnamefont
  {M.~A.}\ \bibnamefont {Pimenta}},\ }\href {\doibase
  10.1103/PhysRevB.76.201401} {\bibfield  {journal} {\bibinfo  {journal} {Phys.
  Rev. B}\ }\textbf {\bibinfo {volume} {76}},\ \bibinfo {pages} {201401}
  (\bibinfo {year} {2007})}\BibitemShut {NoStop}%
\bibitem [{\citenamefont {Kuzmenko}\ \emph
  {et~al.}(2009{\natexlab{b}})\citenamefont {Kuzmenko}, \citenamefont {van
  Heumen}, \citenamefont {van~der Marel}, \citenamefont {Lerch}, \citenamefont
  {Blake}, \citenamefont {Novoselov},\ and\ \citenamefont
  {Geim}}]{Kuzmenko2009}%
  \BibitemOpen
  \bibfield  {author} {\bibinfo {author} {\bibfnamefont {A.~B.}\ \bibnamefont
  {Kuzmenko}}, \bibinfo {author} {\bibfnamefont {E.}~\bibnamefont {van
  Heumen}}, \bibinfo {author} {\bibfnamefont {D.}~\bibnamefont {van~der
  Marel}}, \bibinfo {author} {\bibfnamefont {P.}~\bibnamefont {Lerch}},
  \bibinfo {author} {\bibfnamefont {P.}~\bibnamefont {Blake}}, \bibinfo
  {author} {\bibfnamefont {K.~S.}\ \bibnamefont {Novoselov}}, \ and\ \bibinfo
  {author} {\bibfnamefont {A.~K.}\ \bibnamefont {Geim}},\ }\href {\doibase
  10.1103/PhysRevB.79.115441} {\bibfield  {journal} {\bibinfo  {journal} {Phys.
  Rev. B}\ }\textbf {\bibinfo {volume} {79}},\ \bibinfo {pages} {115441}
  (\bibinfo {year} {2009}{\natexlab{b}})}\BibitemShut {NoStop}%
\bibitem [{\citenamefont {Li}\ \emph {et~al.}(2009)\citenamefont {Li},
  \citenamefont {Henriksen}, \citenamefont {Jiang}, \citenamefont {Hao},
  \citenamefont {Martin}, \citenamefont {Kim}, \citenamefont {Stormer},\ and\
  \citenamefont {Basov}}]{Li2009}%
  \BibitemOpen
  \bibfield  {author} {\bibinfo {author} {\bibfnamefont {Z.~Q.}\ \bibnamefont
  {Li}}, \bibinfo {author} {\bibfnamefont {E.~A.}\ \bibnamefont {Henriksen}},
  \bibinfo {author} {\bibfnamefont {Z.}~\bibnamefont {Jiang}}, \bibinfo
  {author} {\bibfnamefont {Z.}~\bibnamefont {Hao}}, \bibinfo {author}
  {\bibfnamefont {M.~C.}\ \bibnamefont {Martin}}, \bibinfo {author}
  {\bibfnamefont {P.}~\bibnamefont {Kim}}, \bibinfo {author} {\bibfnamefont
  {H.~L.}\ \bibnamefont {Stormer}}, \ and\ \bibinfo {author} {\bibfnamefont
  {D.~N.}\ \bibnamefont {Basov}},\ }\href {\doibase
  10.1103/PhysRevLett.102.037403} {\bibfield  {journal} {\bibinfo  {journal}
  {Phys. Rev. Lett.}\ }\textbf {\bibinfo {volume} {102}},\ \bibinfo {pages}
  {037403} (\bibinfo {year} {2009})}\BibitemShut {NoStop}%
\bibitem [{\citenamefont {{Castro Neto}}\ \emph {et~al.}(2009)\citenamefont
  {{Castro Neto}}, \citenamefont {Guinea}, \citenamefont {Peres}, \citenamefont
  {Novoselov},\ and\ \citenamefont {Geim}}]{CastroNeto2009}%
  \BibitemOpen
  \bibfield  {author} {\bibinfo {author} {\bibfnamefont {A.~H.}\ \bibnamefont
  {{Castro Neto}}}, \bibinfo {author} {\bibfnamefont {F.}~\bibnamefont
  {Guinea}}, \bibinfo {author} {\bibfnamefont {N.~M.~R.}\ \bibnamefont
  {Peres}}, \bibinfo {author} {\bibfnamefont {K.~S.}\ \bibnamefont
  {Novoselov}}, \ and\ \bibinfo {author} {\bibfnamefont {A.~K.}\ \bibnamefont
  {Geim}},\ }\href {\doibase 10.1103/RevModPhys.81.109} {\bibfield  {journal}
  {\bibinfo  {journal} {Rev. Mod. Phys.}\ }\textbf {\bibinfo {volume} {81}},\
  \bibinfo {pages} {109} (\bibinfo {year} {2009})}\BibitemShut {NoStop}%
\bibitem [{\citenamefont {McCann}\ and\ \citenamefont
  {Koshino}(2013)}]{McCann2013}%
  \BibitemOpen
  \bibfield  {author} {\bibinfo {author} {\bibfnamefont {E.}~\bibnamefont
  {McCann}}\ and\ \bibinfo {author} {\bibfnamefont {M.}~\bibnamefont
  {Koshino}},\ }\href {\doibase 10.1088/0034-4885/76/5/056503} {\bibfield
  {journal} {\bibinfo  {journal} {Rep. Prog. Phys.}\ }\textbf {\bibinfo
  {volume} {76}},\ \bibinfo {pages} {056503} (\bibinfo {year}
  {2013})}\BibitemShut {NoStop}%
\bibitem [{\citenamefont {Castro}\ \emph {et~al.}(2006)\citenamefont {Castro},
  \citenamefont {Novoselov}, \citenamefont {Morozov}, \citenamefont {Peres},
  \citenamefont {dos Santos}, \citenamefont {Nilsson}, \citenamefont {Guinea},
  \citenamefont {Geim},\ and\ \citenamefont {Castro~Neto}}]{Castro2007}%
  \BibitemOpen
  \bibfield  {author} {\bibinfo {author} {\bibfnamefont {E.~V.}\ \bibnamefont
  {Castro}}, \bibinfo {author} {\bibfnamefont {K.~S.}\ \bibnamefont
  {Novoselov}}, \bibinfo {author} {\bibfnamefont {S.~V.}\ \bibnamefont
  {Morozov}}, \bibinfo {author} {\bibfnamefont {N.~M.~R.}\ \bibnamefont
  {Peres}}, \bibinfo {author} {\bibfnamefont {J.~M. B.~L.}\ \bibnamefont {dos
  Santos}}, \bibinfo {author} {\bibfnamefont {J.}~\bibnamefont {Nilsson}},
  \bibinfo {author} {\bibfnamefont {F.}~\bibnamefont {Guinea}}, \bibinfo
  {author} {\bibfnamefont {A.~K.}\ \bibnamefont {Geim}}, \ and\ \bibinfo
  {author} {\bibfnamefont {A.~H.}\ \bibnamefont {Castro~Neto}},\ }\href
  {\doibase 10.1103/PhysRevLett.99.216802} {\bibfield  {journal} {\bibinfo
  {journal} {Phys. Rev. Lett.}\ }\textbf {\bibinfo {volume} {99}},\ \bibinfo
  {pages} {216802} (\bibinfo {year} {2006})}\BibitemShut {NoStop}%
\bibitem [{\citenamefont {Xiao}\ \emph {et~al.}(2010)\citenamefont {Xiao},
  \citenamefont {Chang},\ and\ \citenamefont {Niu}}]{Xiao2010}%
  \BibitemOpen
  \bibfield  {author} {\bibinfo {author} {\bibfnamefont {D.}~\bibnamefont
  {Xiao}}, \bibinfo {author} {\bibfnamefont {M.-C.}\ \bibnamefont {Chang}}, \
  and\ \bibinfo {author} {\bibfnamefont {Q.}~\bibnamefont {Niu}},\ }\href
  {\doibase 10.1103/RevModPhys.82.1959} {\bibfield  {journal} {\bibinfo
  {journal} {Rev. Mod. Phys.}\ }\textbf {\bibinfo {volume} {82}},\ \bibinfo
  {pages} {1959} (\bibinfo {year} {2010})}\BibitemShut {NoStop}%
\bibitem [{\citenamefont {Hauss{\"{u}}hl}(2007)}]{Haussuhl2007}%
  \BibitemOpen
  \bibfield  {author} {\bibinfo {author} {\bibfnamefont {S.}~\bibnamefont
  {Hauss{\"{u}}hl}},\ }\href {\doibase 10.1002/9783527621156} {\emph {\bibinfo
  {title} {Physical Properties of Crystals}}}\ (\bibinfo  {publisher}
  {Wiley-VCH Verlag GmbH},\ \bibinfo {address} {Weinheim, Germany},\ \bibinfo
  {year} {2007})\BibitemShut {NoStop}%
\bibitem [{\citenamefont {Yang}\ and\ \citenamefont {Xie}(1995)}]{Yang1995}%
  \BibitemOpen
  \bibfield  {author} {\bibinfo {author} {\bibfnamefont {X.-L.}\ \bibnamefont
  {Yang}}\ and\ \bibinfo {author} {\bibfnamefont {S.-W.}\ \bibnamefont {Xie}},\
  }\href {\doibase 10.1364/AO.34.006130} {\bibfield  {journal} {\bibinfo
  {journal} {Appl. Opt.}\ }\textbf {\bibinfo {volume} {34}},\ \bibinfo {pages}
  {6130} (\bibinfo {year} {1995})}\BibitemShut {NoStop}%
\bibitem [{\citenamefont {Ashcroft}\ and\ \citenamefont
  {Mermin}(1976)}]{Ashcroft1976}%
  \BibitemOpen
  \bibfield  {author} {\bibinfo {author} {\bibfnamefont {N.~W.}\ \bibnamefont
  {Ashcroft}}\ and\ \bibinfo {author} {\bibfnamefont {N.~D.}\ \bibnamefont
  {Mermin}},\ }\href@noop {} {\emph {\bibinfo {title} {Solid state physics}}}\
  (\bibinfo  {publisher} {Thomson Learning, Inc.},\ \bibinfo {address}
  {London},\ \bibinfo {year} {1976})\BibitemShut {NoStop}%
\bibitem [{\citenamefont {Mahan}(2000)}]{Mahan2000}%
  \BibitemOpen
  \bibfield  {author} {\bibinfo {author} {\bibfnamefont {G.~D.}\ \bibnamefont
  {Mahan}},\ }\href {\doibase 10.1007/978-1-4757-5714-9} {\emph {\bibinfo
  {title} {Many-Particle Physics}}},\ \bibinfo {edition} {3rd}\ ed.\ (\bibinfo
  {publisher} {Springer US},\ \bibinfo {address} {Boston, MA},\ \bibinfo {year}
  {2000})\BibitemShut {NoStop}%
\bibitem [{\citenamefont {Marder}(2010)}]{Marder2010}%
  \BibitemOpen
  \bibfield  {author} {\bibinfo {author} {\bibfnamefont {M.~P.}\ \bibnamefont
  {Marder}},\ }\href {\doibase 10.1002/9780470949955} {\emph {\bibinfo {title}
  {Condensed Matter Physics}}},\ \bibinfo {edition} {2nd}\ ed.\ (\bibinfo
  {publisher} {John Wiley \& Sons, Inc.},\ \bibinfo {address} {Hoboken, NJ,
  USA},\ \bibinfo {year} {2010})\BibitemShut {NoStop}%
\bibitem [{\citenamefont {{Das Sarma}}\ \emph {et~al.}(2011)\citenamefont {{Das
  Sarma}}, \citenamefont {Adam}, \citenamefont {Hwang},\ and\ \citenamefont
  {Rossi}}]{DasSarma2011}%
  \BibitemOpen
  \bibfield  {author} {\bibinfo {author} {\bibfnamefont {S.}~\bibnamefont {{Das
  Sarma}}}, \bibinfo {author} {\bibfnamefont {S.}~\bibnamefont {Adam}},
  \bibinfo {author} {\bibfnamefont {E.~H.}\ \bibnamefont {Hwang}}, \ and\
  \bibinfo {author} {\bibfnamefont {E.}~\bibnamefont {Rossi}},\ }\href
  {\doibase 10.1103/RevModPhys.83.407} {\bibfield  {journal} {\bibinfo
  {journal} {Rev. Mod. Phys.}\ }\textbf {\bibinfo {volume} {83}},\ \bibinfo
  {pages} {407} (\bibinfo {year} {2011})}\BibitemShut {NoStop}%
\bibitem [{\citenamefont {S{\"{a}}yn{\"{a}}tjoki}\ \emph
  {et~al.}(2013)\citenamefont {S{\"{a}}yn{\"{a}}tjoki}, \citenamefont
  {Karvonen}, \citenamefont {Riikonen}, \citenamefont {Kim}, \citenamefont
  {Mehravar}, \citenamefont {Norwood}, \citenamefont {Peyghambarian},
  \citenamefont {Lipsanen},\ and\ \citenamefont {Kieu}}]{Saynatjoki2013}%
  \BibitemOpen
  \bibfield  {author} {\bibinfo {author} {\bibfnamefont {A.}~\bibnamefont
  {S{\"{a}}yn{\"{a}}tjoki}}, \bibinfo {author} {\bibfnamefont {L.}~\bibnamefont
  {Karvonen}}, \bibinfo {author} {\bibfnamefont {J.}~\bibnamefont {Riikonen}},
  \bibinfo {author} {\bibfnamefont {W.}~\bibnamefont {Kim}}, \bibinfo {author}
  {\bibfnamefont {S.}~\bibnamefont {Mehravar}}, \bibinfo {author}
  {\bibfnamefont {R.~A.}\ \bibnamefont {Norwood}}, \bibinfo {author}
  {\bibfnamefont {N.}~\bibnamefont {Peyghambarian}}, \bibinfo {author}
  {\bibfnamefont {H.}~\bibnamefont {Lipsanen}}, \ and\ \bibinfo {author}
  {\bibfnamefont {K.}~\bibnamefont {Kieu}},\ }\href {\doibase
  10.1021/nn4042909} {\bibfield  {journal} {\bibinfo  {journal} {ACS Nano}\
  }\textbf {\bibinfo {volume} {7}},\ \bibinfo {pages} {8441} (\bibinfo {year}
  {2013})}\BibitemShut {NoStop}%
\bibitem [{\citenamefont {Autere}\ \emph {et~al.}(2017)\citenamefont {Autere},
  \citenamefont {Ryder}, \citenamefont {S{\"{a}}yn{\"{a}}tjoki}, \citenamefont
  {Karvonen}, \citenamefont {Amirsolaimani}, \citenamefont {Norwood},
  \citenamefont {Peyghambarian}, \citenamefont {Kieu}, \citenamefont
  {Lipsanen}, \citenamefont {Hersam},\ and\ \citenamefont {Sun}}]{Autere2017}%
  \BibitemOpen
  \bibfield  {author} {\bibinfo {author} {\bibfnamefont {A.}~\bibnamefont
  {Autere}}, \bibinfo {author} {\bibfnamefont {C.~R.}\ \bibnamefont {Ryder}},
  \bibinfo {author} {\bibfnamefont {A.}~\bibnamefont {S{\"{a}}yn{\"{a}}tjoki}},
  \bibinfo {author} {\bibfnamefont {L.}~\bibnamefont {Karvonen}}, \bibinfo
  {author} {\bibfnamefont {B.}~\bibnamefont {Amirsolaimani}}, \bibinfo {author}
  {\bibfnamefont {R.~A.}\ \bibnamefont {Norwood}}, \bibinfo {author}
  {\bibfnamefont {N.}~\bibnamefont {Peyghambarian}}, \bibinfo {author}
  {\bibfnamefont {K.}~\bibnamefont {Kieu}}, \bibinfo {author} {\bibfnamefont
  {H.}~\bibnamefont {Lipsanen}}, \bibinfo {author} {\bibfnamefont {M.~C.}\
  \bibnamefont {Hersam}}, \ and\ \bibinfo {author} {\bibfnamefont
  {Z.}~\bibnamefont {Sun}},\ }\href {\doibase 10.1021/acs.jpclett.7b00140}
  {\bibfield  {journal} {\bibinfo  {journal} {J. Phys. Chem. Lett.}\ }\textbf
  {\bibinfo {volume} {8}},\ \bibinfo {pages} {1343} (\bibinfo {year}
  {2017})}\BibitemShut {NoStop}%
\bibitem [{\citenamefont {Alexander}\ \emph {et~al.}(2017)\citenamefont
  {Alexander}, \citenamefont {Savostianova}, \citenamefont {Mikhailov},
  \citenamefont {Kuyken},\ and\ \citenamefont {{Van
  Thourhout}}}]{Alexander2017}%
  \BibitemOpen
  \bibfield  {author} {\bibinfo {author} {\bibfnamefont {K.}~\bibnamefont
  {Alexander}}, \bibinfo {author} {\bibfnamefont {N.~A.}\ \bibnamefont
  {Savostianova}}, \bibinfo {author} {\bibfnamefont {S.~A.}\ \bibnamefont
  {Mikhailov}}, \bibinfo {author} {\bibfnamefont {B.}~\bibnamefont {Kuyken}}, \
  and\ \bibinfo {author} {\bibfnamefont {D.}~\bibnamefont {{Van Thourhout}}},\
  }\href {\doibase 10.1021/acsphotonics.7b00559} {\bibfield  {journal}
  {\bibinfo  {journal} {ACS Photon.}\ } (\bibinfo {year} {2017}),\
  10.1021/acsphotonics.7b00559}\BibitemShut {NoStop}%
\bibitem [{\citenamefont {Ruzicka}\ \emph {et~al.}(2010)\citenamefont
  {Ruzicka}, \citenamefont {Wang}, \citenamefont {Werake}, \citenamefont
  {Weintrub}, \citenamefont {Loh},\ and\ \citenamefont {Zhao}}]{Ruzicka2010}%
  \BibitemOpen
  \bibfield  {author} {\bibinfo {author} {\bibfnamefont {B.~A.}\ \bibnamefont
  {Ruzicka}}, \bibinfo {author} {\bibfnamefont {S.}~\bibnamefont {Wang}},
  \bibinfo {author} {\bibfnamefont {L.~K.}\ \bibnamefont {Werake}}, \bibinfo
  {author} {\bibfnamefont {B.}~\bibnamefont {Weintrub}}, \bibinfo {author}
  {\bibfnamefont {K.~P.}\ \bibnamefont {Loh}}, \ and\ \bibinfo {author}
  {\bibfnamefont {H.}~\bibnamefont {Zhao}},\ }\href {\doibase
  10.1103/PhysRevB.82.195414} {\bibfield  {journal} {\bibinfo  {journal} {Phys.
  Rev. B}\ }\textbf {\bibinfo {volume} {82}},\ \bibinfo {pages} {195414}
  (\bibinfo {year} {2010})}\BibitemShut {NoStop}%
\bibitem [{\citenamefont {Tielrooij}\ \emph {et~al.}(2013)\citenamefont
  {Tielrooij}, \citenamefont {Song}, \citenamefont {Jensen}, \citenamefont
  {Centeno}, \citenamefont {Pesquera}, \citenamefont {{Zurutuza Elorza}},
  \citenamefont {Bonn}, \citenamefont {Levitov},\ and\ \citenamefont
  {Koppens}}]{Tielrooij2013}%
  \BibitemOpen
  \bibfield  {author} {\bibinfo {author} {\bibfnamefont {K.~J.}\ \bibnamefont
  {Tielrooij}}, \bibinfo {author} {\bibfnamefont {J.~C.~W.}\ \bibnamefont
  {Song}}, \bibinfo {author} {\bibfnamefont {S.~A.}\ \bibnamefont {Jensen}},
  \bibinfo {author} {\bibfnamefont {A.}~\bibnamefont {Centeno}}, \bibinfo
  {author} {\bibfnamefont {A.}~\bibnamefont {Pesquera}}, \bibinfo {author}
  {\bibfnamefont {A.}~\bibnamefont {{Zurutuza Elorza}}}, \bibinfo {author}
  {\bibfnamefont {M.}~\bibnamefont {Bonn}}, \bibinfo {author} {\bibfnamefont
  {L.~S.}\ \bibnamefont {Levitov}}, \ and\ \bibinfo {author} {\bibfnamefont
  {F.~H.~L.}\ \bibnamefont {Koppens}},\ }\href {\doibase 10.1038/nphys2564}
  {\bibfield  {journal} {\bibinfo  {journal} {Nat. Phys.}\ }\textbf {\bibinfo
  {volume} {9}},\ \bibinfo {pages} {248} (\bibinfo {year} {2013})}\BibitemShut
  {NoStop}%
\bibitem [{\citenamefont {Woodward}\ \emph {et~al.}(2016)\citenamefont
  {Woodward}, \citenamefont {Murray}, \citenamefont {Phelan}, \citenamefont
  {de~Oliveira}, \citenamefont {Runcorn}, \citenamefont {Kelleher},
  \citenamefont {Li}, \citenamefont {de~Oliveira}, \citenamefont {Fechine},
  \citenamefont {Eda},\ and\ \citenamefont {de~Matos}}]{Woodward2016}%
  \BibitemOpen
  \bibfield  {author} {\bibinfo {author} {\bibfnamefont {R.~I.}\ \bibnamefont
  {Woodward}}, \bibinfo {author} {\bibfnamefont {R.~T.}\ \bibnamefont
  {Murray}}, \bibinfo {author} {\bibfnamefont {C.~F.}\ \bibnamefont {Phelan}},
  \bibinfo {author} {\bibfnamefont {R.~E.~P.}\ \bibnamefont {de~Oliveira}},
  \bibinfo {author} {\bibfnamefont {T.~H.}\ \bibnamefont {Runcorn}}, \bibinfo
  {author} {\bibfnamefont {E.~J.~R.}\ \bibnamefont {Kelleher}}, \bibinfo
  {author} {\bibfnamefont {S.}~\bibnamefont {Li}}, \bibinfo {author}
  {\bibfnamefont {E.~C.}\ \bibnamefont {de~Oliveira}}, \bibinfo {author}
  {\bibfnamefont {G.~J.~M.}\ \bibnamefont {Fechine}}, \bibinfo {author}
  {\bibfnamefont {G.}~\bibnamefont {Eda}}, \ and\ \bibinfo {author}
  {\bibfnamefont {C.~J.~S.}\ \bibnamefont {de~Matos}},\ }\href {\doibase
  10.1088/2053-1583/4/1/011006} {\bibfield  {journal} {\bibinfo  {journal} {2D
  Mater.}\ }\textbf {\bibinfo {volume} {4}},\ \bibinfo {pages} {011006}
  (\bibinfo {year} {2016})}\BibitemShut {NoStop}%
\bibitem [{\citenamefont {Boyd}\ \emph {et~al.}(2014)\citenamefont {Boyd},
  \citenamefont {Shi},\ and\ \citenamefont {{De Leon}}}]{Boyd2014}%
  \BibitemOpen
  \bibfield  {author} {\bibinfo {author} {\bibfnamefont {R.~W.}\ \bibnamefont
  {Boyd}}, \bibinfo {author} {\bibfnamefont {Z.}~\bibnamefont {Shi}}, \ and\
  \bibinfo {author} {\bibfnamefont {I.}~\bibnamefont {{De Leon}}},\ }\href
  {\doibase 10.1016/j.optcom.2014.03.005} {\bibfield  {journal} {\bibinfo
  {journal} {Opt. Commun.}\ }\textbf {\bibinfo {volume} {326}},\ \bibinfo
  {pages} {74} (\bibinfo {year} {2014})}\BibitemShut {NoStop}%
\bibitem [{\citenamefont {Le}\ and\ \citenamefont {{Di Cecca}}(1991)}]{Le1991}%
  \BibitemOpen
  \bibfield  {author} {\bibinfo {author} {\bibfnamefont {H.~Q.}\ \bibnamefont
  {Le}}\ and\ \bibinfo {author} {\bibfnamefont {S.}~\bibnamefont {{Di
  Cecca}}},\ }\href {\doibase 10.1364/OL.16.000901} {\bibfield  {journal}
  {\bibinfo  {journal} {Opt. Lett.}\ }\textbf {\bibinfo {volume} {16}},\
  \bibinfo {pages} {901} (\bibinfo {year} {1991})}\BibitemShut {NoStop}%
\bibitem [{\citenamefont {Pu}\ \emph {et~al.}(2016)\citenamefont {Pu},
  \citenamefont {Ottaviano}, \citenamefont {Semenova},\ and\ \citenamefont
  {Yvind}}]{Pu2016a}%
  \BibitemOpen
  \bibfield  {author} {\bibinfo {author} {\bibfnamefont {M.}~\bibnamefont
  {Pu}}, \bibinfo {author} {\bibfnamefont {L.}~\bibnamefont {Ottaviano}},
  \bibinfo {author} {\bibfnamefont {E.}~\bibnamefont {Semenova}}, \ and\
  \bibinfo {author} {\bibfnamefont {K.}~\bibnamefont {Yvind}},\ }\href
  {\doibase 10.1364/OPTICA.3.000823} {\bibfield  {journal} {\bibinfo  {journal}
  {Optica}\ }\textbf {\bibinfo {volume} {3}},\ \bibinfo {pages} {823} (\bibinfo
  {year} {2016})}\BibitemShut {NoStop}%
\end{thebibliography}%

\end{document}